\documentclass[reprint, amsmath, amssymb,aps,prd,floatfix, eqsecnum, longbibliography]{revtex4-1}

\usepackage{amsmath, graphicx,natbib, bm ,array, subcaption, hyperref}
\hypersetup{colorlinks = true,allcolors = blue }

\newcolumntype{L}{>{$}l<{$}}
\newcolumntype{C}{>{$}c<{$}}
\newcolumntype{R}{>{$}r<{$}}

\DeclareMathOperator\sgn{sgn}
\DeclareMathOperator\sgnleft{\overleftarrow{\text{sgn}}}
\DeclareMathOperator\sgnright{\overrightarrow{\text{sgn}}}

\begin{document}

\newcommand{\pd}{\partial}
\newcommand{\beq}{\begin{equation}}
\newcommand{\eeq}{\end{equation}}
\newcommand{\bseq}{\begin{subequations}}
\newcommand{\eseq}{\end{subequations}}

\newcommand{\coeffI}{Z_I}
\newcommand{\coeffP}{Z_P}
\newcommand{\partition}[1]{\underline{\mathbf{#1}}} 
\newcommand{\pole}{z}
\newcommand{\gmax}{N_{\text{leads}}}
\newcommand{\fermifn}{f}

\title{Many-body wavefunctions for quantum impurities out of equilibrium. II. Charge fluctuations}

\author{Adrian B. Culver}
 \email[Present address: Mani L. Bhaumik Institute for Theoreti- cal Physics, Department of Physics and Astronomy, University of California, Los Angeles, California 90095, USA; ]{adrianculver@physics.ucla.edu}
\author{Natan Andrei}
 \email{natan@physics.rutgers.edu}
\affiliation{Center for Materials Theory, 
Department of Physics and Astronomy, Rutgers University, Piscataway, New Jersey 08854, USA
}

\begin{abstract}
    We extend the general formalism discussed in the previous paper [A. B. Culver and N. Andrei, \href{https://doi.org/10.1103/PhysRevB.103.195106}{Phys. Rev. B {\bf 103}, 195106 (2021)}]  to two models with charge fluctuations: the interacting resonant level model and the Anderson impurity model.  In the interacting resonant level model, we find the exact time-evolving wavefunction and calculate the steady state impurity occupancy to leading order in the interaction.  In the Anderson impurity model, we find the nonequilibrium steady state for small or large Coulomb repulsion $U$, and we find that the steady state current to leading order in $U$ agrees with a Keldysh perturbation theory calculation.
\end{abstract}

\maketitle

\section{Introduction}\label{sec: Introduction}

A quench, or sudden change in a system's Hamiltonian, is a useful way of probing nonequilibrium physics, in particular the  nonequilibrium steady state that may occur in the long-time limit (with the system size taken to infinity first).  In this paper we extend our wavefunction formalism for quench dynamics and nonequilibrium steady states \cite{CulverAndrei_PRB1} to quantum impurity models with charge fluctuations, focusing on the interacting resonant level model and the Anderson impurity model. In the former case, we find the exact time-evolving wavefunction; in the latter case, we find the nonequilibrium steady state for large or small Coulomb repulsion.  We then use these wavefunctions to compute some physical quantities.  This computation leads to complex mathematical expressions which require us to expand in some parameter to make them accessible in the thermodynamic limit.

Let us recall the basic setup of our quench, as described in our previous paper \cite{CulverAndrei_PRB1}.  The system consists of a quantum impurity coupled to any number of leads (reservoirs of electrons), which are held at arbitrary temperatures and chemical potentials.  The leads themselves are noninteracting; it is the coupling between the leads and the impurity that makes this a many-body problem.  Prior to $t=0$, the impurity is decoupled, and the system is in a very simple state: a Fermi sea in each lead filling up to the chemical potential (or more generally, a finite-temperature Fermi distribution in each lead).  The quench at $t=0$ consists of turning on the coupling between the impurity and the leads.

Previously, we used our formalism for calculating the many-body wavefunction and expectation values in the Kondo model, in which the quantum impurity has only a fluctuating spin.  Here we present an extension of our formalism to models in which both the spin and the charge of the impurity can fluctuate.  We set up a general formalism, focusing in particular on two models.  The first is the interacting resonant level model (IRL), in which the impurity is a spinless fermion $d^\dagger$ that has tunneling and Coulomb interaction with any number of leads:
\begin{multline}
    H_{\text{IRL}} = - i \int_{-L/2}^{L/2}dx\ \sum_{\gamma=1}^{\gmax} \psi_{\gamma}^\dagger(x) \frac{d}{dx} \psi_{\gamma}(x) + \epsilon d^\dagger d \\
    + \sum_{\gamma=1}^{\gmax}\left[\frac{v}{\sqrt{\gmax}} \psi_{\gamma}^\dagger(0) d + \text{ h.c.}\right]\\
    + U\sum_{\gamma=1}^{\gmax} \psi_{\gamma}^\dagger(0)\psi_{\gamma}(0) d^\dagger d.\label{eq: H multilead IRL}
\end{multline}
With a view towards universal low energy physics, we have followed the usual steps of taking the wide-band limit and ``unfolding'' the leads, resulting in a one-dimensional model with linear dispersion.  We have also assumed equal tunneling and Coulomb interaction strength for each lead.

The second model we focus on is the Anderson impurity model (AIM), again with any number of leads with equal tunneling to the dot:
\begin{multline}
    H_{\text{AIM}} = -i \int_{-L/2}^{L/2 } dx\ \sum_{\gamma=1}^{\gmax} \psi_{\gamma a}^\dagger(x) \frac{d}{dx}\psi_{\gamma a}(x) + \epsilon d_a^\dagger d_a \\
    + \sum_{\gamma=1}^{\gmax} \left[\frac{v}{\sqrt{\gmax}} \psi_{\gamma a}^\dagger(0) d_a + \text{ h.c.} \right]+ U n_{\uparrow} n_{\downarrow}.\label{eq: H multilead AIM}
\end{multline}
In this case, the impurity is a single spin-$1/2$ fermion $d_a^\dagger$ with a Coulomb energy cost $U$ to having both spins present.  We again are considering the wide-band limit.

The nonequilibrium physics of these models has been studied by a great variety of approaches, usually in the case $\gmax=2$ that is most relevant to transport experiments.  The IRL has been studied by, for instance, a nonequilibrium version of the Bethe ansatz \cite{MehtaAndrei, *MehtaEtAl_erratum}; perturbative, NRG, and Anderson-Yuval Coulomb gas methods \cite{BordaEtAl}; Hershfield density matrix \cite{Doyon}; conformal field theory and integrability \cite{BoulatSaleur}; time-dependent DMRG and integrability \cite{BoulatSaleurSchmitteckert}; and functional renormalization group and real-time renormalization group in frequency space \cite{KarraschEtAl}. There is still more literature on the AIM out of equilibrium; the reader may see the references in Ref. \cite{AshidaEtAl} for an extensive list (that also includes work on the related nonequilibrium Kondo problem).

In the quench setup we consider, the first challenge is to find the many-body wavefunction following the quench.  One of our main results is the exact solution of this problem in the case of the multilead IRL.  We show that in the long-time limit, the time-evolving wavefunction becomes a nonequilibrium steady state (NESS): a solution of the time-independent Schrodinger equation with the boundary condition of incoming plane waves (the Fermi seas of the leads).  In addition, the time-independent version of our formalism yields this NESS directly, without following the quench dynamics.  In the AIM, we use the time-independent formalism to find the NESS in the limits of small $U$ (which we use as a check, comparing with Keldysh perturbation theory) and $U\to\infty$.

NESS wavefunctions for the IRL and AIM have previously been obtained by Nishino and collaborators \cite{NishinoEtAl_PRLonIRL, NishinoEtAl_PRBonIRL,NishinoHatanoOrdonez,ImamuraEtAl}; our more general approach recovers some of their results in special cases.  We discuss these special cases in more detail below.

We emphasize that the NESS wavefunctions in this paper differ in an essential way from Bethe ansatz wavefunctions.  The key point is that Bethe ansatz wavefunctions are well-suited to quantization on a ring with periodic boundary conditions, which is most natural for equilibrium problems (one can enumerate states and calculate the partition function).  In contrast, the NESS wavefunctions in this paper are simple on the ``incoming'' side $(x<0)$ and complicated on the ``outgoing'' side ($x>0$)---they are scattering ``in'' states.  These wavefunctions permit the evaluation of observables directly in steady state nonequilibrium, without the need to follow the real-time dynamics that establish the steady state.

The paper is organized as follows.  In Sec. \ref{sec: Wavefunction formalism for charge fluctuations}, we present our wavefunction formalism for models with charge fluctuations.  This presentation begins with the noninteracting resonant level model as a warmup, then proceeds to the one-lead IRL as the first nontrivial application of our approach.  Then, the approach is presented in a more formal and general way, in both time-dependent and time-independent forms, and results are presented for the multilead IRL and the multilead AIM.  In Sec. \ref{sec: Evaluation of observables}, we use our wavefunctions to calculate observables.  We calculate the impurity occupancy in the IRL at leading order in $U$, verifying that the steady state equilibrium answer agrees with a calculation in the literature and presenting some new results in steady state nonequilibrium.  We also calculate the steady-state current in the two-lead AIM for small $U$ (obtaining an answer that we have verified with Keldysh perturbation theory) and for $U\to\infty$ with small $\Delta \equiv \frac{1}{2}|v|^2$ (recovering a scaling law well known from the equilibrium case).  We conclude in Sec. \ref{sec: Conclusion and outlook} with a summary and outlook.

\section{Wavefunction formalism for charge fluctuations}\label{sec: Wavefunction formalism for charge fluctuations}

We present a reformulation of the many-body Schrodinger equation (time-dependent or time-independent) that allows us to calculate wavefunctions in the IRL and AIM.  Our formalism takes care of much of the combinatorial bookkeeping involved in solving for an $N$-body wavefunction in order to isolate the hard part of the interacting problem, which we find is a certain family of differential equations that we call ``inverse problems.''  The equivalence of the many-body Schrodinger equation to these inverse problems holds under fairly general conditions; in some one-dimensional quantum impurity models with linear spectrum, the inverse problems can be solved in closed form.

We present our formalism first in a specific example: the one-lead IRL.  We warm up in Sec. \ref{sec: Noninteracting case: The resonant level model} with the noninteracting resonant level, which provides a starting point for our calculations in both the IRL and the AIM.  In Sec. \ref{sec: Time-evolving wavefunction of the one-lead IRL}, we present the time-evolving wavefunction of the one-lead IRL.  This example motivates the more general formalism for time evolution that we set up in Sec. \ref{sec: General formalism}; we also give a brief account of the time-independent version of the formalism in Sec. \ref{sec: Time-independent formalism}.  In Sec. \ref{sec: Time-evolving wavefunction of the multilead IRL}, we present the time-evolving wavefunction of the multilead IRL, and in Sec. \ref{sec: NESS of the multilead AIM for small or large U} we present the NESS of the multilead AIM for small or large $U$.  

\subsection{Noninteracting case: The resonant level model}\label{sec: Noninteracting case: The resonant level model}

We consider first the one-lead RLM:
\begin{multline}
    H^{(0)} = - i \int_{-L/2}^{L/2}dx\ \psi^\dagger(x) \frac{d}{dx} \psi(x) + \epsilon d^\dagger d \\
    +\left[ v \psi^\dagger(0) d + \text{ h.c.}\right].\label{eq: one-lead RLM}
\end{multline}
We use the following notation throughout the rest of the paper:
\beq
    \Delta = \frac{1}{2}|v|^2,\ \pole = \epsilon - i\Delta,\ \mathcal{T}(k) = \frac{2\Delta}{k-z}. \label{eq: RLM basic notation}
\eeq
We begin by defining the time evolution of the momentum creation operators $c_k^\dagger\equiv \frac{1}{\sqrt{L}}\int_{-L/2}^{L/2}dx\ e^{i k x} \psi^\dagger(x)$ as follows:
\beq
    c_k^\dagger(t) \equiv e^{-i H^{(0)} t } c_k^\dagger e^{i H^{(0)} t}. 
\eeq
The point is that, since $H^{(0)}$ annihilates the empty state $|0\rangle$, the time evolution of an initial state with arbitrary momenta is given by $e^{-i H^{(0)} t} c_{k_N}^\dagger \dots c_{k_1}^\dagger |0\rangle =c_{k_N}^\dagger(t) \dots c_{k_1}^\dagger(t) |0\rangle $.  This same approach was used by Gurvitz in noninteracting Floquet models \cite{Gurvitz}; our approach will be to use the $c_k^\dagger(t)$ operators as a basic ingredient in constructing the wavefunction in an interacting model.  We emphasize that our calculation is done in the Schrodinger picture. 

A straightforward calculation yields the following explicit form in the regime of interest ($0\le t < L/2$):
\begin{multline}
    c_k^\dagger(t) = e^{-i k t}c_k^\dagger
    + \frac{1}{\sqrt{L}}\int dx\ F_k(t-x)\\
    \times \left( \Theta(0<x<t) \psi^\dagger(x) + \frac{i}{v} \delta(x) d^\dagger \right),\label{eq: ckdagger(t) RLM}
\end{multline}
where
\beq
    F_k(t) =  - i  \mathcal{T}(k)\left( e^{-ik t} - e^{-i \pole t} \right).\label{eq: Fk RLM}
\eeq
Let us make one comment on this solution.  Due to the linearity of the spectrum, the wavefunction is discontinuous at $x=0$, and one therefore needs a prescription to make sense of $\delta(x)$ multiplying a discontinuous function.  Here and in all wavefunction calculations in this paper, the prescription we use is to average the two limits of the discontinuous function as $x\to 0^\pm$.  This has the effect of replacing, e.g., $\Theta(x)\delta(x) \to \frac{1}{2} \delta(x)$.  This prescription has been used successfully in equilibrium calculations with the Bethe ansatz \cite{AndreiFuruyaLowenstein}.

In the infinite time limit, the $c_k^\dagger(t)$ operators create scattering ``in'' states: that is, states with an incoming plane wave ($e^{i kx}$ for $x<0$).  This infinite time limit must be taken in a particular sense, removing a trivial overall phase factor and taking the limit \emph{pointwise}: A limit is reached at each point $x$ but not uniformly for all $x$.  We send $L\to\infty$ before $t\to\infty$, removing the prefactor $1/\sqrt{L}$ to convert from Kronecker delta normalization to Dirac delta normalization.  The result is
\bseq
\begin{align}
    &c_{k,\text{in}}^\dagger \equiv \lim_{t\to\infty} e^{i k t} \lim_{L\to\infty} \sqrt{L} \Biggr( \int dx\ \{ \psi(x), c_k^\dagger(t) \} \psi^\dagger(x) \notag \\
    &\qquad\qquad\qquad\qquad\qquad\qquad\qquad+\{ d ,c_k^\dagger(t) \} d^\dagger \Biggr) \\
    &=c_k^\dagger + \int dx\ F_{k,\text{in}}(x) \left( \Theta(0<x) \psi^\dagger(x) + \frac{i}{v}\delta(x)d^\dagger\right),\label{eq: ckindagger RLM}
\end{align}
\eseq
where $c_k^\dagger=\int dx\ e^{ikx} \psi^\dagger(x)$ in the second line (i.e., Dirac normalized), and
\beq
    F_{k,\text{in}}(x) = -i \mathcal{T}(k) e^{i k x}.
\eeq
From the electron part of the wavefunction $c_{k,\text{in}}^\dagger|0\rangle$, we can see that $\mathcal{T}(k)$ is the bare $\mathcal{T}$ matrix for a single electron crossing the impurity.  The corresponding bare $\mathcal{S}$ matrix is $\mathcal{S}= 1 - i\mathcal{T}(k) = \frac{k -\epsilon - i\Delta}{k-\epsilon + i \Delta}$, in agreement with Bethe ansatz.

We proceed to the simplest multilead RLM, in which an arbitrary number of leads indexed by $\gamma = 1,\dots, \gmax$ all tunnel to the dot with the same tunneling coefficient:
\begin{multline}
    H^{(0)} = - i \int_{-L/2}^{L/2}dx\ \sum_{\gamma=1}^{\gmax} \psi_{\gamma}^\dagger(x) \frac{d}{dx} \psi_{\gamma}(x) + \epsilon d^\dagger d \\
    +\sum_{\gamma=1}^{\gmax}\left[  \frac{v}{\sqrt{\gmax}} \psi_{\gamma}^\dagger(0) d + \text{ h.c.}\right].
\end{multline}
After a unitary rotation, the Hamiltonian separates into $\gmax-1$ free fermion fields and a copy of the RLM, with the latter involving the ``even'' combination $c_{ek} = \frac{1}{\sqrt{\gmax}}\sum_{\beta=1}^{\gmax}c_{\beta k}$.  The time evolution of operators is straightforward in this rotated basis, seeing as the free fermion fields evolve by phase factors and the RLM field evolves according to Eq. \eqref{eq: ckdagger(t) RLM}.  Rotating back to the original basis, we obtain
\begin{multline}
    c_{\gamma k}^\dagger(t) = e^{-i k t}c_{\gamma k}^\dagger + \frac{1}{\gmax\sqrt{L}}\int dx\ F_k(t-x) \\
    \times \left( \Theta(0<x<t)
    \sum_{\beta=1}^{\gmax} \psi_{\beta}^\dagger(x) + \frac{i\sqrt{\gmax}}{v} \delta(x) d^\dagger \right),\label{eq: ckdagger(t) mcRLM}
\end{multline}
where $c_{\gamma k}^\dagger(t) =\frac{1}{\sqrt{L}}\int_{-L/2}^{L/2} dx\ e^{i kx} \psi_{\gamma}^\dagger(x)$.  Taking the long-time limit in the same way as in the one-lead case yields
\begin{multline}
    c_{\gamma k,\text{in}}^\dagger = c_{\gamma k}^\dagger + \frac{1}{\gmax}\int dx\ F_{k,\text{in} }(x) \\
    \times \left( \Theta(0<x)
    \sum_{\beta=1}^{\gmax} \psi_{\beta}^\dagger(x) + \frac{i\sqrt{\gmax}}{v} \delta(x) d^\dagger \right),\label{eq: cgammakindagger mcRLM}
\end{multline}
where the momentum creation operators here are Dirac normalized [$c_{\gamma k}^\dagger =\int dx\ e^{i kx} \psi_{\gamma}^\dagger(x)$].

\subsection{Time-evolving wavefunction of the one-lead IRL}\label{sec: Time-evolving wavefunction of the one-lead IRL}

To the one-lead resonant level model Hamiltonian $H^{(0)}$ of the previous section, we add a Coulomb interaction between the dot and the charge density at $x=0$ to arrive at the one-lead IRL:
\bseq
    \begin{align}
        H^{(1)} &= U \psi^\dagger(0)\psi(0)d^\dagger d,\\
        H &= H^{(0)} + H^{(1)}.
    \end{align}
\eseq
We present the exact time-evolving wavefunction of this model given an initial state $c_{k_N}^\dagger \dots c_{k_1}^\dagger|0\rangle$ with arbitrary momenta.  We use this model to illustrate a more general method which is detailed in the next section.

We wish to find the following time-dependent wavefunction:
\beq
    |\Psi(t) \rangle \equiv e^{-i H t} \left( \prod_{j=1}^N c_{k_j}^\dagger \right) |0 \rangle.
\eeq
Equivalently, we need to solve the differential equation
\beq
    \left( H - i \frac{d}{dt} \right)|\Psi(t) \rangle =0,
\eeq
with the initial condition
\beq
    |\Psi(t=0) \rangle = c_{k_N}^\dagger \dots c_{k_1}^\dagger |0 \rangle.\label{eq: initial condition IRL}
\eeq
In the noninteracting case ($U=0$), the full solution is given by a product of the time-dependent creation operators of the RLM [Eq. \eqref{eq: ckdagger(t) RLM}]:
\beq
    |\Psi^0(t)\rangle = \left(  \prod_{j=1}^N c_{k_j}^\dagger(t) \right)|0\rangle.
\eeq
The method we introduce is a way of systematically adding a finite number of correction terms to $|\Psi^0(t)\rangle$ to form the full solution $|\Psi(t)\rangle$ for arbitrary coupling $U$.  A basic ingredient in the solution is a set of ``crossing states'' $|\Phi_{k_1 \dots k_n}(t)\rangle$, which are called such because they are built from single-particle $\mathcal{T}$-matrices for electrons crossing the origin---both the RLM $\mathcal{T}$ matrix $\mathcal{T}(k)$ [Eq. \eqref{eq: RLM basic notation}] and the $\mathcal{T}$ matrix $\mathcal{T}_U$ for a single electron scattering off a potential $U\delta(x)$ (though such a potential is not present in the Hamiltonian).  The full solution $|\Psi(t)\rangle$ is built from $c_k^\dagger(t)$ operators acting on crossing states.  

We begin by defining two operators that, roughly speaking, measure the failure of the $c_k^\dagger(t)$ operators to describe the full time evolution:
\bseq
    \begin{align}
        A_k(t) &\equiv [ H, c_k^\dagger(t)] - i \frac{\pd }{\pd t} c_k^\dagger(t),\\
        B_{k_1 k_2}(t) &\equiv \{A_{k_2}(t), c_{k_1}^\dagger(t)\}.
    \end{align}
\eseq
A short calculation yields these operators in explicit form.  The first is
\begin{multline}
    A_k(t) 
    = \frac{1}{\sqrt{L}}U d^\dagger \psi^\dagger(0)\\
    \times \left[-i \mathcal{T}(k) \left( e^{-i kt} - e^{-i zt} \right)\left( \frac{i}{v} \psi(0) - \frac{1}{2} d \right)- e^{-i k t} d \right],
\end{multline}
and the second is the antisymmetrization of a ``reduced'' operator:
\beq
    B_{k_1 k_2} (t) = B_{k_1 k_2}^{(\text{red})}(t) - B_{k_2 k_1}^{(\text{red})}(t),\label{eq: B in terms of Bred IRL}
\eeq
where
\beq
    B_{k_1 k_2}^{(\text{red})}(t) = - \frac{U}{Lv} \mathcal{T}(k_1) \left( e^{-i k_1 t}- e^{-i \pole t} \right)e^{-i k_2 t}   d^\dagger \psi^\dagger(0).
\eeq
The reduced operator is not uniquely defined, since one can add any symmetric function, but this is a convenient choice.

We note two properties of these operators for later reference:
\begin{itemize}
    \item Any $A(t)$ annihilates the empty state: 
    \beq
        A_k(t) |0\rangle= 0.\label{eq: A annihilates 0 IRL} 
    \eeq
    \item Any $B(t)$ commutes with any momentum creation operator:
    \beq
        [ B_{k_1 k_2}(t), c_{k_3}^\dagger(t)] = 0.\label{eq: B commutes IRL}
    \eeq
\end{itemize}
The case of $N=1$ is noninteracting: With $|\Psi^0(t)\rangle = c_{k_1}^\dagger(t)|0\rangle$, we have $(H-i\frac{d}{dt})|\Psi^0(t)\rangle= A_{k_1}(t) |0\rangle$, which vanishes due to \eqref{eq: A annihilates 0 IRL}.  We present the cases of $N=2$, $3$, and $4$ in detail, then proceed to general $N$.

\subsubsection{Two electrons}
For $N=2$, the freely-evolving state is $|\Psi^0(t)\rangle = c_{k_2}^\dagger(t)c_{k_1}^\dagger(t)|0\rangle$.  Bringing $(H- i\frac{d}{dt})$ past the momentum operators to annihilate the empty state yields
\bseq
\begin{align}
    \left(H - i \frac{d}{d t} \right) |\Psi^0(t)\rangle &= A_{k_2}(t)c_{k_1}^\dagger(t) | 0 \rangle
    + c_{k_2}^\dagger(t) A_{k_1}(t) | 0 \rangle\\
    &= B_{k_1 k_2}(t) |0\rangle,
\end{align}
\eseq
where we used Eq. \eqref{eq: A annihilates 0 IRL}.  The $N=2$ problem thus reduces to constructing a state $|\Phi_{k_1 k_2}(t)\rangle$ which is the ``inverse of $B_{k_1 k_2}(t)|0\rangle$'' in the following precise sense:
\bseq
    \begin{align}
        \left(H - i \frac{d}{dt}\right) |\Phi_{k_1 k_2}(t)\rangle &= - B_{k_1 k_2}(t) |0\rangle, \label{eq: Phi2 Schrod IRL}\\
        |\Phi_{k_1 k_2}(t=0)\rangle &= 0. \label{eq: Phi2 initial condition IRL}
    \end{align}
\eseq
Given such a state (which we explicitly construct below), the $N=2$ solution is immediate:
\beq
    |\Psi(t) \rangle = |\Psi^0(t)\rangle + |\Psi^2(t)\rangle,
\eeq
where $|\Psi^2(t)\rangle = |\Phi_{k_1 k_2}(t)\rangle$.  The point of these manipulations is that the state $|\Phi_{k_1 k_2}(t)\rangle$ will appear again in the solution for larger $N$.

Recalling Eq. \eqref{eq: B in terms of Bred IRL}, we write
\beq
    |\Phi_{k_1 k_2}(t)\rangle \equiv |\chi_{k_1 k_2}(t)\rangle -|\chi_{k_2 k_1}(t)\rangle,\label{eq: Phi2 in terms of chi2 IRL}
\eeq
where the unsymmetrized crossing state $|\chi_{k_1 k_2}(t)\rangle$ is required to vanish at $t=0$ and satisfy
\bseq
\begin{align}
    &\left(H - i \frac{d}{dt}\right) |\chi_{k_1 k_2}(t)\rangle = - B_{k_1 k_2}^{(\text{red})}(t) |0\rangle\\
    &=\frac{U}{Lv} \mathcal{T}(k_1) \left( e^{-i k_1 t}- e^{-i \pole t} \right)e^{-i k_2 t}   d^\dagger \psi^\dagger(0) |0\rangle.\label{eq: chi2 Schrod IRL}
\end{align}
\eseq
We make the following ansatz for the unsymmetrized crossing state:
\begin{multline}
    |\chi_{k_1 k_2}(t) \rangle = \frac{1}{L}  \int dx_1 dx_2\ F_{k_1 k_2}(t,x_1,x_2) \\
    \times \biggr[  \Theta(0<x_2<x_1<t) \psi^\dagger(x_2) \\
    + \frac{i}{v} \delta(x_2)\Theta(0<x_1 <t) d^\dagger \biggr] \psi^\dagger(x_1)|0\rangle,\label{eq: n=2 ansatz IRL}
\end{multline}
where we use the notation $\Theta(x_n < \dots < x_1) = \Theta(x_1 - x_2) \dots \Theta(x_n- x_{n-1})$, and where $F_{k_1 k_2}$ is a smooth function to be determined below.  Since an ansatz of similar form occurs throughout our calculations in both the IRL and AIM, we describe it in some detail. 

The state $|\chi_{k_1 k_2}(t)\rangle$ should vanish outside the forward ``light cone,'' seeing as the effect of the quench travels rightward from the origin at the Fermi velocity (which we have set to unity).  The ordering $x_2<x_1$ is a convenience and no loss of generality.  The state vanishes at $t=0$ by construction; to see this, we note that the overlap of $|\chi_{k_1 k_2}(t=0) \rangle$ with any reasonable state yields an integral of the form $\int_{-L/2}^{L/2}dx_1 dx_2\ \Theta(0<x_2<x_1<0)X(x_1,x_2)$ with some nonsingular function $X$, and this integral vanishes.  (In other words, the position space wavefunction of $|\chi_{k_1 k_2}(t=0) \rangle$ is nonsingular and vanishes everywhere except on a set of measure zero and hence is equivalent to the identically zero function.)

The only part of Eq. \eqref{eq: n=2 ansatz IRL} remaining that requires explanation is the impurity-electron part of the wavefunction, i.e., $d^\dagger \psi^\dagger(x_1)$.  This term is chosen so that acting on it with the tunneling term of $H$ produces an exact cancellation with the action of the kinetic term minus $i\frac{d}{dt}$ on the Heaviside function in the electron-electron part, i.e., $\psi^\dagger(x_2) \psi^\dagger(x_1)$.  In particular, we have
\begin{multline}
    \int dx_1 dx_2\ F_{k_1 k_2}(t,x_1,x_2) \biggr[ -i \left(\frac{\pd }{\pd t} + \frac{\pd}{\pd x_1} +\frac{\pd}{\pd x_2} \right)  \\
    \times \Theta(0<x_2<x_1<t) \biggr] \psi^\dagger(x_2)\psi^\dagger(x_1)|0\rangle\\
    + v \psi^\dagger(0) d \int dx_1dx_2\ F_{k_1 k_2}(t,x_1,x_2) \frac{i}{v} \delta(x_2)\\
    \times \Theta(0<x_1<t)d^\dagger \psi^\dagger(x_1)|0\rangle = 0. \label{eq: convenient cancellation}
\end{multline}
This cancellation is desirable because we want $(H-i\frac{d}{dt})|\chi_{k_1 k_2}(t)\rangle$ to be of the form $\psi^\dagger(0)d^\dagger |0\rangle$ in order to match the right-hand side of Eq. \eqref{eq: chi2 Schrod IRL}.  Proceeding, we find
\begin{widetext}
\begin{multline}
    \left(H - i \frac{d}{dt} \right)|\chi_{k_1 k_2}(t) \rangle = \frac{1}{L} \Biggr\{ \int dx_1 dx_2\ \left[ -i \left(\frac{\pd }{\pd t} + \frac{\pd}{\pd x_1} +\frac{\pd}{\pd x_2} \right)F_{k_1 k_2}(t,x_1,x_2) \right]\\
    \Theta(0<x_2<x_1<t)\psi^\dagger(x_2)\psi^\dagger(x_1)\\
    + \frac{i}{v} \int dx_1\ \left[ \left( -i \frac{\pd}{\pd t} -i  \frac{\pd}{\pd x_1}  +\pole \right)F_{k_1 k_2}(t,x_1,0)  \right]
    \Theta(0<x_1<t) d^\dagger \psi^\dagger(x_1)\\
    +\frac{i}{v}\left( -i + \frac{1}{2} U \right) F_{k_1 k_2}(t,0,0) d^\dagger \psi^\dagger(0) \Biggr\} |0\rangle,\label{eq: action on chi2 IRL}
\end{multline}
\end{widetext}
where the averaging prescription has been used [see the comment below Eq. \eqref{eq: Fk RLM}] to replace  $\delta(x_2)\Theta(0<x_2<x_1<t) \to \frac{1}{2}\delta(x_2)\Theta(0<x_1<t)$.  Comparing to Eq. \eqref{eq: chi2 Schrod IRL}, we see that it suffices for the function $F_{k_1 k_2}$ to satisfy the following three requirements:
\bseq
\begin{align}
    &\left(\frac{\pd }{\pd t} + \frac{\pd}{\pd x_1} +\frac{\pd}{\pd x_2} \right)F_{k_1 k_2}(t,x_1,x_2) = 0, \label{eq: first requirement n=2 IRL}\\
    &\left[ -i \left(\frac{\pd}{\pd t} + \frac{\pd}{\pd x_1} \right) +\pole \right]F_{k_1 k_2}(t,x_1,0)= 0, \label{eq: second requirement n=2 IRL}\\
    &\left( 1 + \frac{i}{2} U \right) F_{k_1 k_2}(t,0,0) = U\ \mathcal{T}(k_1)\left( e^{-i k_1 t} - e^{-i \pole t} \right)\notag\\
    & \qquad \qquad \qquad \qquad \qquad \qquad  \qquad \times  e^{-i k_2 t}. \label{eq: third requirement n=2 IRL}
\end{align}
\eseq
The first requirement will hold if $F_{k_1k_2}$ is a function of the coordinate differences only $(t-x_1, t-x_2,x_1-x_2)$, while the second requirement will hold if $F_{k_1 k_2}(t,x_1,0)$ is a function of $t-x_1$ times $e^{-i \pole x_1}$.  From the third requirement, we can then read off
\bseq
\begin{align}
    F_{k_1 k_2}(t,x_1,x_2) &= \mathcal{T}_U \mathcal{T}(k_1)\left( e^{-i k_1 (t -x_1)}- e^{-i \pole(t-x_1)}   \right) \notag \\
    &\qquad \times e^{-i k_2 (t-x_1)}e^{-i \pole(x_1- x_2)},\\ \label{eq: F n=2 IRL}
    \text{where }\mathcal{T}_U &= \frac{U}{1+ i U/2}.
\end{align}
\eseq
The quantity $\mathcal{T}_U$ is exactly the $\mathcal{T}$ matrix for a single electron, with linear spectrum, scattering on a potential $U \delta(x)$.  Recalling the antisymmetrization in Eq. \eqref{eq: Phi2 in terms of chi2 IRL}, we see that $|\Phi_{k_1 k_2}(t)\rangle$ is built from the free $\mathcal{T}$-matrices $T_U$, $\mathcal{T}(k_1)$, and $\mathcal{T}(k_2)$.  This is why we refer to $|\Phi_{k_1 k_2}(t)\rangle$ as a ``crossing state.''  

\subsubsection{Three electrons.}

For $N=3$, the freely-evolving state is $|\Psi^0(t)\rangle = c_{k_3}^\dagger(t)c_{k_2}^\dagger(t)c_{k_1}^\dagger(t)|0\rangle$, and we find
\bseq
\begin{align}
    &\left(H - i \frac{d}{dt} \right) |\Psi^0(t)\rangle =A_{k_3}(t) c_{k_2}^\dagger(t) c_{k_1}^\dagger(t) | 0 \rangle\notag\\
    &+ c_{k_3}^\dagger(t) A_{k_2}(t) c_{k_1}^\dagger(t) | 0 \rangle
    +  c_{k_3}^\dagger(t)c_{k_2}^\dagger(t)  A_{k_1}(t)| 0 \rangle\\
    &\qquad =  c_{k_3}^\dagger(t) B_{k_1 k_2}(t)- c_{k_2}^\dagger(t)B_{k_1 k_3}(t) |0\rangle\notag \\
    &\qquad \qquad \qquad \qquad \qquad + c_{k_1}^\dagger(t) B_{k_2 k_3}(t) | 0 \rangle,\label{eq: leftover terms N=3 IRL}
\end{align}
\eseq
where we used Eq. \eqref{eq: A annihilates 0 IRL} and Eq. \eqref{eq: B commutes IRL}.  To cancel these leftover terms, we reuse the same crossing state $|\Phi_{k_1 k_2}(t)\rangle$ that appeared in the two electron case, defining
\begin{multline}
    |\Psi^2(t)\rangle = c_{k_3}^\dagger(t) |\Phi_{k_1 k_2}(t)\rangle  -c_{k_2}^\dagger(t) |\Phi_{k_1 k_3}(t)\rangle \\
    +c_{k_1}^\dagger(t) |\Phi_{k_2 k_3}(t)\rangle .
\end{multline}
The point is that, if we bring $(H-i\frac{d}{dt})$ to the right of the $c_k^\dagger(t)$ operators in $|\Psi^2(t)\rangle$, then by the condition Eq. \eqref{eq: Phi2 Schrod IRL} that the crossing state satisfies, we obtain exactly what we need to cancel the leftover terms on the right-hand side of Eq. \eqref{eq: leftover terms N=3 IRL}.  Bringing $(H-i\frac{d}{dt})$ to the right generates new commutators:
\begin{multline}
    \left( H-  i\frac{d}{dt} \right) \left( |\Psi^0(t)\rangle + |\Psi^2(t)\rangle \right)= A_{k_3}(t) |\Phi_{k_1 k_2}(t)\rangle\\ -A_{k_2}(t) |\Phi_{k_1 k_3}(t)\rangle +
    A_{k_1}(t) |\Phi_{k_2 k_3}(t)\rangle.
\end{multline}
We are thus presented with a new ``inverse problem,'' namely to find a state $|\Phi_{k_1 k_2 k_3}(t)\rangle$ that satisfies
\bseq
    \begin{align}
        &\left(H - i \frac{d}{dt} \right) |\Phi_{k_1 k_2 k_3}(t) \rangle = -\Biggr( A_{k_3}(t) |\Phi_{k_1 k_2}(t)\rangle\notag\\
        &\qquad - A_{k_2}(t) |\Phi_{k_1 k_3}(t)\rangle + A_{k_1}(t) |\Phi_{k_2 k_3}(t)\rangle\Biggr),\\
        &|\Phi_{k_1 k_2 k_3}(t =0) \rangle = 0. 
    \end{align}
\eseq
Given such a state, the full solution is $|\Psi(t)\rangle = |\Psi^0(t)\rangle+|\Psi^2(t)\rangle+|\Psi^3(t)\rangle$, where $|\Psi^3(t)\rangle = |\Phi_{k_1 k_2 k_3}(t)\rangle.$  This exhibits the pattern that continues to all $N$: the states $|\Psi^1(t)\rangle,\dots, |\Psi^{N-1}(t)\rangle$ are built from crossing states that have been encountered already (up to $N-1$), while $|\Psi^N(t)\rangle$ requires a new crossing state.

It is again convenient to write the new crossing state as an antisymmetrized sum over permutations:
\begin{multline}
    |\Phi_{k_1 k_2 k_3}(t)\rangle = \sum_{\sigma \in \text{Sym}(3)}(\sgn \sigma) |\chi_{k_{\sigma_1} k_{\sigma_2} k_{\sigma_3}}(t)\rangle,
\end{multline}
where the unsymmetrized crossing state $|\chi_{k_1 k_2 k_3}(t)\rangle$ must vanish at $t=0$ and satisfy [recall Eq. \eqref{eq: Phi2 in terms of chi2 IRL}]
\bseq
\begin{align}
    &\left(H - i \frac{d}{dt} \right) |\chi_{k_1 k_2 k_3}(t) \rangle = - A_{k_3}(t) |\chi_{k_1 k_2}(t)\rangle \label{eq: Phi3 Schrod IRL}\\
    &\qquad =\frac{1}{L^{3/2}} \frac{i}{v} U \int dx_1\ F_{k_1 k_2}(t,x_1,0) \notag\\
    &\qquad \qquad \times e^{-i k_3 t}
    \Theta(0<x_1<t) d^\dagger \psi^\dagger(0) \psi^\dagger(x_1)|0\rangle.
\end{align}
\eseq
To find the unsymmetrized crossing state, we extend our previous ansatz \eqref{eq: n=2 ansatz IRL} to include another electron:
\begin{multline}
    |\chi_{k_1 k_2 k_3}(t) \rangle = \frac{1}{L^{3/2}}  \int dx_1 dx_2 dx_3\ F_{k_1 k_2 k_3}(t,x_1,x_2,x_3) \\
    \times \biggr[  \Theta(0<x_3<x_2<x_1<t) \psi^\dagger(x_3) \\
    + \frac{i}{v} \delta(x_3)\Theta(0<x_2<x_1 <t) d^\dagger \biggr] \psi^\dagger(x_2) \psi^\dagger(x_1)|0\rangle.\label{eq: n=3 ansatz IRL}
\end{multline}
We require that $F_{k_1 k_2 k_3}(t,x_1,x_2,x_3)$ is a function of coordinate differences only and that $F_{k_1 k_2 k_3}(t,x_1,x_2,0)$ equals $e^{-i \pole x_1}$ times a function of $t-x_1$ and $t-x_2$; then we obtain (see Appendix \ref{sec: Full calculation of nth crossing state} for the full calculation)
\begin{multline}
    \left(H - i \frac{d}{dt}\right)|\chi_{k_1 k_2 k_3}(t)\rangle = \frac{1}{L^{3/2}}\frac{i}{v}\left(-i + \frac{1}{2}U \right)\\
    \times \int dx_1\ F_{k_1 k_2 k_3}(t,x_1,0,0) \Theta(0<x_1<t)\\
    \times d^\dagger \psi^\dagger(0) \psi^\dagger(x_1)|0\rangle.
\end{multline}
Thus, $F_{k_1 k_2 k_3}$ must also satisfy
\begin{multline}
    \left(- i + \frac{1}{2} U \right)F_{k_1 k_2 k_3}(t,x_1,0,0) = U F_{k_1 k_2}(t,x_1,0)\\
    \times e^{-i k_3 t} .
\end{multline}
We can build a suitable function using the $n=2$ solution:
\begin{multline}
    F_{k_1 k_2 k_3}(t,x_1,x_2,x_3) = i \mathcal{T}_U F_{k_1 k_2}(t,x_1,x_3)\\
    \times 
    e^{-i k_3 (t -x_2)}.
\end{multline}

\subsubsection{Four electrons.}
This is a sufficient number to illustrate all properties of the general $N$ solution.  For $N=4$, the freely-evolving state is $|\Psi^0(t)\rangle = c_{k_4}^\dagger(t)c_{k_3}^\dagger(t)c_{k_2}^\dagger(t)c_{k_1}^\dagger(t)|0\rangle$, and following the same steps as before yields
\beq
    |\Psi(t) \rangle = |\Psi^0(t) \rangle+|\Psi^2(t) \rangle+|\Psi^3(t) \rangle + |\Psi^4(t) \rangle,
\eeq
where
\begin{widetext}
    \bseq
        \begin{align}
            |\Psi^2(t) \rangle &= c_{k_4}^\dagger(t) c_{k_3}^\dagger(t) |\Phi_{k_1 k_2}(t) \rangle - c_{k_4}^\dagger(t) c_{k_2}^\dagger(t) |\Phi_{k_1 k_3}(t) \rangle +c_{k_3}^\dagger(t) c_{k_2}^\dagger(t) |\Phi_{k_1 k_4}(t) \rangle \notag\\
            &\qquad \qquad + c_{k_4}^\dagger(t) c_{k_1}^\dagger(t) |\Phi_{k_2 k_3}(t) \rangle - c_{k_3}^\dagger(t) c_{k_1}^\dagger(t) |\Phi_{k_2 k_4}(t) \rangle + c_{k_2}^\dagger(t) c_{k_1}^\dagger(t) |\Phi_{k_3 k_4}(t) \rangle,   \\
            |\Psi^3(t)\rangle &= c_{k_4}^\dagger(t) |\Phi_{k_1 k_2 k_3}(t)\rangle - c_{k_3}^\dagger(t) |\Phi_{k_1 k_2 k_4}(t)\rangle +c_{k_2}^\dagger(t) |\Phi_{k_1 k_3 k_4}(t)\rangle- c_{k_1}^\dagger(t) |\Phi_{k_2 k_3 k_4}(t)\rangle,\\
            |\Psi^4(t)\rangle &= |\Phi_{k_1 k_2 k_3k_4} (t) \rangle,
        \end{align}
    \eseq
    where $|\Phi_{k_1 k_2 k_3 k_4}(t)\rangle$ is a new crossing state, which must vanish at $t=0$ and satisfy
    \begin{multline}
        \left(H - i \frac{d}{dt} \right) |\Phi_{k_1 k_2 k_3 k_4}(t) \rangle = -\Biggr( B_{k_3 k_4}(t) |\Phi_{k_1 k_2}(t) \rangle - B_{k_2 k_4}(t) |\Phi_{k_1 k_3}(t) \rangle +B_{k_2 k_3}(t) |\Phi_{k_1 k_4}(t) \rangle \\
        + B_{k_1 k_4}(t) |\Phi_{k_2 k_3}(t) \rangle - B_{k_1 k_3}(t) |\Phi_{k_2 k_4}(t) \rangle + B_{k_1 k_2}(t) |\Phi_{k_3 k_4}(t) \rangle \Biggr)\\
        -\Biggr( A_{k_4}(t) |\Phi_{k_1 k_2 k_3}(t)\rangle - A_{k_3}(t) |\Phi_{k_1 k_2 k_4}(t)\rangle +A_{k_2}(t) |\Phi_{k_1 k_3 k_4}(t)\rangle- A_{k_1}(t) |\Phi_{k_2 k_3 k_4}(t)\rangle \Biggr).
    \end{multline}
\end{widetext}
There are two types of terms that $(H-i\frac{d}{dt})|\Phi_{k_1 k_2 k_3 k_4}(t)\rangle$ must cancel: the $B(t)$ terms, which come from bringing $H- i\frac{d}{dt}$ past the creation operators in $|\Psi^2(t)\rangle$, and the $A(t)$ terms, which come from the same process in $|\Psi^3(t)\rangle$.  We deal with these separately by introducing two types of unsymmetrized crossing states, $|\chi_{k_1 k_2 | k_3 k_4}(t)\rangle$ and $|\chi_{k_1 k_2 k_3 k_4}(t)\rangle$ (the four momenta are either separated by a vertical line, or not), that are each required to vanish at $t=0$.  We write the full crossing state as an antisymmetrization:
\begin{multline}
    |\Phi_{k_1 k_2 k_3k_4} (t) \rangle = \sum_{\sigma \in \text{Sym}(4)}( \sgn \sigma) \biggr( |\chi_{k_{\sigma_1} k_{\sigma_2} | k_{\sigma_3}k_{\sigma_4}} (t) \rangle\\
    +|\chi_{k_{\sigma_1} k_{\sigma_2} k_{\sigma_3}k_{\sigma_4}} (t) \rangle \biggr).
\end{multline}
Then it suffices for the unsymmetrized crossing states to satisfy
\begin{multline}
    \left(H-i\frac{d}{dt}\right) |\chi_{k_1 k_2| k_3 k_4}(t)\rangle = 
    - B_{k_3 k_4}^{(\text{red})}(t) |\chi_{k_1 k_2}(t)\rangle = \\
    \frac{1}{L^2} \frac{U}{v} \int dx_1 dx_2\ F_{k_1 k_2}(t,x_1,x_2) \mathcal{T}(k_3)\left( e^{-i k_3 t}- e^{-i \pole t} \right)\notag\\
    \times e^{-i k_4 t} \Theta(0<x_2<x_1<t)d^\dagger \psi^\dagger(0) \psi^\dagger(x_2)\psi^\dagger(x_1)|0\rangle,
\end{multline}
and
\begin{multline}
    \left(H-i\frac{d}{dt}\right) |\chi_{k_1 k_2 k_3 k_4}(t)\rangle = 
    - A_{k_4}(t) |\chi_{k_1 k_2 k_3}(t)\rangle =\\
    \frac{1}{L^2}\frac{i}{v} U \int dx_1 dx_2\ F_{k_1 k_2 k_3}(t,x_1,x_2,0)  e^{-i k_4 t}\\
    \times \Theta(0<x_2<x_1<t) d^\dagger \psi^\dagger(0) \psi^\dagger(x_2)\psi^\dagger(x_1) |0\rangle.
\end{multline}
Extending Eq. \eqref{eq: n=3 ansatz IRL} to one more electron, we make the following ansatz:
\begin{widetext}
\begin{multline}
    |\chi_{k_1 k_2 | k_3 k_4 }(t) \rangle = \frac{1}{L^2}  \int dx_1 dx_2 dx_3 dx_4\ F_{k_1 k_2 | k_3 k_4}(t,x_1,x_2,x_3,x_4)  \biggr[  \Theta(0<x_4<x_3<x_2<x_1<t) \psi^\dagger(x_4) \\
    + \frac{i}{v} \delta(x_4)\Theta(0<x_3<x_2<x_1 <t) d^\dagger \biggr] \psi^\dagger(x_3) \psi^\dagger(x_2) \psi^\dagger(x_1)|0\rangle,\label{eq: n=4 ansatz IRL}
\end{multline}
\end{widetext}
with the same ansatz for $|\chi_{k_1 k_2  k_3 k_4}(t)\rangle$ (with $F_{k_1 k_2 | k_3 k_4}$ replaced by $F_{k_1 k_2 k_3 k_4}$).  We require that each $F(t,x_1,x_2,x_3,x_4)$ is a function of coordinate differences only and that each  $F(t,x_1,x_2,x_3,0)$ is of the form $e^{-i \pole x_j}$ times a function of $t-x_1,t-x_2$, and $t-x_3$; then (see Appendix \ref{sec: Full calculation of nth crossing state} for the full calculation) 
\begin{multline}
    \left(H - i \frac{d}{dt} \right)|\chi(t)\rangle = \frac{1}{L^2}\frac{i}{v} \left( -i + \frac{1}{2} U \right) \\
    \times F(t,x_1,x_2,0,0) \Theta(0<x_2<x_1<t)\\ \times d^\dagger \psi^\dagger(0)\psi^\dagger(x_2)\psi^\dagger(x_1)|0\rangle,
\end{multline} 
where $|\chi(t)\rangle$ and $F$ each have the subscript $(k_1,k_2| k_3,k_4)$ or $(k_1,k_2, k_3, k_4)$.  Comparing, we see that the two $F$ functions must satisfy
\begin{multline}
    \left( 1 + \frac{i}{2} U \right) F_{k_1 k_2 | k_3 k_4 }(t,x_1,x_2,0,0) = \\
    U F_{k_1 k_2}(t,x_1,x_2) \mathcal{T}(k_3)\left( e^{-i k_3 t}- e^{-i \pole t} \right)e^{-i k_4 t},
\end{multline}
and
\begin{multline}
    \left( -i + \frac{1}{2} U \right) F_{k_1 k_2 k_3 k_4 }(t,x_1,x_2,0,0) =\\
    U F_{k_1 k_2 k_3}(t,x_1,x_2,0)  e^{-i k_4 t}.
\end{multline}
The solutions are
\begin{multline}
    F_{k_1 k_2 | k_3 k_4}(t,x_1,x_2,x_3,x_4) = F_{k_1 k_2}(t,x_1,x_2)\\
    \times F_{k_3 k_4}(t,x_3,x_4),
\end{multline}
and
\begin{multline}
    F_{k_1 k_2 k_3 k_3}(t,x_1, x_2,x_3,x_4) =
    \left(i\mathcal{T}_U\right)^2\\
    \times F_{k_1 k_2}(t,x_1,x_4) e^{-i k_3 (t-x_2)} e^{-i k_4 (t-x_3)}.
\end{multline}

\subsubsection{Solution for general N.}
From the above calculations, the pattern has emerged.  The full wavefunction is a sum over subsets of the initial $N$ momenta; the chosen subset is put into a crossing state, which is then acted on by a product of $c_k^\dagger(t)$ operators that have the remaining momenta.  Each crossing state $|\Phi_{k_1 \dots k_n}(t)\rangle$ is the antisymmetrization of unsymmetrized crossing states in which the $n$ momenta are separated into any number of ``cells'' of length two or greater---for instance, $|\Phi_{k_1 k_2 k_3 k_4 k_5 k_6}(t)\rangle$ would include $|\chi_{k_1 k_2 k_3 k_4 k_5 k_6}(t)\rangle$, $|\chi_{k_1 k_2 | k_3 k_4 k_5 k_6}(t)\rangle$, $|\chi_{k_1 k_2 k_3 k_4 | k_5 k_6}(t)\rangle$, and $|\chi_{k_1 k_2 | k_3 k_4| k_5 k_6}(t)\rangle$ (all antisymmetrized).  Each new cell is associated with a $B^{(\text{red})}(t)$ operator, while the $A(t)$ operator extends the last cell by one.  The unsymmetrized crossing states are given by the $n$-electron generalization of Eq. \eqref{eq: n=4 ansatz IRL}.  Each new cell leads to a $F_{k_1 k_2}(t,x_1,x_2)$-type term, and any cell can be extended by changing the second $x$ coordinate of the $F_{k_1 k_2}$ function to the last coordinate of the cell and multiplying by a factor of the form $i\mathcal{T}_U e^{-i k_3 (t-x_2)}$.  For example,
\begin{multline}
    F_{k_1 k_2 k_3 |  k_4 k_5 k_6 k_7 }(t,x_1,x_2,x_3,x_4,x_5,x_6,x_7) = \\
    i\mathcal{T}_U F_{k_1 k_2}(t,x_1,x_3) e^{-i k_3 (t-x_2)}\\
    \times \left(i\mathcal{T}_U\right)^2 F_{k_4 k_5}(t,x_4,x_7) e^{-i k_6 (t-x_5)} e^{-i k_7 (t-x_6)}.
\end{multline}
We now present this solution in more detail, leaving the proof to Appendix \ref{sec: Full calculation of nth crossing state}.  For general $N$, we have
\begin{multline}
        |\Psi(t) \rangle = |\Psi^0(t)\rangle + \sum_{n=2}^N \sum_{1\le m_1 < \dots < m_n \le N} \\
        \times (-1)^{m_1+\dots+m_n + 1}
        \left( \prod_{\substack{ j=1 \\ j \ne m_{\ell}\ \forall \ell }}^N c_{k_j}^\dagger(t) \right) \\
        \times|\Phi_{k_{m_1} \dots k_{m_n}} (t) \rangle,\label{eq: wavefn construction IRL}
\end{multline}
where the terms in the summation over $n$ are exactly the $|\Psi^2(t)\rangle$, $|\Psi^3(t)\rangle$, etc. states discussed above.  The sign factor comes from bringing the quantum numbers $(k_{m_1},\dots,k_{m_n})$ to the left of the full list $(k_1,\dots,k_N)$.  To define the crossing states, we first write $\mathcal{A}$ as a shorthand for complete antisymmetrization in momenta---i.e., $\mathcal{A}\  X(k_1,\dots,k_n) = \sum_{\sigma \in \text{Sym}(n) }(\sgn \sigma) X(k_{\sigma_1}, \dots, k_{\sigma_n})$ for any function $X$.  Then we have
\begin{widetext}
    \beq
        |\Phi_{k_1\dots k_n}(t) \rangle = \mathcal{A} \sum_{s=1}^{n/2} \sum_{\substack{2\le j_1 < \dots < j_s \le n \\ j_s =n,\ \text{each } j_{m+1} - j_m \ge 2 } } |\chi_{k_1 \dots k_{j_1} | \dots | k_{j_{s-1}+1 } \dots k_n  } (t)\rangle,
    \eeq
    where
    \begin{multline}
        |\chi_{k_1 \dots k_{j_1} | \dots | k_{j_{s-1} + 1} \dots k_n}(t)\rangle = \frac{1}{L^{n/2}}\int dx_1\dots dx_n\ F_{k_1 \dots k_{j_1} | \dots | k_{j_{s-1} + 1} \dots k_n}(t,x_1,\dots, x_n)\\
        \times \biggr[ \Theta(0<x_n<\dots < x_1< t) \psi^\dagger(x_n)
        + \frac{i}{v} \delta(x_n)\Theta(0<x_{n-1}<\dots < x_1<t) d^\dagger \biggr] \psi^\dagger(x_{n-1}) \dots \psi^\dagger(x_1)|0\rangle.\label{eq: n ansatz IRL} 
    \end{multline}
\end{widetext}
Before constructing the function $F$ for a general number of cells $s$, we first define it in the special case of $s=1$, i.e., a single cell:
\begin{multline}
    F_{k_1 \dots k_n}(t,x_1,\dots,x_n)= \left( i \mathcal{T}_U \right)^{n-2} F_{k_1 k_2}(t,x_1,x_n) \\
    \times \prod_{j=2}^n\ e^{-i k_j(t-x_{j-1})},\label{eq: single-celled F IRL}
\end{multline}
where $F_{k_1 k_2}(t,x_1,x_2)$ is given in Eq. \eqref{eq: F n=2 IRL}.  Then, the function for general $s\ge1$ is a product of single-celled functions:
\begin{multline}
    F_{k_1 \dots k_{j_1} | \dots | k_{j_{s-1} + 1} \dots k_n}(t,x_1,\dots,x_n) =\\
    \prod_{m=1}^s F_{k_{j_{m-1} + 1} \dots k_{j_m}  }  (t, x_{j_{m-1}+1},\dots, x_{j_m}  ), \label{eq: F product of cells IRL}
\end{multline}
where $j_0\equiv 1$ and $j_s\equiv n$.  This completes the construction of the general crossing state, and thus the full many-body wavefunction.

\subsubsection{The NESS.}  In the long-time limit, the time-evolving wavefunction becomes a NESS: a solution to the time-independent Schrodinger equation with the boundary condition of incoming plane waves with momenta $k_1,\dots,k_N$.  As mentioned above Eq. \eqref{eq: ckindagger RLM}, this long-time limit must be taken in a pointwise sense, removing a trivial phase factor and rescaling by $L$ appropriately (see also a similar calculation in the Kondo model in the previous paper \cite{CulverAndrei_PRB1})
\beq
    \langle x |\Psi_{\text{NESS}} \rangle =  \lim_{\substack{t\to\infty, L\to \infty \\ t \ll L}} L^{N/2} e^{i E t} \langle x | \Psi(t) \rangle,
\eeq
where $E= k_1 + \dots + k_N$ and $|x\rangle = \psi^\dagger(x_N) \dots \psi^\dagger(x_1) |0\rangle$.  The overlap of the NESS with a basis state with the dot occupied is obtained similarly.  The result of taking this limit in the IRL wavefunction can essentially be read off by deleting the factors of $L$ and time-dependent phases, sending $t\to\infty$ in the Heaviside functions, and removing all terms in the $F$ functions that decay exponentially in time.  For completeness, we now provide the NESS explicitly.

Define the following time-independent version of the basic function \eqref{eq: F n=2 IRL} that appeared in the time-dependent solution:
\beq
F_{k_1 k_2,\text{in}}(x_1,x_2) = \mathcal{T}_U \mathcal{T}(k_1)e^{i (k_1+k_2)x_1}  e^{-i \pole(x_1- x_2)}.\\ \label{eq: F n=2 IRL NESS}
\eeq
Then, define the time-independent version of the single-celled function \eqref{eq: single-celled F IRL}: 
\begin{multline}
    F_{k_1 \dots k_n,\text{in} }(x_1,\dots,x_n)= \left( i \mathcal{T}_U \right)^{n-2} F_{k_1 k_2,\text{in} }(x_1,x_n) \\
    \times \prod_{j=2}^n\ e^{i k_j x_{j-1}}.\label{eq: single-celled F IRL NESS}
\end{multline}
The time-independent function $F$ for an arbitrary number of cells is then defined as in Eq. \eqref{eq: F product of cells IRL}.  We can then write the NESS wavefunction as follows:
\begin{widetext}
\beq
    |\Psi_{\text{NESS}} \rangle = \left( \prod_{j=1}^N c_{k_j,\text{in}}^\dagger \right) |0\rangle + \sum_{n=2}^N \sum_{1\le m_1 < \dots < m_n \le N}  (-1)^{m_1+\dots+m_n + 1}
        \left( \prod_{\substack{ j=1 \\ j \ne m_{\ell}\ \forall \ell }}^N c_{k_j,\text{in}}^\dagger \right)|\Phi_{k_{m_1} \dots k_{m_n} ,\text{in}} \rangle,
\eeq
with
\beq
        |\Phi_{k_1\dots k_n,\text{in}} \rangle = \mathcal{A} \sum_{s=1}^{n/2} \sum_{\substack{2\le j_1 < \dots < j_s \le n \\ j_s =n,\ \text{each } j_{m+1} - j_m \ge 2 } } |\chi_{k_1 \dots k_{j_1} | \dots | k_{j_{s-1}+1 } \dots k_n ,\text{in} }\rangle,
    \eeq
    where
    \begin{multline}
        |\chi_{k_1 \dots k_{j_1} | \dots | k_{j_{s-1} + 1} \dots k_n,\text{in}}\rangle =\int dx_1\dots dx_n\ F_{k_1 \dots k_{j_1} | \dots | k_{j_{s-1} + 1} \dots k_n ,\text{in} }(x_1,\dots, x_n)\\
        \times \biggr[ \Theta(0<x_n<\dots < x_1) \psi^\dagger(x_n)
        + \frac{i}{v} \delta(x_n)\Theta(0<x_{n-1}<\dots < x_1) d^\dagger \biggr] \psi^\dagger(x_{n-1}) \dots \psi^\dagger(x_1)|0\rangle.
    \end{multline}
\end{widetext}
Applying the time-independent version of our formalism (see Sec. \ref{sec: Time-independent formalism}) confirms that $|\Psi_{\text{NESS}}\rangle$ is an energy eigenstate with energy $E = k_1 + \dots + k_N$.  Alternatively, the time-independent formalism can be used to find $|\Psi_{\text{NESS}}\rangle$ directly, without following the time evolution; the calculation is very similar to the time-dependent case.

\subsection{General formalism}\label{sec: General formalism}
We now generalize the calculation of the previous section to a method that can be applied to a broader class of problems.  The key point is to write the many-body wavefunction as a sum of time-dependent creation operators acting on crossing states, then to identify the ``inverse problems'' that the crossing states must solve in order for the Schrodinger equation to be satisfied.  This takes care of much of the bookkeeping and isolates the hard part of the interacting problem, namely the calculation of the crossing states. 

We consider a Hilbert space consisting of any states produced by fermionic ``field operators'' $c_\alpha^\dagger$ acting on an empty state $|0\rangle$ that is annihilated by any field operator.  (Note that $\alpha$ is a label for any quantum numbers; $d^\dagger$ in the IRL counts as a ``field operator.'') 
We wish to find the time evolution of an initial state with arbitrary quantum numbers $\alpha_1,\dots,\alpha_N$:
\beq
    |\Psi(t) \rangle \equiv e^{-i H t} c_{\alpha_N}^\dagger \dots c_{\alpha_1}^\dagger |0 \rangle.
\eeq
Equivalently, we need to solve the differential equation
\beq
    \left( H - i \frac{d}{dt} \right)|\Psi(t) \rangle =0,
\eeq
with the initial condition
\beq
    |\Psi(t= 0) \rangle = \left( \prod_{j=1}^N c_{\alpha_j}^\dagger \right) |0 \rangle.\label{eq: initial condition}
\eeq
The starting point of the construction is to find the time-evolving operators that would describe the full time evolution in the absence of interaction.  We take the Hamiltonian to be
\beq
    H = H^{(0)} + H^{(1)},
\eeq
where the time evolution of the field operators under $H^{(0)}$ is assumed to be known: 
\beq
    c_\alpha^\dagger(t) \equiv e^{-i H^{(0)} t} c_\alpha^\dagger e^{i H^{(0)} t},\label{eq: cdagger(t) simplest}
\eeq
and where both $H^{(0)}$ and $H^{(1)}$ annihilate the empty state:
\beq
    H^{(0)} |0\rangle = H^{(1)} |0\rangle = 0. \label{eq: H annihilates 0}
\eeq
Thus, in the noninteracting case ($H^{(1)}=0$), the full solution is given by a product of $c_\alpha^\dagger(t)$ operators:
\beq
    |\Psi^0(t)\rangle \equiv \left(\prod_{j=1}^N  c_{\alpha_j }^\dagger(t) \right) |0\rangle.\label{eq: Psi0(t)}
\eeq
The time-evolving state $|\Psi^0(t)\rangle$ satisfies the initial condition \eqref{eq: initial condition}; each term that we will add to it in order to reach the full solution (with $H^{(1)}$ included) will be required to vanish at $t=0$.  We define
\bseq
    \begin{align}
        &A_\alpha(t) \equiv [ H, c_\alpha^\dagger(t)] - i \frac{\pd }{\pd t} c_\alpha^\dagger(t) = [H^{(1)}, c_{\alpha}^\dagger(t)],\\
        &B_{\alpha_1 \alpha_2}(t) \equiv \{A_{\alpha_2}(t), c_{\alpha_1}^\dagger(t)\}.
    \end{align}
\eseq
Generalizing from the IRL, we assume that these operators have the following properties:
\begin{itemize}
    \item Any $A(t)$ annihilates the empty state: 
    \beq
        A_{\alpha}(t) |0\rangle= 0.\label{eq: A annihilates 0} 
    \eeq
    \item Any $B(t)$ commutes with any field creation operator:
    \beq
        [ B_{\alpha_1 \alpha_2}(t), c_{\alpha_3}^\dagger(t)] = 0.\label{eq: B commutes}
    \eeq
\end{itemize}
When $H^{(0)}$ is quadratic, the $c_{\alpha}^\dagger(t)$ operators are linear combinations of field operators; then the above conditions are met whenever the interaction $H^{(1)}$ is a sum of terms of the form $c_{\alpha_1}^\dagger c_{\alpha_2}^\dagger c_{\alpha_1'}c_{\alpha_2'}$ [since we have, schematically, $A(t) \sim c^\dagger c^\dagger c$ and $B(t) \sim c^\dagger c^\dagger$].  Thus, the formalism of this section can in principle be applied to a fairly general class of number-conserving Hamiltonians with a quartic interaction term.  In particular, we have not yet specialized to one-dimensional quantum impurity problems with linearized spectrum.  These additional restrictions seem to become necessary when we seek exact solutions to the differential equations for the crossing states.

It is straightforward to show that $B_{\alpha_1 \alpha_2}(t)$ is antisymmetric under exchange of the quantum numbers $\alpha_1$ and $\alpha_2$ \footnote{We have $B_{\alpha_1 \alpha_2}(t) + B_{\alpha_2 \alpha_1}(t) = [H, \{ c_{\alpha_1}^\dagger(t), c_{\alpha_2}^\dagger(t) \} ] -i \frac{\pd}{\pd t} \{ c_{\alpha_1}^\dagger(t), c_{\alpha_2}^\dagger(t) \} =0 $.}; hence, it can be written in terms of a ``reduced'' operator:
\beq
    B_{\alpha_1 \alpha_2}(t) = B_{\alpha_1 \alpha_2}^{(\text{red})}(t) - B_{\alpha_2 \alpha_1}^{(\text{red})}(t). 
\eeq
While $B_{\alpha_1 \alpha_2}^{(\text{red})}(t) = \frac{1}{2}B_{\alpha_1 \alpha_2}(t)$ is always an option, it can happen that the calculation is simpler with a different choice (as we saw in the IRL).  

Our approach will be to bring $H$ past all of the $c_\alpha^\dagger(t)$ operators to its right at the cost of commutators [$A_\alpha(t)$ operators], then to bring each $A_\alpha(t)$ to the right of the remaining $c_{\alpha}^\dagger(t)$ operators at the cost of anticommutators [$B_{\alpha_1 \alpha_2}(t)$ operators]; then each $B(t)$ can be brought to the right due to Eq. \eqref{eq: B commutes}.  The IRL calculation in the previous section provides explicit examples of these manipulations for $N=2,3,4$.  We now give a summary of the general $N$ case, leaving the proof to Appendix \ref{sec: Proof of general formalism}.

We commute $H$ past each $c_\alpha^\dagger(t)$ operator to find
\bseq
    \begin{align}
        &\left(  H - i \frac{d }{dt} \right) |\Psi^0(t)\rangle = \sum_{m_2=1}^N c_{\alpha_N}^\dagger(t) \dots\notag\\
        &\qquad \left( [H, c_{\alpha_{m_2}}^\dagger(t)] - i \frac{\pd}{\pd t} c_{\alpha_{m_2}}^\dagger(t)\right) \dots c_{\alpha_1}^\dagger(t) |0\rangle \\
        &=\sum_{1\le m_1 < m_2 \le N} (-1)^{m_1+m_2+1} \left( \prod_{j =1,j \ne m_1,m_2}^n c_{\alpha_j}^\dagger(t)\right)\notag\\
        &\qquad \qquad \qquad \qquad \qquad \qquad  \times B_{\alpha_{m_1}\alpha_{m_2}}(t) |0\rangle.\label{eq: leftover from Psi0}
    \end{align}
\eseq
To cancel this, we define a state $|\Psi^2 (t)\rangle$ as
\begin{multline}
    |\Psi^2(t)\rangle = \sum_{1\le m_1 < m_2 \le N} (-1)^{m_1+m_2+1} \\
    \times \left( \prod_{j =1,j \ne m_1,m_2}^n c_{\alpha_j}^\dagger(t)\right) |\Phi_{\alpha_{m_1}\alpha_{m_2}}(t)\rangle,
\end{multline}
where the crossing state $|\Phi_{\alpha_1 \alpha_2}(t)\rangle$ vanishes at $t=0$ and satisfies
\beq
    \left(H - i\frac{d}{dt} \right) |\Phi_{\alpha_1 \alpha_2}(t)\rangle = - B_{\alpha_1 \alpha_2}(t) |0\rangle.\label{eq: Phi2 Schrod}
\eeq
The point is that if $H-i\frac{d}{dt}$ were to act only on $|\Phi_{\alpha_1 \alpha_2}(t)\rangle$, then $(H- i\frac{d}{dt})|\Psi^{(2)}(t)\rangle$ would cancel the right-hand side of Eq. \eqref{eq: leftover from Psi0}.  To reach the $|\Phi_{\alpha_1 \alpha_2}(t)\rangle$ state, though, $H - i\frac{d}{dt}$ must commute past each $c_\alpha^\dagger(t)$ operator; we therefore obtain
\begin{multline}
    \left( H - i \frac{d}{d t} \right) \left(|\Psi^0(t)\rangle + |\Psi^2(t)\rangle \right) =\\
    \sum_{1\le m_1 < m_2 < m_3\le N} (-1)^{m_1 + m_2 +m_3 + 1}\\
    \times 
    \left( \prod_{\substack{j=1 \\ j\ne m_1, m_2,m_3 }}^N c_{\alpha_j }^\dagger(t) \right) \Biggr( A_{\alpha_{m_3}}(t) |\Phi_{\alpha_{m_1}\alpha_{m_2} }(t) \rangle\\ -A_{\alpha_{m_2}}(t) |\Phi_{\alpha_{m_1}\alpha_{m_3} }(t) \rangle
    +A_{\alpha_{m_1}}(t) |\Phi_{\alpha_{m_2}\alpha_{m_3} }(t) \rangle\Biggr). \label{eq: leftover from Psi0 + Psi2}
\end{multline}
Note that this equation has a similar structure to Eq. \eqref{eq: leftover from Psi0}, but with $N-3$ of the $c_\alpha^\dagger(t)$ operators appearing instead of $N-2$.  To cancel the new leftover terms, we need a new crossing state $|\Phi_{\alpha_1 \alpha_2 \alpha_3}(t)\rangle$---that vanishes at $t=0$ and satisfies Eq. \eqref{eq: Phi3 Schrod IRL} with each $k_j$ replaced by $\alpha_j$---from which we can construct $|\Psi^3(t)\rangle$ to cancel the right-hand side of Eq. \eqref{eq: leftover from Psi0 + Psi2}.  This results in new terms to cancel, in which at most $N-4$ of the $c_{\alpha}^\dagger(t)$ operators appear in any particular term; we build $|\Psi^4(t)\rangle$ from a new crossing state $|\Phi_{\alpha_1 \alpha_2 \alpha_3 \alpha_4}(t)\rangle$, and so on.  This process terminates when we reach $|\Psi^N(t)\rangle$ and all $N$ of the $c_\alpha^\dagger(t)$ operators are eliminated.

Let us state the general result (proven in Appendix \ref{sec: Proof of general formalism}).  The full time-evolving wavefunction can be written as
\begin{multline}
        |\Psi(t) \rangle = |\Psi^0(t)\rangle + \sum_{n=2}^N \sum_{1\le m_1 < \dots < m_n \le N} \\
        \times (-1)^{m_1+\dots+m_n + 1}
        \left( \prod_{\substack{ j=1 \\ j \ne m_{\ell}\ \forall \ell }}^N c_{\alpha_j}^\dagger(t) \right) \\
        \times|\Phi_{\alpha_{m_1} \dots \alpha_{m_n}} (t) \rangle,\label{eq: wavefn construction}
\end{multline}
where the terms in the summation over $n$ are exactly the $|\Psi^2(t)\rangle$, $|\Psi^3(t)\rangle$, etc. states discussed above.  The crossing states are antisymmetrizations of unsymmetrized crossing states in which the quantum numbers are separated into cells of length two or greater.  Writing $\mathcal{A}$ as a shorthand for complete antisymmetrization of $\alpha_j$ quantum numbers---i.e., $\mathcal{A}\  X(\alpha_1,\dots,\alpha_n) = \sum_{\sigma \in \text{Sym}(n) }(\sgn \sigma) X(\alpha_{\sigma_1}, \dots, \alpha_{\sigma_n})$ for any function $X$---we claim that the following requirements are sufficient for Eq. \eqref{eq: wavefn construction} to satisfy the time evolution problem: 
\begin{widetext}
\beq
    |\Phi_{\alpha_1\dots\alpha_n}(t) \rangle = \mathcal{A} \sum_{s=1}^{n/2} \sum_{\substack{2\le j_1 < \dots < j_s \le n \\ j_s =n,\ \text{each } j_{m+1} - j_m \ge 2 } } |\chi_{\alpha_1 \dots \alpha_{j_1} | \dots | \alpha_{j_{s-1}+1 } \dots \alpha_{j_n}  } (t)\rangle,\label{eq: crossing state as antisymmetrization}
\eeq
where
\begin{multline}
    \left(H - i \frac{d}{d t} \right)|\chi_{\alpha_1 \dots \alpha_{j_1} | \dots | \alpha_{j_{s-1}+1 } \dots \alpha_n  } (t)\rangle =
    \begin{cases}
        -B_{\alpha_{n-1} \alpha_n}^{(\text{red})}(t)|\chi_{\alpha_1 \dots \alpha_{j_1} | \dots | \alpha_{j_{s-2}+1 } \dots \alpha_{n-2}} (t)\rangle &  j_{s-1} = n-2\\
        - A_{\alpha_n}(t) |\chi_{\alpha_1 \dots \alpha_{j_1} | \dots | \alpha_{j_{s-1}+1 } \dots \alpha_{n-1}} (t)\rangle & n\ge 3 \text{ and } j_{s-1} \le n-3,\\
    \end{cases}\label{eq: general aux state Schrod}
\end{multline}
and
\beq
    |\chi_{\alpha_1 \dots \alpha_{j_1} | \dots | \alpha_{j_{s-1}+1 } \dots \alpha_n  } (t=0 )\rangle =0. \label{eq: general aux state initial condition}
\eeq
\end{widetext}
Throughout, $j_0 \equiv 0$ and the sum over $s$ goes over only integer values.  We set $|\chi(t)\rangle \equiv |0\rangle$ so that for $n=2$, Eq. \eqref{eq: general aux state Schrod} recovers Eq. \eqref{eq: Phi2 Schrod} after antisymmetrizing.

Thus, we have transformed the original many-body Schrodinger equation to the problem of finding states that satisfy Eq. \eqref{eq: general aux state Schrod} and Eq. \eqref{eq: general aux state initial condition}.

\subsection{Time-independent formalism}\label{sec: Time-independent formalism}
It is convenient in some problems to solve for the infinite time limit of the wavefunction directly, without following the detailed time evolution.  Here, we present the formalism of the previous section in a time-independent form.  Although it is not strictly necessary, we formulate the entire discussion in terms of scattering theory.

As in standard scattering theory, the passage from the time-dependent to the time-independent picture results in the initial condition in time (at $t=0$ in our setup, usually $t=-\infty$ in scattering theory) becoming a time-independent boundary 
condition in space (e.g., incoming plane waves).

We write the Hamiltonian as $H = H^{(0)} + H^{(1)} = h + \mathcal{V}$, where $h$ is the noninteracting Hamiltonian from the point of view of scattering theory.  That is, $h$ describes the propagation of plane waves, not including any tunneling to the impurity or scattering off a potential.  For instance, $h= -i \int dx\ \psi^\dagger(x)\frac{d}{dx}\psi(x)$ in the IRL.  (Note that we work in infinite volume.)  We write $h= \int d\alpha\ E_\alpha c_{\alpha}^\dagger c_{\alpha}$, where the $c_\alpha$ operators are Dirac normalized and where the integral over $\alpha$ can also include a sum over discrete quantum numbers.  The term $H^{(0)}$ contains $h$ and any other quadratic terms, and $H^{(1)}$ contains interaction terms.

The Lippmann-Schwinger equation for scattering ``in'' states is $|\Psi_{\text{in}} \rangle = |\Psi\rangle + \frac{1}{h - E + i \eta}\mathcal{V}|\Psi_{\text{in}}\rangle$, where $|\Psi \rangle \equiv c_{\alpha_N}^\dagger\dots c_{\alpha_1}^\dagger |0\rangle$ is an eigenstate of $h$ with energy $E\equiv E_{\alpha_1} + \dots +E_{\alpha_N}$.  This is equivalent to the time-independent Schrodinger equation
\beq
    \left( H - E \right) |\Psi_\text{in} \rangle = 0,
\eeq
with the boundary condition of incoming plane waves with quantum numbers $\alpha_1,\dots,\alpha_N$.  

The noninteracting Hamiltonian $H^{(0)}$, which includes \emph{quadratic} terms such as impurity tunneling and potential scattering, has a set of scattering operators $c_{\alpha,\text{in}}^\dagger$ that satisfy
\beq
    [H^{(0)}, c_{\alpha,\text{in}}^{ \dagger}] -E_\alpha c_{\alpha,\text{in}}^{ \dagger} = 0
\eeq
and that create scattering ``in'' states corresponding to $c_{\alpha}^\dagger$.  [See Eq. \eqref{eq: ckindagger RLM} for these operators in the case of the RLM.]  The solution to the Lippman-Schwinger equation in the special case of no interaction ($H^{(1)} = 0$) is given by a product of these operators:
\beq
    |\Psi_{\text{in}}^0\rangle =
    c_{\alpha_N, \text{in}}^{\dagger} \dots c_{\alpha_1,\text{in}}^{ \dagger} |0\rangle.
\eeq
To include the interaction term $H^{(1)}$, we proceed in much the same way as in the time-dependent case.  The main point is to isolate the core difficulty of the interacting problem, which is in this case to find time-independent crossing states satisfying the appropriate ``inverse problems.''  We begin by defining time-independent versions of the $A$ and $B$ operators:
\bseq
\begin{align}
    A_{\alpha,\text{in}} &= [H, c_{\alpha,\text{in}}^{\dagger} ] - E_\alpha c_{\alpha,\text{in}}^{\dagger}, \\
    B_{\alpha_1 \alpha_2, \text{in}} &= \{ A_{\alpha_2,\text{in}},c_{\alpha_1,\text{in}}^{\dagger} \} = B_{\alpha_1 \alpha_2, \text{in}}^{(\text{red})} - B_{\alpha_2 \alpha_1,\text{in}}^{(\text{red})}.
\end{align}
\eseq
As in the time-dependent case, we assume that $H^{(0)}|0\rangle = H^{(1)}|0\rangle = A_{\alpha,\text{in}}|0\rangle = 0$ and that $B_{\alpha_1 \alpha_2, \text{in}}$ commutes with any $c_{\alpha,\text{in}}^\dagger$.  The same manipulations yield an exact reformulation of the Lippmann-Schwinger equation.  We have the following representation of the wavefunction [the time-independent version of Eq. \eqref{eq: wavefn construction}]
\begin{multline}
        |\Psi_{\text{in}} \rangle = |\Psi_{\text{in}}^0\rangle + \sum_{n=2}^N \sum_{1\le m_1 < \dots < m_n \le N} \\
        \times (-1)^{m_1+\dots+m_n + 1}
        \left( \prod_{\substack{ j=1 \\ j \ne m_{\ell}\ \forall \ell }}^N c_{\alpha_j,\text{in} }^{ \dagger} \right) \\
        \times|\Phi_{\alpha_{m_1} \dots \alpha_{m_n}, \text{in}} \rangle,
\end{multline}
where the crossing states are given by
\begin{widetext}
\beq
    |\Phi_{\alpha_1\dots\alpha_n,\text{in}} \rangle = \mathcal{A} \sum_{s=1}^{n/2} \sum_{\substack{2\le j_1 < \dots < j_s \le n \\ j_s =n,\ \text{each } j_{m+1} - j_m \ge 2 } } |\chi_{\alpha_1 \dots \alpha_{j_1} | \dots | \alpha_{j_{s-1}+1 } \dots \alpha_{j_n},\text{in}  }\rangle.
\eeq
The unsymmetrized crossing states must satisfy
\begin{multline}
    \left(H - \sum_{\ell=1}^n E_{\alpha_{\ell}} \right)|\chi_{\alpha_1 \dots \alpha_{j_1} | \dots | \alpha_{j_{s-1}+1 } \dots \alpha_n ,\text{in} }\rangle =
    \begin{cases}
        -B_{\alpha_{n-1} \alpha_n,\text{in}}^{(\text{red})}|\chi_{\alpha_1 \dots \alpha_{j_1} | \dots | \alpha_{j_{s-2}+1 } \dots \alpha_{n-2},\text{in}}\rangle &   j_{s-1} = n-2\\
        - A_{\alpha_n,\text{in}} |\chi_{\alpha_1 \dots \alpha_{j_1} | \dots | \alpha_{j_{s-1}+1 } \dots \alpha_{n-1},\text{in}}\rangle & n\ge 3,\ j_{s-1} \le n-3
    \end{cases}
\end{multline}
\end{widetext}
where $j_0 \equiv 1$, $|\chi_{ , \text{in}}\rangle = |0\rangle$, and the sum over $s$ goes over only integer values; also, each $|\chi_{\alpha_1 \dots \alpha_{j_1} | \dots | \alpha_{j_{s-1}+1 } \dots \alpha_n ,\text{in} } \rangle$ must satisfy the boundary condition of having \emph{no} plane waves coming in from infinity (since the incoming $\alpha_1,\dots,\alpha_N$ quantum numbers are already accounted for in $|\Psi_{\alpha_1 \dots \alpha_N,\text{in}}^0\rangle$).  This last condition is the time-independent analog of the initial condition that crossing states vanish at $t=0$ [Eq. \eqref{eq: general aux state initial condition}].

While we have specified incoming boundary conditions, the entire procedure carries through with any other choice of boundary conditions (e.g., outgoing).  In principle, the formalism may even apply to the problem of finding energy eigenstates in a finite-volume system.

\subsection{Time-evolving wavefunction of the multilead IRL}\label{sec: Time-evolving wavefunction of the multilead IRL}
As another application of our general formalism, we find the exact time-evolving wavefunction of the simplest version of the multilead IRL, in which each lead has the same tunneling and Coulomb interaction with the dot:
\bseq
    \begin{align}
        H^{(0)} &= - i \int_{-L/2}^{L/2}dx\ \sum_{\gamma=1}^{\gmax} \psi_{\gamma}^\dagger(x) \frac{d}{dx} \psi_{\gamma}(x) + \epsilon d^\dagger d \notag\\
        &\qquad  +\left[ \sum_{\gamma=1}^{\gmax}\frac{v}{\sqrt{\gmax}} \psi_{\gamma}^\dagger(0) d + \text{ h.c.}\right],\\
        H^{(1)} &= U\sum_{\gamma=1}^{\gmax} \psi_{\gamma}^\dagger(0)\psi_{\gamma}(0) d^\dagger d, \label{eq: H1 mcIRL}\\
        H &= H^{(0)} + H^{(1)}.
    \end{align}
\eseq
Before presenting our results, we recall some prior work from Nishino \emph{et al}. \cite{NishinoEtAl_PRLonIRL, NishinoEtAl_PRBonIRL, NishinoHatanoOrdonez, NishinoEtAl_doubledot1, NishinoEtAl_doubledot2}.  References \cite{NishinoEtAl_PRLonIRL}, \cite{NishinoEtAl_PRBonIRL}, and \cite{NishinoHatanoOrdonez} present the NESS wavefunction in the two-lead case.  The results seem to agree with ours for $N=2,3$ electrons (when we take the steady state limit of the wavefunction that we find below); while Nishino \emph{et al}. obtained the general $N$ case as well, it is not written explicitly.  Reference \cite{NishinoHatanoOrdonez} allows the tunnelings and Coulomb interactions to be lead dependent, which is a more general case than we consider here; also, Refs. \cite{NishinoEtAl_doubledot1} and \cite{NishinoEtAl_doubledot2} present NESS wavefunctions for $N=2,3$ electrons with two-leads and two quantum dots.  We expect that our formalism should also apply to these other variations of the IRL.  

We present the exact time evolution of an arbitrary initial state with the dot unoccupied:
\beq
    |\Psi_{\gamma_1 k_1 \dots \gamma_N k_N}(t) \rangle \equiv e^{-i H t} c_{\gamma_N k_N}^\dagger \dots c_{\gamma_1 k_1}^\dagger |0\rangle,
\eeq
where the lead indices $\gamma_j$ and momenta $k_j$ are arbitrary.  Since the calculation is similar to the one-lead case (Sec. \ref{sec: Time-evolving wavefunction of the one-lead IRL}), we present only the main points (see Ref. \cite{Culver_thesis} for details).

The $A(t)$ and $B(t)$ operators of the multilead model are found to be
\begin{multline}
    A_{\gamma k}(t) = -\frac{U}{\sqrt{L}}\biggr[ e^{-i k t} d^\dagger \psi_\gamma^\dagger(0) d + \frac{1}{\gmax} F_k(t) \sum_{\beta=1}^{\gmax}  \\
    \times d^\dagger \psi_{\beta}^\dagger(0) \left( \frac{1}{2}d - \frac{i\sqrt{\gmax}}{v}\psi_{\beta}(0)  \right)\biggr],
\end{multline}
and
\beq
    B_{\gamma_1 k_1 \gamma_2 k_2}^{(\text{red})}(t) = -\frac{i}{\sqrt{\gmax}} \frac{U}{Lv} F_{k_1}(t) e^{-i k_2 t} d^\dagger \psi_{\gamma_2}^\dagger(0).
\eeq
The conditions that we need in order to apply the general formalism [Eqs. \eqref{eq: H annihilates 0}, \eqref{eq: A annihilates 0}, \eqref{eq: B commutes}] are easily verified.  Thus, the wavefunction takes the general form of Eqs. \eqref{eq: wavefn construction} and \eqref{eq: crossing state as antisymmetrization}, with the generic quantum number $\alpha$ replaced by $\gamma k$, and we only need to specify the unsymmetrized crossing states that solve the inverse problems of the model [Eqs. \eqref{eq: general aux state Schrod} and \eqref{eq: general aux state initial condition}].

As we mentioned in Sec. \ref{sec: Noninteracting case: The resonant level model}, the noninteracting Hamiltonian $H^{(0)}$ separates under rotation into $\gmax - 1$ free fermion fields and a single copy of the one-lead RLM.  This separation breaks down once the interaction term $H^{(1)}$ is included.  However, it turns out that some ingredients of the one-lead solution can be reused.  By similar calculations as in the one-lead case, we find that the first unsymmetrized crossing state is the following generalization of Eq. \eqref{eq: n=2 ansatz IRL}:
\begin{multline}
    |\chi_{\gamma_1 k_1 \gamma_2 k_2}(t)\rangle = \frac{1}{\gmax L}\int dx_1 dx_2\ F_{k_1 k_2}(t,x_1,x_2)\\ \times \biggr[ \Theta(0<x_2<x_1<t) \sum_{\beta=1}^{\gmax}\psi_\beta^\dagger(x_2)\\
    + \frac{i\sqrt{\gmax}}{v}\delta(x_2)\Theta(0<x_1<t)d^\dagger \biggr] \psi_{\gamma_2}^\dagger(x_1)|0\rangle,\label{eq: chi n=2 mcIRL}
\end{multline}
where $F_{k_1 k_2}$ is the same function as in the one-lead case, Eq. \eqref{eq: F n=2 IRL}.  For the general case---an unsymmetrized crossing state with $n\ge 2 $ quantum numbers---we find
\begin{widetext}
\begin{multline}
    |\chi_{\gamma_1 k_1 \dots \gamma_{j_1} k_{j_1} | \dots | \gamma_{j_{s-1} + 1} k_{j_{s-1} + 1} \dots \gamma_n k_n}(t)\rangle =\\
    \frac{1}{L^{n/2}}\int dx_1\dots dx_n\ \sum_{\beta_1,\dots,\beta_n=1}^{\gmax} F_{\gamma_1 k_1 \dots \gamma_{j_1} k_{j_1} | \dots | \gamma_{j_{s-1} +1}k_{j_{s-1} + 1} \dots \gamma_n k_n}^{\beta_1 \dots \beta_n}(t,x_1,\dots, x_n)\\
    \times \biggr[ \Theta(0<x_n<\dots < x_1< t) \psi_{\beta_n}^\dagger(x_n)
    + \frac{i}{\sqrt{\gmax} v} \delta(x_n)\Theta(0<x_{n-1}<\dots < x_1<t) d^\dagger \biggr] \psi_{\beta_{n-1}}^\dagger(x_{n-1}) \dots \psi_{\beta_1}^\dagger(x_1)|0\rangle, \label{eq: n ansatz mcIRL}
\end{multline}
\end{widetext}
where the function $F$, now with lead indices, is defined as follows.  In the simplest case of a single cell ($s=1$), we define
\begin{multline}
    F_{\gamma_1 k_1 \dots \gamma_n k_n  }^{\beta_1 \dots \beta_n}(t,x_1,\dots,x_n)=\\
    \frac{1}{\gmax}F_{k_1 \dots k_n  }(t,x_1,\dots,x_n)\prod_{\ell=1}^{n-1}\delta_{\gamma_{\ell+1}}^{\beta_\ell},
\end{multline}
where $F$ on the right-hand side is as in the one-lead solution [Eq. \eqref{eq: single-celled F IRL}].  Then the full solution, with an arbitrary number of cells, is given by a product
\begin{multline}
    F_{\gamma_1 k_1 \dots \gamma_{j_1} k_{j_1} | \dots | \gamma_{j_{s-1}+ 1 } k_{j_{s-1} + 1} \dots \gamma_n k_n }^{\beta_1 \dots \beta_n }(t,x_1,\dots, x_n) = \\
    \prod_{m=1}^s F_{\gamma_{j_{m-1} + 1} k_{j_{m-1} + 1} \dots \gamma_{j_m} k_{j_m}  }^{\beta_{j_{m-1}+1}  \dots \beta_{j_m} }  (t, x_{j_{m-1}+1},\dots, x_{j_m}  ),
\end{multline}
where $j_0 \equiv 0$ and $j_s \equiv n$.

\subsection{NESS of the multilead AIM for small or large \texorpdfstring{$U$}{U}.}\label{sec: NESS of the multilead AIM for small or large U}
In this section, we apply the time-independent version of our formalism (Sec. \ref{sec: Time-independent formalism}) to the multilead AIM, considered directly in the infinite volume limit:
\bseq
    \begin{align}
        H^{(0)} &= - i \int dx\ \sum_{\gamma=1}^{\gmax} \psi_{\gamma a}^\dagger(x) \frac{d}{dx} \psi_{\gamma a}(x) + \epsilon d_a^\dagger d_a \notag\\
        &  +\sum_{\gamma=1}^{\gmax}\left[\frac{v}{\sqrt{\gmax}}  \psi_{\gamma a}^\dagger(0) d_a + \text{ h.c.}\right],\\
        H^{(1)} &= U n_{\uparrow}n_{\downarrow},\\
        H_{\text{finite }U} &= H^{(0)} + H^{(1)}.\label{eq: HfiniteU mcAIM}
    \end{align}
\eseq
In Ref. \cite{ImamuraEtAl}, Imamura \emph{et al}. find the two electron NESS in the one-lead case---a result that we reproduce below and extend to arbitrary $N$ electrons in the cases of small and large $U$.

In the limit $U\to\infty$, it is convenient to use the auxiliary boson technique \cite{Coleman}, according to which we write the following Hamiltonian:
\begin{multline}
    H_{\text{infinite } U} =  - i \int dx\ \sum_{\gamma=1}^{\gmax} \psi_{\gamma a}^\dagger(x) \frac{d}{dx} \psi_{\gamma a}(x) +\epsilon d_a^\dagger d_a 
    \\
    + \sum_{\gamma=1}^{\gmax}\left[ \frac{v}{\sqrt{\gmax}}  \psi_{\gamma a}^\dagger(0) b^\dagger  d_a   + \text{ h.c.} \right],\label{eq: H multilead AIM infU}
\end{multline}
which has a conserved charge $Q \equiv b^\dagger b + d_a^\dagger d_a$; working in the subspace $Q=1$ imposes the constraint that the dot cannot be doubly occupied, which is equivalent to sending $U\to\infty$ in $H_{\text{finite }U}$.

In either case ($U$ finite or infinite), the same unitary transformation as in Sec. \ref{sec: Noninteracting case: The resonant level model} isolates the interacting sector of the model:
\bseq
\begin{align}
    &H_{e,\text{finite }U} =  - i \int dx\  \psi_{ea}^\dagger(x) \frac{d}{dx} \psi_{e a}(x) +\epsilon d_a^\dagger d_a \notag\\
    &+\left[ v \psi_{e a}^\dagger(0) d + \text{ h.c.}\right]+  U n_{\uparrow}n_{\downarrow} ,\\
    &H_{e,\text{infinite }U}  =  - i \int dx\  \psi_{ea}^\dagger(x) \frac{d}{dx} \psi_{e a}(x) +\epsilon d_a^\dagger d_a \notag \\
    &\qquad \qquad \qquad \qquad + \left[ v \psi_{ea}^\dagger(0) b^\dagger  d_a   + \text{ h.c.} \right].\label{eq: He AIM infU}
\end{align}
\eseq
The model thus decouples into a copy of the one-lead AIM and $N-1$ free fermions.  We can use this decoupling to show that the crossing states that we need for the multilead model are related to the crossing states of the even sector by simple prefactors (see our previous paper \cite{CulverAndrei_PRB1} for a similar calculation in the Kondo model with $\gmax=2$):
\begin{multline}
    |\Phi_{\gamma_1 k_1 a_1 \dots \gamma_n k_n a_n,\text{in}}\rangle =\\
    \left( \frac{1}{\gmax}\right)^{n/2} |\Phi_{e k_1 a_1 \dots e k_n a_n,\text{in}}\rangle.
\end{multline}
Thus, it suffices to solve the scattering problem with incoming ``even'' plane waves.  We can then reuse the same crossing states to read off the solution to the scattering problem with incoming plane waves in the original multilead basis.

\paragraph{Solution for small $U$.}

In the finite $U$ case, a short calculation yields
\beq
    A_{e k a,\text{in}} \equiv [H_e, c_{e k a,\text{in}}^\dagger ] - k c_{e k a,\text{in}}^\dagger =   \frac{U}{v}\mathcal{T}(k) d_a^\dagger d_b^\dagger d_b,\label{eq: Ae Anderson}
\eeq
and
\bseq
\begin{align}
    B_{e k_1 a_1  e k_2 a_2,\text{in}} \equiv \{ A_{e k_2 a_2,\text{in} }, c_{e k_1 a_1,\text{in} }^\dagger \} \\
    = B_{e k_1 a_1  e k_2 a_2,\text{in}}^{(\text{red})} - B_{e k_2 a_2  e k_1 a_1,\text{in}}^{(\text{red})}, 
\end{align}
\eseq
where
\beq
    B_{e k_1 a_1  e k_2 a_2,\text{in}}^{(\text{red})} = \frac{1}{2v^2} U \mathcal{T}(k_1)  \mathcal{T}(k_2) P_{- a_1 a_2}^{\ b_1 b_2}d_{b_2}^\dagger d_{b_1}^\dagger,\label{eq: Bered Anderson} 
\eeq
and where $P_- = \frac{1}{2} \left( I - P \right)$ is the antisymmetric spin projection operator ($I_{a_1 a_2}^{b_1 b_2} \equiv \delta_{a_1}^{b_1}\delta_{a_2}^{b_2}$, $P_{a_1 a_2}^{b_1 b_2}\equiv  \delta_{a_1}^{b_2}\delta_{a_2}^{b_1}$).

Our task is to find a state $|\chi_{e k_1 a_1 e k_2 a_2,\text{in}} \rangle$ that has no incoming plane waves and that satisfies:
\bseq
\begin{align}
    &\left( H - k_1 - k_2\right) |\chi_{e k_1 a_1 e k_2 a_2,\text{in} }\rangle = - B_{e k_1 a_1 e k_2 a_2, \text{in}}^{({\text{red}})}|0\rangle\\
    &= -\frac{U}{2v^2} \mathcal{T}(k_1)\mathcal{T}(k_2) P_{- a_1 a_2}^{\ b_1 b_2}d_{b_2}^\dagger d_{b_1}^\dagger|0\rangle.\label{eq: inverse problem n=2 AIM}
\end{align}
\eseq
Given such a state, the solution to the two electron scattering problem is
\begin{multline}
    |\Psi_{e k_1 a_1 e k_2 a_2,\text{in} }\rangle = c_{e k_2 a_2,\text{in}}^\dagger c_{e k_1 a_1,\text{in}}^\dagger|0\rangle \\
    + |\chi_{e k_1 a_1 e k_2 a_2,\text{in} }\rangle  - |\chi_{e k_2 a_2 e k_1 a_1,\text{in} }\rangle.  
\end{multline}
We make the ansatz:
\begin{multline}
    |\chi_{e k_1 a_1 e k_2 a_2,\text{in}} \rangle =  \int dx_1 dx_2\ F_{e k_1 a_1 e k_2 a_2}^{b_1 b_2}(x_1,x_2) \\
    \times \biggr[ \Theta(0< x_2 < x_1) \psi_{e b_2}^\dagger(x_2)\psi_{e b_1}^\dagger(x_1) \\
    + \frac{i}{v}\delta(x_2) \Theta(0< x_1) d_{b_2}^\dagger \psi_{e b_1}^\dagger(x_1) \\
    -\frac{1}{2 v^2} \delta(x_1)\delta(x_2) d_{b_2}^\dagger d_{b_1}^\dagger \biggr] |0\rangle, \label{eq: ansatz n=2 AIM}
\end{multline}
where $F_{e k_1 a_1 e k_2 a_2}$ is a smooth function that is determined shortly.  By construction, this ansatz vanishes when any position variable is to the left of the origin; this guarantees that there are no incoming waves from $x=-\infty$.  As the model contains only right movers, there is no possibility of waves coming in from $x=+\infty$; hence, this ansatz does not disturb the scattering boundary condition satisfied by $|\Psi_{\text{in}}^0\rangle$.  Furthermore, this ansatz is chosen so that certain terms that are not of the form we want ($d_{b_2}^\dagger d_{b_1}^\dagger|0\rangle$) cancel automatically when we act on it with $H - k_1 -k_2$ [see Eq. \eqref{eq: convenient cancellation} for a similar calculation in the IRL case].  A straightforward calculation yields
\begin{widetext}
\begin{multline}
    \left( H - k_1 -k_2 \right) |\chi_{e k_1 a_1 e k_2 a_2,\text{in}} \rangle =\\
    \int dx_1 dx_2\ \left\{ \left[ -i \left(\frac{\pd}{\pd x_1} +\frac{\pd}{\pd x_2}  \right) - k_1 -k_2  \right] F_{e k_1 a_1 ek_2 a_2}^{b_1 b_2}(x_1,x_2 ) \right\}\Theta(0<x_2  < x_1 ) \psi_{e b_2}^\dagger(x_2)\psi_{e b_1}^\dagger(x_1) |0\rangle\\
    + \frac{i}{v} \int dx_1\ \left[ \left( -i \frac{\pd}{\pd x_1}  - k_1 -k_2 +\pole \right) F_{e k_1 a_1 ek_2 a_2}^{b_1 b_2}(x_1,0 ) \right]  \Theta(0<x_1  ) d_{b_2}^\dagger \psi_{e b_1}^\dagger(x_1) |0\rangle\\
    -\frac{1}{2v} \left( F_{e k_1 a_1 ek_2 a_2}^{b_1 b_2 }(0,0) + F_{e k_1 a_1 ek_2 a_2}^{b_2 b_1 }(0,0) \right) \Theta(0<x_1 ) d_{b_2}^\dagger \psi_{e b_1}^\dagger(0)  |0\rangle\\
    -\frac{1}{2 v^2} \left( - k_1-k_2 +2  \pole + U \right) F_{e k_1 a_1 ek_2 a_2}^{b_1 b_2 }(0,0) d_{b_2}^\dagger d_{b_1}^\dagger |0\rangle.\label{eq: H-E on chi2 Anderson}
\end{multline}
\end{widetext}
To get the desired result $\left( H - k_1 -k_2 \right) |\chi_{e k_1 a_1 e k_2 a_2,\text{in}} \rangle  = - B_{e k_1 a_1  e k_2 a_2,\text{in}}^{(\text{red})}|0\rangle$, we require that the first three terms of Eq. \eqref{eq: H-E on chi2 Anderson} all vanish and that the fourth matches Eq. \eqref{eq: inverse problem n=2 AIM}; this leads to the following requirements on the function $F$:
\bseq
\begin{align}
    &\left[ -i \left(\frac{\pd}{\pd x_1} +\frac{\pd}{\pd x_2}  \right) - k_1 -k_2  \right] F_{e k_1 a_1 ek_2 a_2}^{b_1 b_2} =0 ,\\
    &\left( -i \frac{\pd}{\pd x_1}  - k_1 -k_2 +\pole \right) F_{e k_1 a_1 ek_2 a_2}^{b_1 b_2}(x_1,0 ),\\
    &F_{e k_1 a_1 ek_2 a_2}^{b_1 b_2}(0,0)+ F_{e k_1 a_1 ek_2 a_2}^{b_2 b_1}(0,0)= 0,\\
    &\left( -k_1 -k_2 +2 \pole + U\right)F_{e k_1 a_1 ek_2 a_2}^{b_1 b_2 }(0,0) =\notag\\
    &\qquad \qquad  \qquad  \qquad  \qquad  U \mathcal{T}(k_1) \mathcal{T}(k_2) P_{- a_1 a_2}^{\ b_1 b_2}.
\end{align}
\eseq
A function that meets these requirements is
\begin{multline}
    F_{e k_1 a_1 e k_2 a_2}^{b_1 b_2}(x_1,x_2) = -  \mathcal{T}(k_1) \mathcal{T}(k_2) \frac{U \mathcal{T}\left(\frac{k_1 + k_2 -U}{2}  \right)}{4\Delta}\\
    \times 
    e^{i(k_1 + k_2) x_1}e^{-i \pole (x_1 -x_2)}P_{- a_1 a_2}^{\ b_1 b_2}.\label{eq: F n=2 AIM} 
\end{multline}
The Schrodinger equation with the boundary condition (of incoming plane waves) can be expected to have a unique solution; hence, this is the answer.

Collecting all terms of the wavefunction, we have exact agreement with the two electron NESS obtained in Ref. \cite{ImamuraEtAl}.  As another check, we have repeated the calculation in finite volume with the time-dependent formalism and found exactly this answer in the steady state and infinite volume limit [where the limit is taken pointwise, with factors of $L$ and the free evolution phase factor removed as in Eq. \eqref{eq: ckindagger RLM}] \cite{Culver_thesis}.

In Eq. \eqref{eq: F n=2 AIM}, we see a similar structure as appeared in the IRL solution: The two electrons are bound together over a distance scale of order $1/\Delta$ [compare to Eq. \eqref{eq: F n=2 IRL}].

\paragraph{Solution for $U\to\infty$.}  We present the final results only; details can be found in Ref. \cite{Culver_thesis}.  We set $H \equiv H_{\text{infinite }U}$ throughout this section.  Our time-independent formalism (Sec. \ref{sec: Time-independent formalism}) carries through straightforwardly with the state $b^\dagger|0\rangle$ replacing $|0\rangle$ and the scattering ``in'' operators given by
\begin{multline}
    c_{\gamma k a,\text{in}}^\dagger \equiv c_{\gamma k a}^\dagger + \frac{1}{\gmax} \int dx\ F_{k,\text{in}}(x) \Biggr[ \Theta(0<x)\\
    \times \sum_{\gamma=1}^{\gmax} \psi_{\gamma a}^\dagger(x) +\frac{i\sqrt{\gmax}}{v} \delta(x) d_a^\dagger b\Biggr].\label{eq: cgammakaindagger AIM infU}
\end{multline}
The unsymmetrized crossing states are given by
\begin{widetext}
\begin{multline}
    |\chi_{e k_1 a_1 \dots e k_{j_1} a_{j_1} | \dots | e k_{j_{s-1} + 1}a_{j_{s-1} + 1} \dots e k_n a_n,\text{in} }\rangle =  \int dx_1\dots dx_n\ F_{e k_1 a_1 \dots e k_{j_1} a_{j_1} | \dots | e k_{j_{s-1} + 1}a_{j_{s-1} + 1} \dots e k_n a_n}^{b_1 \dots b_n}(x_1,\dots,x_n )\\
    \times \biggr[ \Theta(0<x_n < \dots < x_1 ) \psi_{e b_n}^\dagger(x_n) 
    + \frac{i}{v} \delta(x_n)\Theta(0<x_{n-1} < \dots < x_1 ) d_{b_n}^\dagger b  \biggr]\psi_{e b_{n-1}}^\dagger(x_{n-1}) \dots \psi_{e b_1}^\dagger(x_1) b^\dagger |0\rangle,\label{eq: ansatz general n AIM infU}
\end{multline}
\end{widetext}
where the function $F$ is now to be specified.  First, we define $F$ in the special case of $s=1$, i.e., a single cell:
\begin{multline}
    F_{e k_1 a_1 \dots e k_\ell a_\ell}^{b_1 \dots b_\ell}(x_1,\dots, x_\ell) =
    (2i)^{\ell -2} \delta_{a_1}^{c_1}\delta_{c_\ell}^{b_\ell} \mathcal{T}(k_1)e^{i k_1 x_1} \\
    \times \left[ \prod_{m=2}^\ell \mathcal{T}(k_m) P_{- c_{m-1} a_m}^{\ b_{m-1} c_m}  e^{i k_m x_{m-1}} \right] e^{-i \pole(x_1 - x_\ell)}.\label{eq: F single-celled AIM infU}
\end{multline}
It is straightforward to check that $\ell=2$ agrees with the $U\to\infty$ limit of the finite $U$ function, Eq. \eqref{eq: F n=2 AIM}.  The function for general $s\ge1$ is a product of single-celled functions:
\begin{multline}
    F_{e k_1  a_1 \dots e k_{j_1} a_{j_1} | \dots | e a_{j_{s-1} + 1} k_{j_{s-1} + 1} \dots e k_n a_n}^{b_1 \dots b_n}(x_1,\dots,x_n) =\\
    \prod_{m=1}^s F_{e k_{j_{m-1} + 1} a_{j_{m-1} + 1} \dots e k_{j_m} a_{j_m}  }^{b_{j_{m-1}+1} \dots b_{j_m} }  ( x_{j_{m-1}+1},\dots, x_{j_m}  ), \label{eq: F product of cells AIM}
\end{multline}
where $j_0\equiv 1$ and $j_s\equiv n$.  Note that the spin matrices multiply in the same diagonal manner as in the Kondo wavefunction found in our previous paper \cite{CulverAndrei_PRB1}.  Note also that we have a factor of $e^{-i \pole(x_1-x_\ell)}$, indicating that electrons are bound together on a distance scale $1/\Delta$.  The single-celled function describes $\ell$ electrons bound together, and the full function with a general partition has some number of these cells.

\section{Evaluation of observables}\label{sec: Evaluation of observables}
We present some results of using the IRL and AIM wavefunctions to calculate expectation values of observables.  We focus in particular on these expectation values in the steady state, accessed either by taking the long-time limit after the quench or by evaluating directly in the NESS.

Though the wavefunctions presented in the previous section are exact for any fixed number $N$ of electrons, the number of terms grows rapidly with $N$, making the evaluation of observables in the thermodynamic limit a formidable task.  (Note that taking the thermodynamic limit is essential to obtain physical results, since we linearized the spectrum.)  At present, we can calculate this limit only by making an expansion in some parameter.  In the IRL, this parameter is the Coulomb interaction $U$, while in the AIM, it can either be $U$ or (in the limit $U\to\infty$) the tunneling parameter $\Delta$.  In each case, we evaluate an observable to the leading order by keeping just the $|\Psi^0\rangle$ and $|\Psi^2\rangle$ terms of the wavefunction (i.e., putting at most two quantum numbers into a crossing state), which makes the thermodynamic limit tractable.

\subsection{Dot occupancy in the multilead IRL}\label{sec: Dot occupancy in the multilead IRL}

We evaluate the expectation value of the dot occupancy to the leading order in the interaction strength $U$.  We show that the steady state occupancy is a universal function of the external parameter $\epsilon_d$ (which is defined below in terms of the bare parameter $\epsilon$ that appears in the Hamiltonian) and the temperatures and chemical potentials of the leads.  This universal function is parameterized by two RG invariants: $U$ and an emergent energy scale $T_K$.  

We compare our results with the literature and provide a general discussion of the RG flow of the IRL, emphasizing universal aspects.  We then specialize to the zero temperature case.  We verify that our answer agrees with the equilibrium Bethe ansatz result for the occupancy as a function of applied field in the multilead model, and then we present some results in two steady-state nonequilibrium regimes of the two-lead model with the leads separated by a bias voltage.

\subsubsection{Evaluation.}
Our task is to evaluate
\beq
    \langle n_d \rangle_t \equiv \langle \Psi(t) | d^\dagger d |\Psi(t)\rangle,
\eeq
where $|\Psi(t) \rangle = e^{-i H t} |\Psi\rangle$ and $|\Psi\rangle= c_{\gamma_N k_N}^\dagger \dots c_{\gamma_1 k_1}^\dagger |0\rangle$.  Note that $|\Psi\rangle$ is normalized to unity.  The initial quantum numbers are arbitrary for the moment, though we later specialize to the case of a Fermi sea in each lead.

We begin by expanding the wavefunction to first order in $U$:
\beq
    |\Psi(t)\rangle = |\Psi^0(t)\rangle + |\Psi^2(t)\rangle + O(U^2),
\eeq
where
\bseq
    \begin{align}
        |\Psi^0(t)\rangle &= \left( \prod_{j=1}^N c_{\gamma_j k_j}^\dagger(t) \right) |0\rangle, \\
        |\Psi^2(t) \rangle &= \sum_{1\le m_1 < m_2 \le N} (-1)^{m_1 + m_2 +1 } \left( \prod_{\substack{j=1 \\ j\ne m_1,m_2}}^N c_{\gamma_j k_j}^\dagger(t) \right)\notag\\
        &\qquad \times  |\Phi_{\gamma_{m_1} k_{m_1} \gamma_{m_2} k_{m_2}}(t)\rangle.
    \end{align}
\eseq
The occupancy to leading order is therefore
\beq
    \langle n_d \rangle_t = \langle n_d \rangle_t^{(0)} + \langle n_d \rangle_t^{(1)},
\eeq
where $\langle n_d \rangle_t^{(0)} = \langle \Psi^0(t) | d^\dagger d | \Psi^0(t)\rangle$ and $\langle n_d \rangle_t^{(1)} = 2\ \text{Re}\left( \langle \Psi^0(t)| d^\dagger d|\Psi^2(t)\rangle\right)$ (where $|\Psi^2(t)\rangle$ is to be expanded to first order in $U$; we will see that this expansion is simple).  The main tool in the calculation is Wick's theorem combined with the fact that the time-evolving field operators have canonical anticommutation relations: $\{ c_{k'}(t), c_k^\dagger(t)\} = \delta_{k k'}$.  Using these, we obtain the following for the noninteracting (zeroth order in $U$) part of the answer:
\bseq
\begin{align}
    \langle n_d\rangle_t^{(0)} &= \sum_{j=1}^N | \{ d, c_{\gamma_j k_j}(t)\}|^2 \\
    &= \frac{1}{L}\sum_{j=1}^N \left|\frac{i}{v\sqrt{\gmax}}F_{k_j}(t)\right|^2,
\end{align}
\eseq
where $F_k$ is given by Eq. \eqref{eq: Fk RLM}.

Next, we specialize to the case of interest (filled Fermi seas in each lead) and then take the thermodynamic limit.  We describe this step in some detail now, since it occurs again in our subsequent calculations.  Specializing the $N$ initial quantum numbers to describe filled Fermi seas at zero temperature is equivalent to the following replacement: 
\beq
    \sum_{m=1}^N X(\gamma_m,k_m) = \sum_{\gamma=1}^{\gmax} \sum_{k \in \mathcal{K}_\gamma }  X(\gamma,k),\label{eq: sum replacement rule no spin}
\eeq
where $\mathcal{K}_\gamma$ is the set of allowed momenta in lead $\gamma$ (i.e., ranging from $-D$ to $\mu_\gamma$) and $X$ is any function.  If the sum $\sum_{m=1}^N$ comes with a prefactor $1/L$ (as it always will in our calculations), then the sum over momenta in a given lead $\gamma$ becomes an integral $\int_{-D}^D \frac{dk}{2\pi}\ \Theta(\mu_\gamma- k) (\dots)$ in the thermodynamic limit.  To generalize to arbitrary lead temperatures, we replace the step function by the Fermi function $\fermifn_\gamma(k) = \left[ e^{(k-\mu_\gamma)/T_\gamma} +1 \right]^{-1}$.  All together, the prescription for taking the thermodynamic limit and including temperature is
\begin{multline}
    \frac{1}{L}\sum_{m=1}^N X(\gamma_m,k_m) \overset{\text{therm. limit}}{\longrightarrow}\\
    \sum_{\gamma=1}^{\gmax} \int_{-D}^D\frac{dk}{2\pi}\ \fermifn_\gamma(k) X(\gamma,k).
\end{multline}
This generalizes to the higher-order summations we encounter in the interacting case, as well; for instance, a double sum $\frac{1}{L^2} \sum_{m_1,m_2=1}^N X(\gamma_{m_1}k_{m_1}, \gamma_{m_2}k_{m_2})$ becomes $\sum_{\gamma_1,\gamma_2=1}^{\gmax}\int_{-D}^D \frac{dk_1}{2\pi}\frac{dk_2}{2\pi}\ \fermifn_{\gamma_1}(k_1)\fermifn_{\gamma_2}(k_2) X(\gamma_1 k_1, \gamma_2 k_2)$.  We have confirmed the above prescription for generalizing to arbitrary temperatures by setting up the calculation with an initial density matrix and verifying that the same result is obtained \cite{Culver_thesis}. 

Thus, we obtain the noninteracting contribution to the occupancy in the thermodynamic limit:
\beq
    \langle n_d \rangle_t^{(0)} \overset{\text{therm. limit}}{\longrightarrow} \frac{1}{\gmax}\sum_{\gamma=1}^{\gmax}\int_{-D}^D \frac{dk}{2\pi}\ \fermifn_\gamma(k) \frac{|F_k(t)|^2}{2\Delta}.
\eeq
In Ref. \cite{Iyer_thesis}, $\langle n_d\rangle_t^{(0)}$ is calculated in the one-lead model at zero temperature; our result agrees in this special case.  In the steady state (s.s.) limit, we find
\bseq
\begin{align}
    &\langle n_d \rangle_{\text{s.s.}}^{(0)} \equiv \lim_{t\to\infty} \langle n_d\rangle_t^{(0)}\\
    &\qquad =\frac{1}{\gmax}\sum_{\gamma=1}^{\gmax} \int_{-D}^D \frac{dk}{2\pi}\ \fermifn_\gamma(k) \frac{|\mathcal{T}(k)|^2}{2\Delta} \label{eq: occupancy mcRLM}\\
    &= \frac{1}{\gmax}\sum_{\gamma=1}^{\gmax} \int_{-D}^D \frac{dk}{2\pi}\ \fermifn_\gamma(k) \frac{2\Delta}{(k-\epsilon)^2 + \Delta^2}.
\end{align}
\eseq
In the equilibrium limit (equal temperatures and chemical potentials in all leads), we recover the standard occupancy of the RLM.  In the two-lead model at zero temperature with a voltage drop across the leads, our answer agrees with Ref. \cite{KarraschEtAl}. 

We proceed to the leading correction in $U$.  Again using Wick's theorem and the anticommutation relation, and canceling some sign factors, we obtain
\begin{multline}
    \langle n_d\rangle_t^{(1)} = 2 \text{Re} \Biggr\{ \sum_{1 \le m_1 < m_2 \le N} \Bigg[ \{ c_{\gamma_{m_2} k_{m_2}}(t), d^\dagger\} \\
    \times \langle 0 | c_{\gamma_{m_1} k_{m_1}}(t) d | \Phi_{\gamma_{m_1} k_{m_1} \gamma_{m_2} k_{m_2}}(t)\rangle 
    - (m_1 \leftrightarrow m_2) \Bigg]\Biggr\}.
\end{multline}
It is advantageous to consider the ``off-diagonal'' case, in which the quantum numbers on either side of the matrix element that appears in the previous equation are arbitrary.  By fermionic antisymmetry, the matrix element must be the antisymmetrization of some function $\Omega(t;\gamma_1' k_1',\gamma_2' k_2'; \gamma_1 k_1,\gamma_2 k_2)$ as follows:
\begin{multline}
    \{ c_{\gamma_1' k_1'}(t), d^\dagger\} \langle 0|  c_{\gamma_2' k_2'}(t)  d|\Phi_{\gamma_1 k_1 \gamma_2 k_2}(t) \rangle =\\
    \frac{1}{L^2} \sum_{\sigma, \sigma' \in \text{Sym}(2)} (\sgn \sigma)
    (\sgn \sigma')\\
    \times \Omega(t;\gamma_{\sigma_1'}' k_{\sigma_1'}',\gamma_{\sigma_2'}' k_{\sigma_2'}'; \gamma_{\sigma_1} k_{\sigma_1},\gamma_{\sigma_2} k_{\sigma_2}),\label{eq: nd term in terms of Omega IRL}
\end{multline}
where the factor of $1/L^2$ is inserted for the convenience of taking the thermodynamic limit.  The key point is that $\Omega$ (which we write explicitly below) does not depend on $L$, due to the fact that the crossing state vanishes outside the forward ``light cone'' in position space.  Relabelling summation variables, we obtain
\begin{multline}
    \langle n_d \rangle_t^{(1)} = \frac{1}{L^2} \sum_{m_1,m_2=1}^N   \sum_{\sigma \in \text{Sym}(2)} (\sgn \sigma)\\
    \times  2\text{Re}\left[ \Omega(t;\gamma_{m_1}k_{m_1},\gamma_{m_2}k_{m_2}; \gamma_{\sigma_{m_1}} k_{\sigma_{m_1}},\gamma_{\sigma_{m_2}} k_{\sigma_{m_2}})  \right]\\
    \overset{\text{therm. limit}}{\longrightarrow} \sum_{\gamma_1,\gamma_2=1}^{\gmax} \int_{-D}^D \frac{dk_1}{2\pi}\frac{dk_2}{2\pi}\ \fermifn_{\gamma_1}(k_1)\fermifn_{\gamma_2}(k_2) \sum_{\sigma \in \text{Sym}(2)}  \\
    \times  (\sgn \sigma) 2\text{Re}\left[ \Omega(t;\gamma_1 k_1,\gamma_2 k_2; \gamma_{\sigma_1} k_{\sigma_1},\gamma_{\sigma_2} k_{\sigma_2})  \right].\label{eq: nd^1 mcIRL}
\end{multline}
Recall that the crossing state is given by $|\Phi_{\gamma_1 k_1 \gamma_2 k_2}(t)\rangle = |\chi_{\gamma_1 k_1 \gamma_2 k_2}(t)\rangle  - |\chi_{\gamma_2 k_2 \gamma_1 k_1}(t)\rangle $, with $\chi$ given by Eq. \eqref{eq: chi n=2 mcIRL}.  Then from Eq. \eqref{eq: nd term in terms of Omega IRL} and Eq. \eqref{eq: ckdagger(t) RLM}, we can read off
\begin{multline}
    \Omega(t;\gamma_1' k_1',\gamma_2' k_2'; \gamma_1 k_1,\gamma_2 k_2) = \frac{1}{2\gmax\Delta}\\
    \times \int dx_1\ F_{k_1 k_2}(t,x_1,0) 
    \biggr[\delta_{\gamma_1'}^{\gamma_2} e^{i k_1'(t-x_1)} \\
    + \frac{1}{\gmax}F_{k_1'}^*(t-x_1)   \biggr]
    F_{k_2'}^*(t) \Theta(0<x_1<t),\label{eq: Omega as integrals mcIRL}
\end{multline}
where $F_{k_1 k_2}$ is given by Eq. \eqref{eq: F n=2 IRL} with $\mathcal{T}_U \to U$ (since we work to leading order in $U$).

We focus on the steady state limit.  Including the zeroth order answer \eqref{eq: occupancy mcRLM}, we find the following result for the occupancy to first order in $U$ (see Sec. \ref{sec: Dot occupancy in the multilead IRL Appendix} in the Appendix for details):
\begin{widetext}
\bseq
\begin{align}
    \langle n_d \rangle_{\text{s.s.}} &\equiv   \lim_{t\to\infty}  \langle n_d\rangle_t  \\
    &=\langle n_d\rangle_{\text{s.s.}}^{(0)}
    +\frac{U}{2\gmax \Delta} \Biggr\{ \left[\frac{1}{2\pi \Delta}\sum_{\gamma_1=1}^{\gmax} \left( D+ \mu_{\gamma_1}\right) - \langle n_d\rangle_{\text{s.s.}}^{(0)} \right] \sum_{\gamma_2=1}^{\gmax}\int_{-D}^D \frac{dk_2}{2\pi}\  \fermifn_{\gamma_2}(k_2) |\mathcal{T}(k_2)|^2  \text{Re}\left[ \mathcal{T}(k_2)\right]\notag\\
    &\qquad - \frac{1}{\Delta} \sum_{\gamma_1,\gamma_2=1}^{\gmax}\int_{-D}^{D} \frac{dk_1}{2\pi}\frac{dk_2}{2\pi}\  \fermifn_{\gamma_1}(k_1)\fermifn_{\gamma_2}(k_2) \text{Re}\left[ \mathcal{T}(k_1)\right]
    |\mathcal{T}(k_2)|^2 \left(\delta_{\gamma_1}^{\gamma_2} - \frac{1}{2\gmax} |\mathcal{T}(k_2)|^2 \right)\Biggr\} + O(U^2),\label{eq: occupancy bare mcIRL}
\end{align}
\eseq
\end{widetext}
where error terms exponentially small in bandwidth---$O\left(e^{-\frac{D-|\mu_\gamma|}{T}} \right)$---have been dropped. 

As usual in a field theory calculation, this answer diverges as the bandwidth is sent to infinity.  In this case, there is both a linear and a logarithmic divergence.  In the next section, we perform the necessary steps---re-expressing the answer in terms of physical parameters rather than bare parameters---to get a meaningful result.

\subsubsection{Universality in and out of equilibrium}

To obtain universal results, we take the scaling limit, in which all energy scales are much smaller than the bandwidth.

We replace the bare parameter $\epsilon$ by a physical parameter $\epsilon_d$ by making the following shift:
\beq
    \epsilon = \epsilon_d - U \sum_{\gamma=1}^{\gmax} (D+\mu_\gamma)/(2\pi) + U \Delta/2. \label{eq: epsilon shift mcIRL}
\eeq
To explain this, we recall that the interaction term of the IRL would usually take the normal ordered form $H_{\text{conventional}}^{(1)} = \sum_{\gamma=1}^{\gmax} U :\psi_\gamma^\dagger(0) \psi_\gamma(0): \left(d^\dagger d - 1/2\right)$, which corresponds to half-filling in the lattice model.  Relative to our $H^{(1)}$ [Eq. \eqref{eq: H1 mcIRL}], this shifts the dot energy by $U \sum_{\gamma=1}^{\gmax} (D+\mu_\gamma)/(2\pi)$ and introduces a potential scattering term $-\frac{1}{2}U\psi^\dagger(0)\psi(0)$ (there is also an overall energy shift that has no effect).  The point is that with $H_{\text{conventional}}^{(1)}$ as the interaction term, the equilibrium resonance is at $\epsilon_{\text{conventional}}=0$.  Though we can shift our $\epsilon$ easily enough, our calculation does not include the potential scattering term.  We find, however, that at least to the leading order in $U$, the equilibrium resonance can be fixed at $\epsilon_d=0$ by including another shift: the $\Delta$-dependent term in Eq. \eqref{eq: epsilon shift mcIRL}.

In the above argument, we assumed that the normal ordering in $H_{\text{conventional}}^{(1)}$ was done relative to the initial state of the quench (free Fermi seas in each lead with arbitrary chemical potentials), so that $  \psi_{\gamma}^\dagger(0) \psi_{\gamma}(0) - :\psi_{\gamma}^\dagger(0)\psi_{\gamma}(0):\ = \frac{1}{2\pi} \sum_{\gamma=1}^{\gmax} (D +\mu_\gamma )$.  Had we instead done normal ordering relative to the noninteracting \emph{equilibrium} ground state, then all $\mu_\gamma$ would be set to zero in Eq. \eqref{eq: epsilon shift mcIRL} and our answers below would be modified.  We suggest that the prescription we use is the appropriate generalization beyond the equilibrium case.

Working to first order in $U$ and using Eq. \eqref{eq: occupancy mcRLM} and the identity $\frac{\pd}{\pd \epsilon} |\mathcal{T}(k)|^2 = \frac{1}{\Delta} |\mathcal{T}(k)|^2 \text{Re}\ \mathcal{T}(k)$, we obtain
\begin{widetext}
    \begin{multline}
        \langle n_d \rangle_{\text{s.s.}} = \langle n_d\rangle_{\text{s.s.}}^{(0)} + \frac{U}{2\gmax\Delta}\Biggr[ \left( \frac{1}{2}- \langle n_d\rangle_{\text{s.s.}}^{(0)} \right)\sum_{\gamma=1}^{\gmax} \int_{-D}^D \frac{dk}{2\pi}\ \fermifn_\gamma(k) |\mathcal{T}(k)|^2 \text{Re} \left[ \mathcal{T}(k) \right]\\
        - \frac{1}{\Delta} \sum_{\gamma_1,\gamma_2=1}^{\gmax} \int_{-D}^{D} \frac{dk_1}{2\pi} \frac{dk_2}{2\pi}\  \fermifn_{\gamma_1}(k_1)\fermifn_{\gamma_2}(k_2) \text{Re}\left[ \mathcal{T}(k_1) \right] |\mathcal{T}(k_2)|^2 \left(\delta_{\gamma_1}^{\gamma_2} - \frac{1}{2\gmax} |\mathcal{T}(k_2)|^2 \right)\Biggr] +O(U^2) ,\label{eq: occupancy mcIRL}
    \end{multline}
\end{widetext}
where each $\mathcal{T}$ matrix is now evaluated with $\epsilon_d$ instead of the original $\epsilon$ [i.e., $\mathcal{T}(k) = 2\Delta(k- \epsilon_d + i \Delta)^{-1}$], including in the free occupancy $\langle n_d \rangle_{\text{s.s}}^{(0)}$ as given in Eq. \eqref{eq: occupancy mcRLM}.

Since $\mathcal{T}(k)\sim 1/k$ for large $k$, the second line of \eqref{eq: occupancy mcIRL} diverges logarithmically for large $D$.  This encodes the emergence of a universal scale through the Callan-Symanzik equation:
\bseq
\begin{align}
    &\left( D \frac{\pd}{\pd D} + \beta_\Delta \Delta \frac{\pd }{\pd \Delta} \right) \langle n_d \rangle_{\text{s.s.}} = O(1/D), \label{eq: CS eqn mcIRL}\\
    &\qquad \text{where }\beta_\Delta = - \frac{U}{\pi} + O(U^2). \label{eq: beta fn mcIRL}
\end{align}
\eseq
To see that the Callan-Symanzik equation holds, we note
\begin{multline}
    D \frac{\pd}{\pd D} \langle n_d \rangle_{\text{s.s.}} \overset{D\to\infty}{\longrightarrow}
    \frac{U}{2\pi\gmax \Delta} \sum_{\gamma_2=1}^{\gmax}\int_{-\infty}^\infty\frac{dk_2}{2\pi}\ \fermifn_{\gamma_2}(k)\\
    \times |\mathcal{T}(k_2)|^2\left(1 - \frac{1}{2} |\mathcal{T}(k_2)|^2 \right),\label{eq: D term in CS mcIRL}
\end{multline}
which follows from $D\mathcal{T}(\pm D)\overset{D\to\infty}{\longrightarrow} \pm 2 \Delta$ and simple properties of the Fermi function.  Then we obtain the beta function as in Eq. \eqref{eq: beta fn mcIRL} from Eq. \eqref{eq: occupancy mcRLM} and the identity $\Delta \frac{\pd}{\pd \Delta} \left( \frac{1}{\Delta}|\mathcal{T}(k)|^2 \right) = \frac{1}{\Delta} |\mathcal{T}(k)|^2\left(1 - \frac{1}{2} |\mathcal{T}(k)|^2\right)$.

Now that we are focusing on the large bandwidth regime, we can confirm that $\epsilon_d=0$ is the location of the equilibrium resonance in Eq. \eqref{eq: occupancy mcIRL}.  Setting all $\fermifn_\gamma(k) = \fermifn(k)$ (i.e., all chemical potentials $\mu_\gamma$ set to $0$) and $\epsilon_d=0$, we have $\langle n_d\rangle_{\text{s.s.}}^{(0)} = 1/2 +O(1/D)$ and $\int_{-D}^D dk\ \fermifn(k) |\mathcal{T}(k)|^2\left(1 - \frac{1}{2} |\mathcal{T}(k)|^2\right) = O(1/D)$ (shown numerically), so that $\langle n_d\rangle_{\text{s.s.}} =1/2$.

The Callan-Symanzik equation encodes the fact that for large bandwidth, $\langle n_d \rangle_{\text{s.s.}}$ takes a universal form, depending only on the external parameters ($\epsilon_d$ and the temperatures and chemical potentials of the leads) and on two scaling invariants.  The first invariant is an emergent energy scale
\beq
    T_K = \left( 1 -  \frac{U}{\pi}\right) D \left(\frac{\Delta}{D}\right)^{\frac{1}{1+U/\pi}}. \label{eq: TK mcIRL}
\eeq
It can be verified that $\left( D \frac{\pd}{\pd D} + \beta_\Delta \Delta \frac{\pd }{\pd \Delta} \right) T_K = 0$ and that $T_K = \Delta$ for $U=0$.  The $U$-dependent overall scale of $T_K$ is arbitrary (as we discuss in more detail below), and we have chosen it so that the equilibrium susceptibility of the dot at $\epsilon_d=0$ takes the form \cite{KarraschEtAl, CamachoSchmitteckertCarr} $-\frac{\pd}{\pd \epsilon_d}\rvert_{T=\epsilon_d=0}\langle n_d\rangle_{\text{s.s.}} = 1/(\pi T_K)$.  The second scaling invariant is the coupling constant $U$ (or equivalently, the parameter $\alpha$ defined below).  Thus, staying always in the large bandwidth regime from now on, we can write
\beq
    \langle n_d \rangle_{\text{s.s.}} = f_{\text{universal}}\left(U; \left\{ \frac{T_\gamma}{T_K}\right\};\left\{\frac{\mu_\gamma}{T_K}\right\};\frac{\epsilon_d}{T_K}\right),\label{eq: nd is universal mcIRL}
\eeq
where the brackets indicate all the channels: $\left\{\frac{T_\gamma}{T_K}\right\} = (T_1/T_K,\dots, T_{\gmax}/T_K)$, $\left\{\frac{\mu_\gamma}{T_K}\right\} = (\mu_1/T_K,\dots, \mu_{\gmax}/T_K)$.  Below, we evaluate this universal function to leading order in $U$ in a few regimes at zero temperature.  First, we make some general comments on the RG flow of the model and compare our results with the literature.

\subsubsection{RG discussion}
The RG flow of the model is the following:
\bseq
\begin{align}
    \frac{\pd U}{\pd \ln D} & = 0,\\
    \frac{\pd \ln \Delta}{\pd \ln D} &= \beta_\Delta(U) = -\frac{U}{\pi} + O(U^2).
\end{align}
\eseq
The $U$ parameter, which does not flow, determines the direction of flow of $\Delta$ through the beta function $\beta_\Delta(U)$.  If $\beta_\Delta(U)$ is negative, then $\Delta$ increases as the bandwidth $D$ is reduced.  While our calculation is only to first order in $U$, it is known to all orders that $D\frac{\pd U}{\pd D}=0$, i.e., $U$ does not flow.

While the RG flow of the IRL has been studied by many methods, the most direct comparison we can make to the literature is to other works that have found the flow from the evaluation of an expectation value to leading order in $U$.  In particular, previous work on the two-lead IRL driven by bias has found linear and logarithmic divergences in the charge current.  In Ref. \cite{Doyon}, the linear divergences are removed by a redefinition of $\epsilon$ which we expect to be equivalent to what we did above (although an equation is not given).  In Ref. \cite{NishinoEtAl_PRBonIRL}, these divergences are removed by the same shift of $\epsilon$ that we used above (in the special case $\gmax = 2$), albeit without the additional shift that we included to put the resonance at $\epsilon_d=0$.  In either case, the logarithmic divergences are accounted for by the Callan-Symanzik equation, as we did above, with the same result \eqref{eq: beta fn mcIRL} for the beta function at leading order.

For further comparison with the literature, let us rewrite our equation for $T_K$ [Eq. \eqref{eq: TK mcIRL}] in another form:
\bseq
\begin{align}
    T_K &= \left( 1 -  \frac{U}{\pi}\right) D \left(\frac{\Delta}{D}\right)^{\alpha/2},\\
    &\qquad \text{where } \alpha = \frac{2}{1+  U/\pi}.\label{eq: alpha mcIRL}
\end{align}
\eseq
The exponent $\alpha$ in the RG invariant $T_K$ [Eq. \eqref{eq: TK mcIRL}] has been much discussed in the literature.  Various answers for $\alpha$ as a function of $U$ [or as a function of the single particle phase shift, which is $\delta_U = \arctan(U/2)$ in our case] have been found.  Our answer, Eq. \eqref{eq: alpha mcIRL}, agrees with some Bethe ansatz calculations, but not all, and a different answer has been obtained by bosonization.  See Table I in Ref. \cite{CamachoSchmitteckertCarr} for a summary of the literature.  While all calculations agree that $\alpha=2$ for zero coupling (or zero phase shift), there is disagreement already at the first order correction.

For the purpose of calculating universal quantities, the precise dependence of $\alpha$ on the coupling constant is only meaningful within a particular cutoff scheme.   This theory has two RG invariants, which we choose as $T_K$ and $U$, and they determine results by values assigned to them.  The final outputs of a field theory calculation are functions such as $f_{\text{universal}}$ that have RG invariants as inputs.  The numerical values of the RG invariants are not themselves calculable in field theory.  Instead, one fixes the value of the RG invariants by fitting universal functions to data.  One of the advantages of doing a field theory calculation (on what is ultimately a lattice system) is that one has a great freedom to choose a cutoff scheme that makes the calculation of universal functions more convenient; the price one pays is that \emph{only} these universal functions can be compared meaningfully with a lattice system.

One technical caveat is that the functional form of $\alpha$ \emph{does} matter insofar as it determines the possible values $\alpha$ can take.  This point does not seem to arise in the IRL, seeing as all of the forms of $\alpha$ in the literature permit $\alpha$ to range from $-\infty$ to $\infty$ given $U$ ranging from $-\infty$ to $\infty$.  Note that our calculation in this paper is only consistent for $\alpha$ in a narrow range around $\alpha=2$, since we took $U$ to be small; however, the Bethe ansatz result for $\alpha$ is given by the same Eq. \eqref{eq: alpha mcIRL} with no restriction on $U$.

In Ref. \cite{CamachoSchmitteckertCarr}, Camacho \emph{et al}. use bosonization, and hence have a different functional form of $\alpha$ in terms of $U$.  They emphasize, however, that their final answer for $\langle n_d \rangle_{\text{equilibrium}}$ at zero temperature agrees exactly with the Bethe ansatz answer once both are expressed as functions of $\alpha$ and $\epsilon_d/T_K$.  This agrees with our discussion in the previous paragraph.  To disprove our claim, it would be necessary to find another universal function whose form differs between the bosonization and Bethe ansatz calculations, even after the invariants $\alpha$ and $T_k$ are fixed by matching the answers for, e.g., $\langle n_d \rangle_{\text{equilibrium}}$.

A stronger claim of Camacho in Ref. \cite{Camacho_thesis} is that the formula for $\alpha$ in terms of $U$ (or rather, in terms of the phase shift $\delta_U$) is scheme independent, contrary to what we find in Eq. \eqref{eq: alpha mcIRL}.  Though we have not examined the argument in detail, we wonder if the unconventional cutoff schemes employed in this paper and in the Bethe ansatz might somehow be outside the range of cutoff schemes considered in the bosonization calculation of Ref. \cite{Camacho_thesis}.  (These cutoff schemes are unconventional in that the Hamiltonian formally has all energies.)

Similar comments apply to the $U$-dependent prefactor in $T_K$---its precise dependence on $U$ can differ between schemes.

\subsubsection{Evaluation at zero temperature}

We evaluate the steady state occupancy \eqref{eq: occupancy mcIRL} at zero temperature.  We then use RG improvement to extract the universal function \eqref{eq: nd is universal mcIRL} in a few specific regimes.

The standard method for finding a universal function from a perturbative result is RG improvement: One changes the original parameters $(D,\Delta)$ to new parameters $(D', \Delta')$ with the same value of $T_K$, where $D'$ is chosen so as to eliminate large logarithms in the perturbation series.  The net effect is to delete these large logarithms and to replace $\Delta$ by the ``running'' coupling constant $\Delta'$.  Note that this replacement is only valid on the part of the answer that satisfies the Callan-Symanzik equation---thus, one must first take $D$ to be large before applying RG improvement. 

In the zero temperature limit, the momentum integrals in Eq. \eqref{eq: occupancy mcIRL} can all be carried out analytically to yield
\beq
    \langle n_d \rangle_{\text{s.s.}}^{(0)} = \frac{1}{2} -\frac{1}{\gmax \pi} \sum_{\gamma=1}^{\gmax} \arctan\frac{\epsilon_d - \mu_\gamma}{\Delta},
\eeq
and
\begin{widetext}
\begin{multline}
    \langle n_d \rangle_{\text{s.s.}} = \langle n_d\rangle_{\text{s.s.}}^{(0)} + \frac{U}{\gmax \pi} \Biggr[ -\frac{1}{4}\left(\frac{1}{2} - \langle n_d \rangle_{\text{s.s.}}^{(0)}\right) \sum_{\gamma=1}^{\gmax} |\mathcal{T}(\mu_\gamma)|^2 + \sum_{\gamma_1=1}^{\gmax} \Bigg( \frac{1}{\pi \gmax} \sum_{\gamma_2=1}^{\gmax} \frac{\epsilon_d - \mu_{\gamma_2}}{\Delta}|\mathcal{T}(\mu_{\gamma_2})|^2\\
    +\frac{1}{2} - \frac{1}{\pi}\arctan\frac{\epsilon_d - \mu_{\gamma_1}}{\Delta} - \langle n_d \rangle_{\text{s.s.}}^{(0)} \Bigg) \ln \frac{D}{\sqrt{(\epsilon_d-\mu_{\gamma_1})^2 + \Delta^2}} \Biggr].\label{eq: nd zero temp mcIRL} 
\end{multline}
\end{widetext}
Note that there are large logarithms with many different scales involved, so that there is no one choice of $D'$ that will eliminate all of them in the general case (arbitrary chemical potentials $\mu_\gamma$).  We proceed to specialize to some specific regimes in which there are just one or two different large logs to be eliminated.

\paragraph{Equilibrium.}
Setting all chemical potentials to zero, we find
\begin{multline}
    \langle n_d \rangle_{\text{s.s.}} =
    \frac{1}{2} - \frac{1}{\pi} \arctan \frac{\epsilon_d}{\Delta }\\
    +\frac{U}{\pi^2}\frac{\Delta}{\epsilon_d^2 +\Delta^2} \left[\epsilon_d \ln \frac{D}{\sqrt{\epsilon_d^2 + \Delta^2} }- \Delta \arctan\frac{\epsilon_d}{\Delta}   \right].
\end{multline}
The large logarithm is to be eliminated by the self-consistent choice $D' = \sqrt{\epsilon_d^2 + (\Delta')^2}$, which determines the running coupling:
\beq
    \Delta' = \left\{  1 + \frac{U}{\pi} \left[1 - \frac{1}{2} \ln \left(1 + \frac{\epsilon_d^2}{T_K^2}\right) \right] \right\} T_K.
\eeq
We thus obtain a universal answer, valid to leading order in $U$:
\begin{multline}
    \langle n_d \rangle_{\text{s.s.}} =\frac{1}{2} - \frac{1}{\pi}\arctan \frac{\epsilon_d}{T_K} 
    +\frac{U}{2\pi^2(1+\epsilon_d^2/T_K^2)}\\
    \times \left[ 2\left(  \frac{\epsilon_d}{T_K} - \arctan \frac{\epsilon_d}{T_K}\right)-\frac{\epsilon_d}{T_K}\ln \left(1+ \frac{\epsilon_d^2}{T_K^2} \right)  \right],\label{eq: ndeq univ mcIRL}
\end{multline}
which agrees with the leading order expansion of the exact equilibrium result from Bethe ansatz \cite{Ponomarenko} (see Appendix \ref{sec: Equilibrium occupancy of the IRL from the literature}).  This confirms, at least in the zero temperature limit and to this order, that in the long-time limit following the quench, the occupancy thermalizes.

We emphasize that the output of our field theory calculation is a two-parameter \emph{family} of functions of the physical quantity $\epsilon_d$, parameterized by $U$ and $T_K$.  Redefinitions of $U$ and $T_K$ can change the details of the parametrization, but not the full family of functions that is obtained by letting $U$ and $T_K$ range over all allowed values.  We brought our answer to the form \eqref{eq: ndeq univ mcIRL} as a convenient way of showing that the full family of functions agrees with the Bethe ansatz result in the parameter range we consider:  $U$ small (or equivalently, $\alpha$ close to $2$) and $T_K$ arbitrary. 

In the $U$-dependent part of Eq. \eqref{eq: ndeq univ mcIRL}, only the coefficients of the arctangent and logarithm terms have universal meaning.  Replacing the term $U\Delta/2$ by $a U$ (with a varying parameter $a$) in the shift \eqref{eq: epsilon shift mcIRL} controls a term proportional to $1/(1+\epsilon_d^2/T_K^2)$; we took $a=1/2$ to eliminate this term, putting the resonance at $\epsilon_d=0$.  [This choice also puts the resonance at $\epsilon_d =0$ for arbitrary temperature, as we showed below Eq. \eqref{eq: D term in CS mcIRL}.]  Similarly, we can adjust the coefficient of the $(\epsilon_d/T_K)/(1+\epsilon_d^2/T_K^2)$ term in Eq. \eqref{eq: ndeq univ mcIRL} by varying a parameter $b$ in $T_K = [1+ b U]D(\Delta/D)^{2\alpha}$; this term controls the dot susceptibility at $\epsilon_d=0$, and our choice of $b=-1/\pi$ normalizes $T_K$ according to $T_K^{-1} = - \pi \frac{\pd}{\pd \epsilon_d}\rvert_{T=\epsilon_d=0} \langle n_d\rangle_{\text{s.s}}$.

\begin{figure*}[htb]
    \subfloat{%
    \includegraphics[width=\columnwidth]{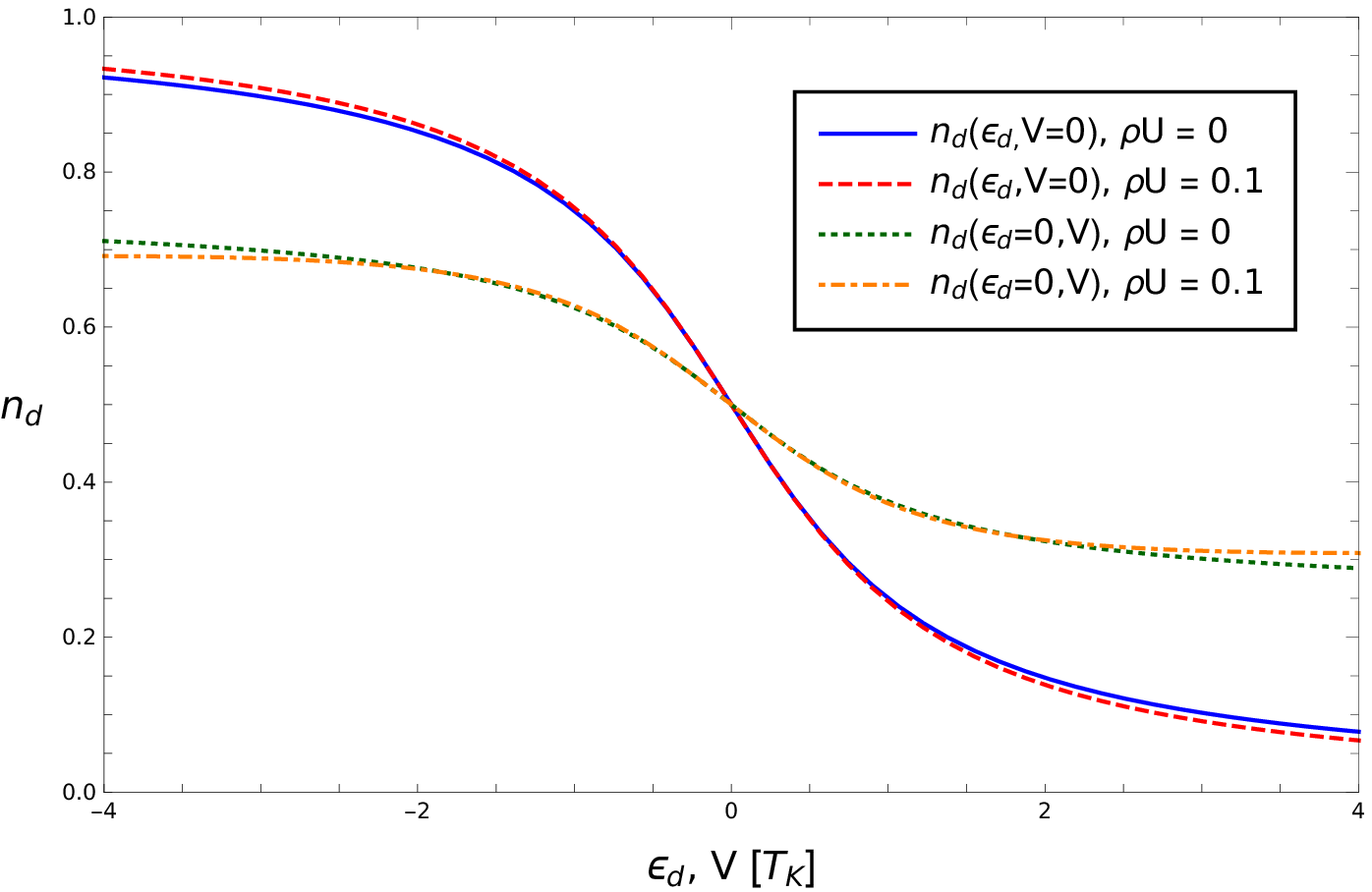}%
    }\hfill
    \subfloat{%
    \includegraphics[width=\columnwidth]{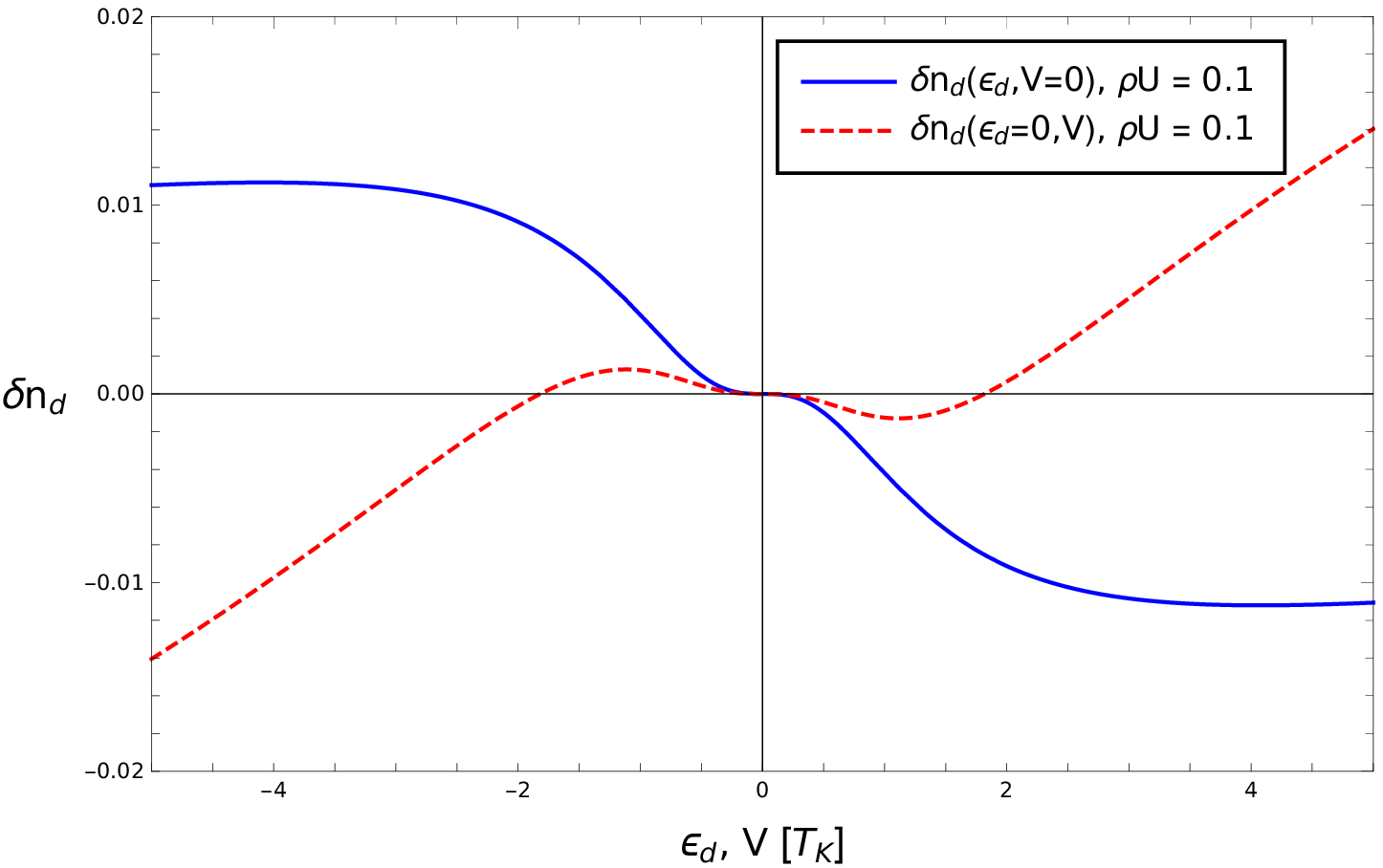}%
    }
    \caption{Left: the steady state occupancy $n_d \equiv \langle n_d \rangle_{\text{s.s.}}$ at zero temperature in the two-lead IRL, either as a function of dot potential $\epsilon_d$ or voltage $V$.  The leads are held at chemical potentials $\mu_1 =0$ and $\mu_2=-V$.  The equilibrium ($V=0$) curves are given by Eq. \eqref{eq: ndeq univ mcIRL} and in fact are independent of the number of leads, in agreement with the Bethe ansatz answer from the literature (Appendix \ref{sec: Equilibrium occupancy of the IRL from the literature}).  The nonequilibrium ($V\ne0$) curves are given by Eq. \eqref{eq: ndnoneq univ 2leadIRL}.  In both cases, we compare the noninteracting occupancy ($\rho U =0$) with the weakly interacting occupancy (first order in $\rho U = 0.1$), where $\rho=1/(2\pi)$ is the density of states per unit length.  Right: the weakly interacting case with the noninteracting occupancy subtracted, i.e., $\delta n_d \equiv n_d - n_d \rvert_{U=0}$.  In equilibrium, $\delta n_d$ reaches  finite limits as $\epsilon_d/T_K \to\pm\infty$.  Out of equilibrium, $|\delta n_d|$ grows logarithmically as $V/T_K \to\pm\infty$, indicating that some resummation of the series in $\rho U$ is needed to make sense of the extremely large voltage regime.}\label{fig: nd and deltand}
\end{figure*}

\paragraph{Out of equilibrium---two leads at $\epsilon_d=0$.}  Consider the two-lead model with the leads separated by a bias voltage $V$ and with the dot potential set to zero---that is, $\gmax=2$, $\mu_1 = 0$, $\mu_2=-V$, and $\epsilon_d=0$.  (The case of arbitrary $\epsilon_d$ is also possible, but messier.)  The occupancy \eqref{eq: nd zero temp mcIRL} contains two large logarithms, $\ln\frac{D}{\Delta}$ and $\ln\frac{D}{\sqrt{\Delta^2 + V^2}}$; we can choose $D'$ to cancel either one, with the same final result (see Fig. \ref{fig: nd and deltand} as well):
\begin{multline}
    \langle n_d\rangle_{\text{s.s.}}  = \frac{1}{2} - \frac{\arctan\frac{V}{T_K}}{2\pi}+ \frac{U}{2\pi^2} \Biggr\{\frac{\frac{V^2}{T_K^2}\arctan \frac{V}{T_K}}{2\left( 1 + \frac{V^2}{T_K^2}\right)} \\
    -\left( \arctan \frac{V}{T_K}- \frac{\frac{V}{T_K}}{1+\frac{V^2}{T_K^2}}\right)\left[ 1 - \frac{1}{4} \ln\left( 1+ \frac{V^2}{T_K^2} \right) \right]   \Biggr\}.\label{eq: ndnoneq univ 2leadIRL}
\end{multline}
The particular numbers that appear in this answer become meaningful once the values of $U$ and $T_K$ are fixed by, e.g., matching the equilibrium answer \eqref{eq: ndeq univ mcIRL} with data.  Note that the contribution of the interaction begins at order $V^2$, beyond linear response.

The leading correction in $U$ in Eq. \eqref{eq: ndnoneq univ 2leadIRL} grows logarithmically with voltage as $V/T_K\to \pm \infty$; this is a consequence of the fact that no choice of $D'$ can cancel both of the large logarithms.  This implies that some resummation of the series in $U$ is needed to make sense of the regime of very large voltage.  We can characterize the scale at which the $U$ series breaks down out of equilibrium as the voltage $V_0$ for which the $U$ correction term ($\delta n_d$ in Fig. \ref{fig: nd and deltand}) equals $1/2$; the result is $V_0 \sim T_K e^{2/(\rho U)}$, where $\rho=1/(2\pi)$ is the density of states per unit length in our convention.  The number $2$ in the exponent is not sharply defined, since we had to make an arbitrary choice for what value of the $U$ correction is large enough to say that the series breaks down.  Though our calculation sends the bandwidth $D\to \infty$, we suggest that this scale $V_0$ could also be significant in the lattice model if it lies in the universal regime, i.e., if $V_0 \ll D_{\text{lattice}}$.  The scale $V_0$ may be connected to the power law dependence on $U$ seen in Ref. \cite{KarraschEtAl}.

\paragraph{Out of equilibrium---two leads close to the particle-hole symmetric point.} We again consider the two-lead model with the leads separated by a bias voltage $V$, this time with $\epsilon_d$ close to halfway between the two chemical potentials.  That is, we set $\mu_1 = \epsilon_d + V/2$ and $\mu_2 = \epsilon_d - V/2 - \delta V$.  For $\delta V=0$, the steady state occupancy is its free value, $1/2$.  Self-consistently setting $D' = \sqrt{(\Delta')^2 + V^2/4}$, we obtain the following correction for small $\delta V$:
\begin{multline}
    \langle n_d\rangle_{\text{s.s.}} =\frac{1}{2} 
    - \frac{1 }{2\pi\left(1 + \frac{V^2}{4 T_K^2} \right)} \Biggr\{ 1 + \frac{U}{\pi}\\
    \times \Bigg[ \frac{\frac{V^2}{4T_K^2} + \frac{1}{2} \left( 1-\frac{V^2}{4T_K^2}\right)\ln\left(1+ \frac{V^2}{4T_K^2}\right)}{1 + \frac{V^2}{4 T_K^2}}\\
    - \frac{V}{2T_K}\arctan \frac{V}{2T_K} \Bigg] \Biggr\}\frac{\delta V}{T_K}.
\end{multline}
As before, this expression is valid for $V \ll T_K e^{1/(2U)}$ (in addition to requiring $U$ and $\delta V/T_K$ to be small).

\subsection{Steady state current in the two-lead AIM}\label{sec: Steady state current in the two-lead AIM}
We evaluate the steady state current in the two-lead AIM in the approximation that no more than two quantum numbers can be in a crossing state.  We see below that this approximation encompasses both the regime of weak coupling (small $U/\Delta$) and strong coupling with weak tunneling ($U\to\infty$ with small $\Delta/|\epsilon-\mu_\gamma|$).  Our result for small $U$ agrees with a calculation that we did using Keldysh perturbation theory (see Appendix \ref{sec: Perturbative check the current in the Anderson model}), and our result for large $U$ reproduces a well-known scaling law.

Throughout this section, $H = H_{\text{finite } U}$ is the two-lead AIM given by Eq. \eqref{eq: HfiniteU mcAIM} (with $\gmax = 2$).  We work directly in the steady state limit, which means in particular that the system size is infinite.  We therefore use Dirac normalized operators: $c_{\gamma k a}^\dagger = \int_{-\infty}^{\infty}dx\ e^{i k x}\psi_{\gamma a}^\dagger(x)$.

\subsubsection{Setup and reduction to an overlap}

The current operator in the AIM for electrons leaving lead $\gamma$ (with $\gamma=1,2$) is well known to be $\widehat{I}_\gamma \equiv \frac{iv }{\sqrt{2}} \psi_{\gamma a}^\dagger(0) d_a + \text{h. c.}$ (see, e.g., Ref. \cite{MeirWingreen}).  Since the two currents are equal and opposite in the steady state ($I_1 =-I_2$), we can consider the symmetrized operator
\beq
    \widehat{I}_{\text{Sym}} = \frac{i}{2\sqrt{2}} v \left( \psi_{1 a}^\dagger(0) - \psi_{2 a}^\dagger(0) \right) d_a + \text{h.c.}
\eeq
Our task is to evaluate this operator in the nonequilibrium steady state.  That is, we wish to evaluate
\beq
    \langle \widehat{I}_{\text{Sym}} \rangle \equiv \mathcal{N}^{-1} \langle \Psi_\text{in} | \widehat{I}_{\text{Sym}} | \Psi_\text{in} \rangle,\label{eq: ISym expectation value def}
\eeq
where the normalization factor $\mathcal{N} \equiv \langle \Psi| \Psi\rangle$ is discussed in more detail below, and where $|\Psi_\text{in}\rangle$ is the Lippmann-Schwinger ``in'' state corresponding to two Fermi seas.  That is,
\bseq
\begin{align}
    &|\Psi_\text{in} \rangle = |\Psi \rangle + \frac{1}{E - h +i \eta}\mathcal{V} |\Psi_{\text{in}}\rangle,\label{eq: Psi_in LS form AIM}\\
    &\text{where }|\Psi \rangle = c_{\gamma_N k_N a_N}^{ \dagger} \dots c_{\gamma_1 k_1 a_1}^{\dagger} |0\rangle,\\
    &h = -i \int dx\ \sum_{\gamma=1,2} \psi_{\gamma a}^\dagger(x) \frac{d}{dx} \psi_{\gamma a}(x),\\
    & \mathcal{V} = H - h.\label{eq: mathcalV AIM}
\end{align}
\eseq
The quantum numbers $(\gamma_j,k_j,a_j)$ are arbitrary for the moment; they will later be specialized to describe two Fermi seas with an applied bias voltage appearing as the difference of the chemical potentials.

To simplify the calculation, we now write the expectation value of the current operator (i.e., a matrix element) as the derivative of an overlap, using an approach that we have presented in more generality in Ref. \cite{Culver_thesis}.  The idea is to add the current operator $\widehat{I}_{\text{Sym}}$ to the Hamiltonian as a source term in such a way that we can read off the wavefunction for the Hamiltonian (with source) from our previous results.

Let $\overline{\phi}$ be a real variable (the bar is a label and does not signify complex conjugation) and consider the following $\overline{\phi}$-dependent Hamiltonian:
\begin{multline}
    \overline{H} = H +  \bigg[ \frac{v}{\sqrt{2}} \left( e^{i \frac{1}{2} \overline{\phi} } -1 \right) \psi_{1a}^\dagger(0) d_a \\
    + \frac{v}{\sqrt{2}} \left( e^{-i \frac{1}{2} \overline{\phi} } -1 \right) \psi_{2a}^\dagger(0) d_a + \text{ h.c.} \bigg].\label{eq: Hbar as H + source}
\end{multline}
Note that setting $\overline{\phi}=0$ recovers the original Hamiltonian.  From here on, an overbar means that a quantity depends on $\overline{\phi}$, and removing the bar corresponds to setting $\overline{\phi}= \phi \equiv 0$.

We are interested in the expectation value of $\widehat{I}_{\text{Sym}}$ in some eigenstate $|\Psi(E)\rangle$ of $H$ with energy $E$.  Since we work in infinite volume, the energy varies continuously, so there is also a family of eigenstates $|\Psi(E')\rangle$ with varying energy $E'$.  Let $|\overline{\Psi}(E)\rangle$ be any $\overline{\phi}$-dependent family of eigenstates of $\overline{H}$ (with energy $E$) such that $|\overline{\Psi}(E) \rangle \rvert_{\overline{\phi} = 0} = |\Psi(E)\rangle$ (a condition that is built-in to our notation).  Then we have the following expression for the unnormalized expectation value:
\begin{multline}
    \langle \Psi(E)| \widehat{I}_{\text{Sym}} |\Psi(E) \rangle =\\
    \lim_{E'\to E}\left( E - E'\right) \frac{\pd}{\pd \overline{\phi}}\biggr \rvert_{\overline{\phi}=0}  \langle \Psi(E') | \overline{\Psi} (E)\rangle.\label{eq: deriv formula first AIM}
\end{multline}
Naively, the right-hand side appears to be zero; however, we find in practice that the $\overline{\phi}$ derivative produces a $1/(E-E')$ pole that cancels the prefactor.

The proof of Eq. \eqref{eq: deriv formula first AIM} follows from noting that  $\widehat{I}_{\text{Sym}} = \frac{\pd}{\pd \overline{\phi}} | _{\overline{\phi} = 0} (\overline{H} -H)$ and dropping the term $\langle \Psi(E') | (\overline{H} - H) \frac{\pd}{\pd \overline{\phi}} |\overline{\Psi}(E)\rangle|_{\overline{\phi}=0}$.  In principle, it must be checked that $\frac{\pd}{\pd \overline{\phi}} |\overline{\Psi}(E)\rangle$ is not too singular as $\overline{\phi}\to 0$; this is not an issue in our calculation below, since the dependence on $\overline{\phi}$ will be analytic.  To avoid any possible issues with order of limits, we will apply Eq. \eqref{eq: deriv formula first AIM} before taking the thermodynamic limit.

The eigenstate of interest is $|\Psi(E)\rangle = |\Psi_{\text{in}}\rangle$, which has energy $E= \sum_{j=1}^N k_j$.  A convenient choice for the $E'$-dependent states $|\Psi(E')\rangle $ is to simply let the momenta vary; thus, we write $|\Psi_{\text{in}}'\rangle$ for same Lippmann-Schwinger state \eqref{eq: Psi_in LS form AIM} with momenta $k_1,\dots,k_N$ replaced by $k_1',\dots,k_N'$.  Then the energy $E' = \sum_{j=1}^N k_j'$ varies continuously.

We have a considerable freedom in constructing the $\overline{\phi}$-dependent states $|\overline{\Psi}(E)\rangle$.  It is convenient to bring the $\overline{\phi}$-dependent Hamiltonian \eqref{eq: Hbar as H + source} to the same form as the original Hamiltonian, allowing us to use the wavefunction already obtained.  To do this, we define a convenient set of $\overline{\phi}$-dependent fields by a unitary transformation:
\bseq
\begin{align}
    &\begin{pmatrix}
        \overline{c}_{1 k a}\\
        \overline{c}_{2 k a}
    \end{pmatrix}
    = \mathcal{U}^\dagger \overline{\mathcal{U}} 
    \begin{pmatrix}
        c_{1k a}\\
        c_{2ka }
    \end{pmatrix}
    ,\\
    &\text{where } \overline{\mathcal{U}} = \frac{1}{\sqrt{2}}
    \begin{pmatrix}
        e^{- i\frac{1}{2} \overline{\phi}} & -e^{i\frac{1}{2} \overline{\phi}} \\
        e^{-i \frac{1}{2} \overline{\phi} } & e^{i \frac{1}{2} \overline{\phi}}
    \end{pmatrix}
    ,\\
    &\text{which implies } \mathcal{U} = \frac{1}{\sqrt{2}}
    \begin{pmatrix}
        1  & -1 \\
        1 & 1
    \end{pmatrix}.
\end{align}
\eseq
Then we have
\begin{multline}
    \overline{H} =  -i \int_{-L/2}^{L/2 } dx\ \sum_{\gamma=1,2} \overline{\psi}_{\gamma a}^\dagger(x) \frac{d}{dx}\overline{\psi}_{\gamma a}(x) + \epsilon d_a^\dagger d_a \\
    + \left[\frac{v}{\sqrt{2}} \sum_{\gamma=1,2}\overline{\psi}_{\gamma a}^\dagger(0) d_a + \text{ h.c.} \right]+ U n_{\uparrow} n_{\downarrow},\label{eq: HBar twoleadAIM}
\end{multline}
which is the same Hamiltonian already considered, with each unbarred electron field replaced by the corresponding barred field.  We know the ``in'' states of this Hamiltonian provided that the incoming plane waves are in the barred basis.  Hence, it is convenient to let the $\overline{\phi}$-dependent family of eigenstates be as in Eqs. \eqref{eq: Psi_in LS form AIM}-\eqref{eq: mathcalV AIM}, with a bar over everything:
\bseq
\begin{align}
    &|\overline{\Psi}(E)\rangle \equiv |\overline{\Psi}_\text{in} \rangle = |\overline{\Psi} \rangle + \frac{1}{E - \overline{h} +i \eta}\overline{\mathcal{V}} |\overline{\Psi}_{\text{in}}\rangle,\\
    &\text{where }|\overline{\Psi}\rangle = \overline{c}_{\gamma_N k_N a_N }^{\dagger} \dots \overline{c}_{\gamma_1 k_1 a_1}^{ \dagger} |0\rangle,\\
    &\overline{h} = -i \int dx\ \sum_{\gamma=1,2} \overline{\psi}_{\gamma a}^\dagger(x) \frac{d}{dx} \overline{\psi}_{\gamma a}(x),\\
    & \overline{\mathcal{V}} = \overline{H} - \overline{h}.
\end{align}
\eseq
By construction, these states satisfy the required condition, namely they reduce to the original state of interest \eqref{eq: Psi_in LS form AIM} at $\overline{\phi}=0$.  Eq. \eqref{eq: deriv formula first AIM} then yields
\beq
    \langle \widehat{I}_{\text{Sym}}\rangle = 
    \mathcal{N}^{-1} \lim_{\text{all } k_j' \to k_j} \left(E - E'\right)    \frac{\pd }{\pd \overline{\phi} }\biggr\rvert_{\overline{\phi} = 0}\langle \Psi_\text{in}' |\overline{\Psi}_\text{in} \rangle.\label{eq: deriv formula second AIM}
\eeq
Thus, the calculation reduces to finding the overlap $\langle \Psi_\text{in}' |\overline{\Psi}_\text{in} \rangle$ for $\overline{\phi}$ near $0$ and $E'$ near $E$.

In the expectation value \eqref{eq: ISym expectation value def}, one may have expected the normalization factor $\mathcal{N}$ to be $\langle \Psi_{\text{in}}| \Psi_{\text{in}}\rangle$; however, comparison with the time-dependent version of the calculation shows that the correct normalization is $\mathcal{N} = [2\pi\delta(0)]^N = \langle \Psi | \Psi\rangle$.  The full overlap $\langle \Psi_{\text{in}}| \Psi_{\text{in}}\rangle$ seems to contain additional delta function terms beyond the noninteracting norm $[2\pi\delta(0)]^N$ (though it could be that these terms have no effect in the thermodynamic limit).

\subsubsection{Evaluation}\label{sec: Evaluation AIM current}
We evaluate the right-hand side of Eq. \eqref{eq: deriv formula second AIM} with the wavefunction truncated so that no more than two quantum numbers can be assigned to a crossing state---that is, $|\overline{\Psi}_\text{in}\rangle = |\overline{\Psi}_\text{in}^0\rangle + |\overline{\Psi}_\text{in}^2 \rangle$ and $_{\text{in}}\langle \Psi'| = {}_{\text{in}}\langle \Psi'^0|  + {}_{\text{in}}\langle \Psi'^2| $.  We work to first order in the crossing state, i.e.,
\beq
    \langle \widehat{I}_{\text{Sym}} \rangle = \langle \widehat{I}_{\text{Sym}} \rangle^{(0,0)} + \langle \widehat{I}_{\text{Sym}} \rangle^{(0,2)} +\langle \widehat{I}_{\text{Sym}} \rangle^{(2,0)},\label{eq: current as 00+02+20 AIM}
\eeq
where
\begin{multline}
    \langle \widehat{I}_{\text{Sym}} \rangle^{(\ell_1,\ell_2)} = [\langle \Psi|\Psi\rangle]^{-1} \lim_{\text{all } k_j' \to k_j} \left(E - E'\right) \\
    \times \frac{\pd }{\pd \overline{\phi} }\biggr\rvert_{\overline{\phi} = 0}\langle \Psi_\text{in}'^{\ell_1} |\overline{\Psi}_\text{in}^{\ell_2} \rangle.
\end{multline}
The term $\langle \widehat{I}_{\text{Sym}} \rangle^{(2,2)}$ is not kept as it involves the product of two crossing states.  We will see below that in the small $U$ regime, expanding in crossings amounts to expanding in $U$, and our calculation is to first order \footnote{More generally, the pattern seems to be that a crossing state with $n$ quantum numbers is $O(U^{n-1})$.}.  For $U\to\infty$, the expansion in crossings appears to be an expansion in powers of $\Delta$.

The terms of the wavefunction that we need are
\beq
    |\overline{\Psi}_{\text{in}}^0\rangle = \left( \prod_{j=1}^N \overline{c}_{\gamma_j k_j a_j ,\text{in}}^\dagger \right) |0\rangle,\label{eq: Psi0Bar AIM}
\eeq
and
\begin{multline}
    |\overline{\Psi}_{\text{in}}^2\rangle = \frac{1}{2} \sum_{1\le m_1 < m_2 \le N} (-1 )^{m_1 +m_2 +1} \\
    \times \left( \prod_{\substack{j=1 \\ j\ne m_1,m_2}}^N \overline{c}_{\gamma_j k_j a_j ,\text{in}}^\dagger  \right) |\overline{\Phi}_{e k_{m_1}a_{m_1} e k_{m_2} a_{m_2} ,\text{in} }\rangle.
\end{multline}
We can take the adjoint, remove the bar, and relabel each $k_j\to k_j'$ to get $\langle \Psi_{\text{in}}'| = \langle \Psi_{\text{in}}'^{0}|+ \langle \Psi_{\text{in}}'^{2}|$.

The first contribution to the current, $\langle \widehat{I}_{\text{Sym}} \rangle^{(0,0)}$, is the noninteracting part, and we find that it agrees with the standard RLM answer.  For $N$ electrons, we obtain (see Appendix \ref{sec: The current in the two-lead AIM})
\beq
    \langle \widehat{I}_{\text{Sym}} \rangle^{(0,0)} = \frac{1}{2\pi \delta(0)}\sum_{m=1}^N \frac{1}{4} (-1)^{\gamma_m-1}  |\mathcal{T}(k_m)|^2 .\label{eq: current N electrons RLM}
\eeq
The Dirac delta term comes from the overlap of two plane waves of equal momenta (e.g., $\{ c_{1k'\uparrow}, c_{1k\uparrow}^\dagger\}$ with $k'=k$); we should thus identify $2\pi \delta(0)$ with the system size $L$ (which is formally infinite).  Taking the arbitrary $N$ quantum numbers to describe two filled Fermi seas replaces
\beq
    \sum_{m=1}^N X(\gamma_m,k_m,a_m) \to \sum_{\gamma=1,2} \sum_{k \in \mathcal{K}_\gamma } \sum_a X(\gamma,k,a),\label{eq: sum replacement rule with spin}
\eeq
where $X$ is any function and $\mathcal{K}_\gamma$ is the set of momenta in the Fermi sea of lead $\gamma$ [spaced by $\delta k \leftrightarrow 1/\delta(0)$ and cut off by $|k| < D$].  We can then generalize to include temperature; see Eq. \eqref{eq: sum replacement rule no spin} and the comments below.  We thus obtain
\beq
    \langle \widehat{I}_{\text{Sym}} \rangle^{(0,0)}\overset{\text{therm. limit}}{\longrightarrow}  \int_{-D}^{D}\frac{dk}{2\pi}\  \left[ \fermifn_1(k) - \fermifn_2(k) \right]  \frac{1}{2}|\mathcal{T}(k)|^2,
\eeq
which is twice the standard spinless RLM answer, as expected from spin degeneracy.

The same identification $1/\delta(0) \leftrightarrow \delta k$ was used by Nishino \emph{et al}.  For further justification, we have repeated the calculation in our time-dependent formalism, which permits us to work in a finite system size $L$ before sending $L\to\infty$; the final result for the current is the same in the steady state limit.  This is similar to how calculations with non-normalizable states in single-particle scattering theory are justified by considering the long-time limit of time-evolving wave packets. 

We proceed to calculate the contribution from the first crossing.  We show only the main steps here, leaving many details in Appendix \ref{sec: The current in the two-lead AIM}.  We only need to calculate $\langle \widehat{I}_{\text{Sym}} \rangle^{(0,2)}$, since $\langle \widehat{I}_{\text{Sym}} \rangle^{(2,0)}$ turns out to be the complex conjugate.  Using Wick's theorem and noting that $\langle \Psi | \Psi\rangle = [2\pi \delta(0)]^N$, we obtain the following after some calculation:
\begin{widetext}
\begin{multline}
    \langle \widehat{I}_{\text{Sym}} \rangle^{(0,2)} = [2\pi \delta(0)]^{-2} \frac{1}{4} \sum_{m_1,m_2=1}^N \lim_{\substack{ k_{m_1}'\to k_{m_1} \\ k_{m_2}'\to k_{m_2} } } (k_{m_1} + k_{m_2} - k_{m_1}' - k_{m_2}' ) \frac{\pd}{\pd \overline{\phi}}\biggr\rvert_{\overline{\phi} =0}\\
    \times \langle 0| c_{\gamma_{m_1} k_{m_1}' a_{m_1},\text{in}} c_{\gamma_{m_2} k_{m_2}' a_{m_2},\text{in}} | \overline{\Phi}_{e k_{m_1} a_{m_1} e k_{m_2} a_{m_2},\text{in}}\rangle,\label{eq: I02 AIM intermediate step}
\end{multline}
\end{widetext}
where we have used the antisymmetry of the operators and crossing state to replace the original sum over $m_1<m_2$ with an unrestricted sum with an extra factor of $1/2$.  After taking the limits $k_{m_1}'\to k_{m_1}$ and $k_{m_2}'\to k_{m_2} $, we again have a summation in which it is clear how to take the thermodynamic limit using \eqref{eq: sum replacement rule with spin} and the identification $\delta k \leftrightarrow 1/\delta(0)$.  Collecting terms, we find the following answer for the current:
\begin{widetext}
\begin{multline}
    \langle \widehat{I}_{\text{Sym}}\rangle = \int_{-D}^D \frac{dk}{2\pi}\ \left[\fermifn_1(k) - \fermifn_2(k) \right] \frac{1}{2}|\mathcal{T}(k)|^2 -\frac{1}{16\Delta^2} \int_{-D}^D \frac{dk_1}{2\pi}\frac{dk_2}{2\pi}\  \left[\fermifn_1(k_1) + \fermifn_2(k_1)\right] \left[\fermifn_1(k_2) - \fermifn_2(k_2) \right]\\
    \times\text{Im}\left\{ U \mathcal{T}\left( \frac{k_1 + k_2 -U}{2} \right) \mathcal{T}^*(k_1)\mathcal{T}(k_2) \left[\mathcal{T}(k_1) + \mathcal{T}(k_2) \right] \right\} + (\text{higher crossings}).\label{eq: current up to first crossing AIM}
\end{multline}
\end{widetext}
What does this ``expansion in crossings'' really mean?  While we cannot give a general answer, we can at least understand this result for the current by examining the limits of small and large $U$.

\subsubsection{Small \texorpdfstring{$U$}{U} regime}

Expanding to first order in $U$ replaces $U \mathcal{T}\left[ (k_1 +k_2 -U)/2\right] \to U \mathcal{T}\left[ (k_1 +k_2 )/2\right]$.  Then, using the simple identities $\mathcal{T}\left[ (k_1 +k_2 )/2\right]\left[ \mathcal{T}(k_1)  + \mathcal{T}(k_2)\right] = 2 \mathcal{T}(k_1)\mathcal{T}(k_2)$ and $\text{Im}\left[ \mathcal{T}(k)^2\right] =  |\mathcal{T}(k)|^2 \text{Re}\left[ \mathcal{T}(k)\right]$, we obtain
\begin{multline}
    \langle \widehat{I}_{\text{Sym}}\rangle = \int_{-D}^D \frac{dk}{2\pi}\ \left[\fermifn_1(k) - \fermifn_2(k) \right] \frac{1}{2}|\mathcal{T}(k)|^2 \\
    + \frac{U}{8 \Delta^2} \int_{-D}^D \frac{dk_1}{2\pi}\frac{dk_2}{2\pi}\ \left[\fermifn_1(k_1) + \fermifn_2(k_1) \right]\left[\fermifn_1(k_2) - \fermifn_2(k_2) \right] \\
    \times |\mathcal{T}(k_1)|^2 |\mathcal{T}(k_2)|^2 \text{Re}\left[\mathcal{T}(k_2) \right] + O(U^2). \label{eq: current small U AIM}
\end{multline}
This calculation mainly serves as a check on our formalism.  We have verified Eq. \eqref{eq: current small U AIM} by calculating the steady state current with Keldysh perturbation theory (see Appendix \ref{sec: Perturbative check the current in the Anderson model}).  Indeed, the agreement also holds if we allow a magnetic field on the dot, i.e., a spin-dependent dot energy $\epsilon_a$ (which modifies the crossing state \cite{Culver_thesis}).

We note that the small $U$ expansion of the AIM has been used in the literature to explore the neighborhood of the strong coupling fixed point of the Kondo model both in and out of equilibrium.  This proceeds by, e.g., assuming the impurity is in a singlet state by a choice of Green's function \cite{HershfieldDaviesWilkins_PRB}, expanding about the Hartree-Fock solution \cite{Matsumoto}, or using a Fermi liquid theory approach \cite{Oguri}.  In contrast, our result \eqref{eq: current small U AIM} describes the AIM itself in the regime of small $U/\Delta$.

Since $\mathcal{T}(k)\sim 1/k$ for large $|k|$, there are no divergences in Eq. \eqref{eq: current small U AIM} as the bandwidth $D$ is sent to infinity.  This is consistent with prior work on the AIM (see, e.g., Ref. \cite{Haldane_pert}).

\subsubsection{Infinite U regime: Expansion in tunneling}
If we instead send $U\to\infty$, then $U \mathcal{T}\left[ (k_1 + k_2 -U)/2 \right] \to -4\Delta$, leaving
\begin{multline}
    \langle \widehat{I}_{\text{Sym}}\rangle = \int_{-D}^D \frac{dk}{2\pi}\ \left[\fermifn_1(k) - \fermifn_2(k) \right] \frac{1}{2}|\mathcal{T}(k)|^2 \\
    +\frac{1}{4\Delta} \int_{-D}^D \frac{dk_1}{2\pi}\frac{dk_2}{2\pi}\  \left[ \fermifn_1(k_1) + \fermifn_2(k_1) \right]\left[\fermifn_1(k_2) - \fermifn_2(k_2) \right]\\
    \times \text{Im}\left\{  \mathcal{T}^*(k_1)\mathcal{T}(k_2) \left[\mathcal{T}(k_1) + \mathcal{T}(k_2) \right] \right\} \\
    + (2 \text{ or more crossings}).\label{eq: current bare infU AIM}
\end{multline}
This expansion in crossings appears to capture the regime of small $\Delta$.  We note first that Eq. \eqref{eq: current bare infU AIM} satisfies the following Callan-Symanzik equation:
\bseq
\begin{align}
    &\left( D\frac{\pd}{\pd D} + \beta_\epsilon  \epsilon \frac{\pd}{\pd \epsilon} \right) \langle \widehat{I}_{\text{Sym}}\rangle = O(1/D), \label{eq: CS eqn AIM}\\
    &\text{where: } \beta_{\epsilon} = -\frac{ \Delta}{\pi \epsilon} + O\left( \frac{\Delta^2}{\epsilon^2} \right).\label{eq: beta fn AIM}
\end{align}
\eseq
To show this, we proceed similarly as in the multilead IRL calculation [see Eq. \eqref{eq: D term in CS mcIRL} and below].  Under $D\frac{\pd}{\pd D}$, the only terms that survive for large bandwidth are those with $k_1$ integrated (since the $k_2$ Fermi functions cancel at $k_2= -\infty$) and a single $\mathcal{T}$ matrix in $k_1$ [since $\mathcal{T}(k) \sim 2\Delta/k$ for large $|k|$].  Thus, we obtain
\begin{multline}
    D\frac{\pd}{\pd D} \langle \widehat{I}_{\text{Sym}}\rangle \overset{D\to\infty}{
    \longrightarrow}\\
    -\frac{1}{2\pi} \int_{-\infty}^\infty\frac{dk_2}{2\pi}\ \left[\fermifn_1(k_2) - \fermifn_2(k_2)\right] \text{Im}\left[ \mathcal{T}(k_2)^2\right].
\end{multline}
Then \eqref{eq: beta fn AIM} follows from the identity $\frac{\pd}{\pd \epsilon} |\mathcal{T}(k)|^2 = \frac{1}{\Delta} |\mathcal{T}(k)|^2 \text{Re}\ \mathcal{T}(k) = -\frac{1}{\Delta} \text{Im}\left[ \mathcal{T}(k)^2\right]$.  The associated scaling invariants are $\Delta$ and
\beq
    \epsilon_d \equiv \epsilon + \frac{\Delta}{\pi} \ln \frac{D}{\Delta},\label{eq: scaling invariant AIM}
\eeq
which is the standard result \cite{Haldane_PRL, Hewson}.

To clarify the meaning of the expansion in crossings, we consider the zero temperature limit with a voltage drop across the leads: $\mu_1=0$ and $\mu_2=-V$.  Then the conductance is given by
\begin{multline}
    \frac{dI}{dV}=\frac{1}{\pi}\frac{\Delta^2}{(\epsilon + V)^2 + \Delta^2}  \biggr[ 1  -\frac{ \Delta(\epsilon+V)}{\pi^2 (\epsilon + V)^2 + \Delta^2}\\
    \times \left(\ln \frac{D}{\sqrt{(\epsilon +V)^2 +\Delta^2}} +\ln \frac{D}{\sqrt{\epsilon^2 +\Delta^2}} + \text{finite}\right) \biggr],
\end{multline}
where the omitted terms are finite as $D\to\infty$ (or involve additional crossings).  It is seen here that the contribution from the first crossing (i.e., two quantum numbers in the crossing state) starts at the third order in $\Delta$, while the RLM contribution is second order.  By further calculation, we find that the next contribution (allowing three quantum numbers to be in crossing states) starts at another order higher ($\Delta^4$).

Strictly speaking, our result should be interpreted as a power series in $\Delta$, meaning that we should keep only up to order $\Delta^3$.  It is interesting to note, however, that when $\langle \widehat{I}_{\text{Sym}}\rangle$ is calculated to the leading order in crossings (as we did above), the Callan-Symanzik equation \eqref{eq: CS eqn AIM} holds \emph{to all orders} in $\Delta$.  Our demonstration of the Callan-Symanzik equation did not expand in $\Delta$.  The expansion in crossings can be thought of as a particular resummation of terms of the $\Delta$ expansion; the fact that the Callan-Symanzik equation holds exactly suggests that this resummation may be a useful one. 

While much work has been done on the infinite-$U$ AIM, the most direct comparison we can make to the literature is to Ref. \cite{SivanWingreen}, in which the current is calculated analytically for $U\to\infty$ up to order $\Delta^3$.  Our result here disagrees beyond the first order in $\Delta$.  In particular, Ref. \cite{SivanWingreen} finds a small Kondo peak beginning to develop at zero bias, which we do not.  However, a true comparison can only be made once both answers are expressed in terms of RG invariants, and the result of Ref. \cite{SivanWingreen} does not seem to have the standard quantity given in Eq. \eqref{eq: scaling invariant AIM} as a scaling invariant.

\section{Conclusion and outlook}\label{sec: Conclusion and outlook}
In this paper, we presented a method for calculating many-body wavefunctions.  We applied the time-dependent version of the  method to find the time-evolving wavefunction for the interacting resonant level model with any number of leads.  We also applied the time-independent version to find the nonequilibrium steady state wavefunction of the Anderson impurity model in the two limits of small $U$ and infinite $U$.  The methods of Bethe ansatz and the integrability properties of the models studied made no obvious appearance in the calculations.

As a preliminary application of these wavefunctions to the evaluation of observables, we found the steady state occupancy of the multilead IRL to leading order in the interaction $U$.  We demonstrated universality in and out of equilibrium, verified our answer in the zero temperature equilibrium limit by comparison with the literature, and presented results out of equilibrium.  In the two-lead AIM, we used the NESS wavefunction to evaluate the steady state current first for small $U$, then for infinite $U$ with small $\Delta$.  This provided an example of how we can calculate observables directly in steady state nonequilibrium without following the time evolution.  Our IRL results can also be obtained this way.

It is our hope that further technology for the evaluation of observables using these wavefunctions can be developed so that some nonperturbative results can be found in the thermodynamic limit.  Also, the general reformulation of the many-body Schrodinger equation that we presented could be of wider use, beyond exact solutions of quantum impurity models.

\begin{acknowledgments}
    We are grateful to Chung-Hou Chung, Piers Coleman, Garry Goldstein, Yashar Komijani, Yigal Meir, Andrew Mitchell, Achim Rosch, and Hubert Saleur for helpful discussions.  We have benefited from working on related problems with Huijie Guan, Paata Kakashvili, Christopher Munson, and Roshan Tourani.  A.B.C. acknowledges support from the Samuel Marateck Fellowship in Quantum Field Theory Physics and the Excellence Fellowship (both from Rutgers University).  This material is based upon work supported by the National Science Foundation under Grant No. 1410583.
\end{acknowledgments}

\appendix

\begin{widetext}

    \section{Notation for calculations}\label{sec: Notation for calculations}
    Throughout the appendices, we use a compressed notation for manipulating lists of indices---see Appendix A of the previous paper \cite{CulverAndrei_PRB1} for details.  The main points are: (1) boldface letters indicate lists of indices, e.g., $\mathbf{m}=(1,3,6,7)$; (2) $\mathcal{I}_j(\mathbf{m})$ indicates the set of all \emph{increasing} lists of length $j$ chosen from $\mathbf{m}$; (3) given $\bm{\ell} \in \mathcal{I}_j(\mathbf{m})$, $\sgnleft \bm{\ell}$  indicates the sign of the permutation that maps $\mathbf{m} \to \bm{\ell}\  \mathbf{m}/\bm{\ell}$, i.e., brings the entries of $\bm{\ell}$ to the left of the list while leaving the remaining entries in order.  We define $\sgnright \bm{\ell}$ similarly (bring $\bm{\ell}$ to the right, instead).
    
    It is also convenient to have a notation for a list divided into smaller parts (``cells'') in various ways.  Given a list $\mathbf{m}$, we define a \emph{partition} of $\mathbf{m}$ to be a separation of the list elements into \emph{cells of length $2$ or greater}.  Partitions are denoted by underlined, boldface letters (typically the letter p, as in $\partition{p}$).  Take $\mathbf{m} = (1,3,6,7)$ as an example; the two partitions of $\mathbf{m}$ are $\partition{p}= (1,3,6,7)$ (one cell) and $\partition{p}=( 1,3 | 6,7)$ (two cells).

    A partition with $s$ cells can be written as $\partition{p}=(\mathbf{p}_1| \dots | \mathbf{p}_s)$, where each $\mathbf{p}_j$ is a list.  Elements of these lists are written as $p_j(\ell)= p_{j \ell} = p(j\ell)$.  The set of all partitions of a list $\mathbf{m}$ is written as $\mathbb{P}(\mathbf{m})$:
    \beq
        \mathbb{P}(\mathbf{m}) = \{ \partition{p} \equiv  (\mathbf{p}_1 | \dots |\mathbf{p}_s )\ |\  1\le s \le |\mathbf{m}|/2,\ (\mathbf{p}_1 , \dots ,\mathbf{p}_s ) = \mathbf{m},\ |\mathbf{p}_j| \ge2 \text{ for all }j \}.
    \eeq
    The set of partitions whose last cell has length $q$ is denoted with a subscript $q$:
    \beq
        \mathbb{P}_q(\mathbf{m}) = \{ \partition{p} \equiv (\mathbf{p}_1|\dots | \mathbf{p}_s) \in \mathbb{P}(\mathbf{m})\ |\ |\mathbf{p}_s| = q \}.
    \eeq
    
    \section{Proof of general formalism}\label{sec: Proof of general formalism}
    
    We show that the wavefunction construction in Sec. \ref{sec: General formalism} satisfies the time-dependent Schrodinger equation.  The demonstration that the time-independent version (Sec. \ref{sec: Time-independent formalism}) satisfies the time-independent Schrodinger equation can be obtained by simple adjustments.
    
    The wavefunction construction [Eq. \eqref{eq: wavefn construction}] can be written in our compressed notation as
    \beq
        |\Psi(t) \rangle= \sum_{n=0}^N|\Psi^n(t) \rangle, \text{ where: }
        |\Psi^n(t) \rangle= \sum_{\mathbf{m} \in \mathcal{I}_n(\mathbf{N}) } \left( \sgnleft\mathbf{m} \right)  c_{\alpha_{\mathbf{N} / \mathbf{m} }}^\dagger(t) |\Phi_{\alpha_{\mathbf{m} }} (t) \rangle,
    \eeq
    where $|\Phi(t)\rangle \equiv |0\rangle$ [so that the $n=0$ term of the sum agrees with the earlier definition Eq. \eqref{eq: Psi0(t)}] and where $|\Phi_{\alpha_1}(t)\rangle \equiv 0$ for any $\alpha_1$ (so that the $n=1$ term of the sum vanishes).  The crossing states [Eq. \eqref{eq: crossing state as antisymmetrization}] become
    \beq
        | \Phi_{\alpha_{\mathbf{m} } }(t) \rangle = \sum_{\sigma \in \text{Sym}(n) } \left( \sgn\sigma \right) \sum_{\partition{p} \in \mathbb{P}(\mathbf{m})}  |\chi_{\alpha_{\partition{p} \circ \sigma} }(t) \rangle,  
    \eeq
    where the unsymmetrized crossing states satisfy [see Eqs. \eqref{eq: general aux state Schrod} and \eqref{eq: general aux state initial condition} and comments below]
    \bseq
    \begin{align}
        \left(  H - i \frac{d}{d t} \right) |\chi_{  \alpha_{\partition{p}} } (t) \rangle &= 
        \begin{cases}
             -B_{ \alpha_{n-1}   \alpha_n  }^{(\text{red})}(t)  |\chi_{ \alpha_{\partition{p} / (n-1,n) } } (t) \rangle& q =2 \\
            -A_{\alpha_n}(t) | \chi_{ \alpha_{\partition{p} / n } } (t) \rangle & 3 \le q \le n,
        \end{cases}\label{eq: general aux state Schrod Appendix}\\
        |\chi_{   \alpha_{\partition{p}}}(t=0) \rangle &= 0,
    \end{align}
    \eseq
    as well as $|\chi(t)\rangle = |0\rangle$ and $|\chi_{\alpha_1}(t)\rangle \equiv 0$.  We wish to show that $(H-i\frac{d}{dt})|\Psi(t)\rangle=0$.  Our first task is to show that the crossing states satisfy the following condition:
    \beq
        \left( H - i \frac{d}{dt }\right) | \Phi_{\alpha_{\mathbf{m} }}(t) \rangle = - \sum_{\bm{\ell} \in \mathcal{I}_1(\mathbf{m}) } \left( \sgnright\bm{\ell} \right) A_{\alpha_{\ell_1} }(t) | \Phi_{\alpha_{\mathbf{m} / \bm{\ell} } }(t) \rangle - \sum_{\bm{\ell} \in \mathcal{I}_2(\mathbf{m}) } \left( \sgnright\bm{\ell} \right) B_{\alpha_{\ell_1} \alpha_{\ell_2} }(t) | \Phi_{\alpha_{\mathbf{m} / \bm{\ell} } }(t) \rangle.\label{eq: Phi condition type B}
    \eeq
    To show this, we note that the sum over all partitions can be separated into sums over partitions with specified length $q$ of the last cell, i.e., $\sum_{\partition{p}  \in \mathbb{P}(\mathbf{m}) } =\sum_{q=2}^n \sum_{\partition{p} \in \mathbb{P}_q(\mathbf{m}) }$.  Separating the $q=2$ term from the others and using Eq. \eqref{eq: general aux state Schrod Appendix}, we obtain
    \begin{multline}
        \left(H - i \frac{d}{dt} \right) | \Phi_{\alpha_{\mathbf{m} }}(t)\rangle= -\sum_{\sigma \in \text{Sym}(n)} (\sgn \sigma) \sum_{\partition{p} \in \mathbb{P}_2(\mathbf{m}) }  B_{\alpha_{p(\sigma_{n-1})} \alpha_{p(\sigma_n)}}^{(\text{red})}(t) | \chi_{\alpha_{\partition{p} / (p(\sigma_{n-1}), p(\sigma_n)  ) }}(t)\rangle \\
        -\sum_{\sigma \in \text{Sym}(n)} (\sgn \sigma) \sum_{q=3}^n \sum_{\partition{p} \in \mathbb{P}_q(\mathbf{m}) }    A_{\alpha_{p(\sigma_n)}}(t) |\chi_{\alpha_{\partition{p} / p(\sigma_n)}  }(t)\rangle.  \label{eq: proof of Phi condition two terms type B}
    \end{multline}
    The two terms on the right-hand side will now be massaged separately.  Relabelling $\sigma_{n-1}\to \ell_1$ and $\sigma_n \to \ell_2$, we obtain
    \bseq
    \begin{align}
        1\text{st term of } \eqref{eq: proof of Phi condition two terms type B}\ &= -\sum_{\bm{\ell} \in \mathcal{I}_2(\mathbf{m}) } \left( \sgnright\bm{\ell}\right) \left( B_{\alpha_{\ell_1} \alpha_{\ell_2} }^{(\text{red})}(t) - B_{\alpha_{\ell_2} \alpha_{\ell_1} }^{(\text{red})}(t) \right)\sum_{\sigma \in \text{Sym}(n-2)} (\sgn \sigma)  \sum_{\partition{p} \in \mathbb{P}(\mathbf{m} / \bm{\ell} ) }  | \chi_{\alpha_{\partition{p} \circ\sigma}}(t) \rangle\\
        &= -\sum_{\bm{\ell} \in \mathcal{I}_2(\mathbf{m}) } \left( \sgnright\bm{\ell}\right) B_{\alpha_{\ell_1} \alpha_{\ell_2} }(t) |\Phi_{\alpha_{\mathbf{m} / \bm{\ell} } }(t)\rangle.
    \end{align}
    \eseq
    In the second term, relabelling $\sigma_n\to\ell_1$ and $q\to q+1$ yields
    \bseq
    \begin{align}
        2\text{nd term of } \eqref{eq: proof of Phi condition two terms type B}
        &= - \sum_{\bm{\ell} \in \mathcal{I}_1(\mathbf{m})} \left( \sgnright\bm{\ell} \right) A_{\alpha_{\ell_1}}(t) \sum_{\sigma \in \text{Sym}(n-1)} (\sgn \sigma) \sum_{q=2}^{n-1} \sum_{\partition{p} \in \mathbb{P}_q(\mathbf{m}/\bm{\ell}) }  |\chi_{\alpha_{\partition{p}\circ\sigma} }(t)\rangle\\
        &=- \sum_{\bm{\ell} \in \mathcal{I}_1(\mathbf{m})} \left( \sgnright\bm{\ell} \right) A_{\alpha_{\ell_1}}(t) | \Phi_{ \alpha_{\mathbf{m} / \bm{\ell} }}(t)\rangle.
    \end{align}
    \eseq
    This completes the proof of Eq. \eqref{eq: Phi condition type B}.

    The next step is to write down a formula for the action of $H - i \frac{d}{dt}$ on a product of $c_{\alpha}^\dagger(t)$ operators.  If $|X(t)\rangle$ is any time-dependent state and $\mathbf{m}$ is any list of indices, then we have
    \begin{multline}
        \left( H - i \frac{d}{dt} \right)   c_{\alpha_\mathbf{m}}^\dagger(t) |X(t)\rangle = \sum_{\bm{\ell} \in \mathcal{I}_1(\mathbf{m})} \left( \sgnleft\bm{\ell}  \right) c_{\alpha_{\mathbf{m} / \bm{\ell}}}^\dagger(t) A_{\alpha_{\ell_1}}(t) | X(t) \rangle\\
        +\sum_{\bm{\ell} \in \mathcal{I}_2(\mathbf{m})} \left( \sgnleft\bm{\ell}  \right) c_{\alpha_\mathbf{m} / \bm{\ell}}^\dagger(t)  B_{\alpha_{\ell_1} \alpha_{\ell_2}}(t) | X(t) \rangle +    c_{\alpha_\mathbf{m}}^\dagger(t)  \left( H - i\frac{d}{d t} \right) |X(t)\rangle. \label{eq: H - i d/dt past prod type B}
    \end{multline}
    Note that we have used the assumption that any $B(t)$ commutes with any $c_\alpha^\dagger(t)$ [Eq. \eqref{eq: B commutes}].  Applying Eq. \eqref{eq: H - i d/dt past prod type B}, we then find
    \bseq
    \begin{align}
        \left( H - i \frac{d}{d t} \right) &|\Psi(t) \rangle =  \sum_{n=0}^N\sum_{\mathbf{m} \in \mathcal{I}_n(\mathbf{N} ) } \left( \sgnleft\mathbf{m} \right)  \left( H - i \frac{d}{dt} \right) c_{\alpha_{\mathbf{N} / \mathbf{m} }}^\dagger(t)  | \Phi_{\alpha_{\mathbf{m} }}(t) \rangle\\
        &= \sum_{n=1}^{N-1} \sum_{\mathbf{m} \in \mathcal{I}_n(\mathbf{N} ) } \left( \sgnleft\mathbf{m} \right) \sum_{\bm{\ell} \in \mathcal{I}_1(\mathbf{N} / \mathbf{m} ) } \left( \sgnleft\bm{\ell} \right)  c_{\alpha_{\mathbf{N} / \mathbf{m} / \bm{\ell} }}^\dagger(t)  A_{\alpha_{\ell_1} }(t) | \Phi_{\alpha_{\mathbf{m}} }(t) \rangle\notag\\
        &+ \sum_{n=0}^{N-2} \sum_{\mathbf{m} \in \mathcal{I}_n(\mathbf{N} ) } \left( \sgnleft\mathbf{m} \right) \sum_{\bm{\ell} \in \mathcal{I}_2(\mathbf{N} / \mathbf{m} ) } \left( \sgnleft\bm{\ell} \right) c_{\alpha_{\mathbf{N} / \mathbf{m} / \bm{\ell}}}^\dagger(t)  B_{\alpha_{\ell_1}\alpha_{\ell_2} }(t) | \Phi_{\alpha_{\mathbf{m}} }(t) \rangle    \notag\\
        &\qquad + \sum_{n=2}^N\sum_{\mathbf{m} \in \mathcal{I}_n(\mathbf{N} ) } \left( \sgnleft\mathbf{m} \right) c_{\alpha_{\mathbf{N} / \mathbf{m}}}^\dagger(t)  \left( H - i \frac{d}{dt} \right) | \Phi_{\alpha_{\mathbf{m}} }(t) \rangle. \label{eq: Schrod proof three terms to cancel type B} 
    \end{align}
    \eseq
    Note that in the first term, we dropped the $n=N$ part of the sum, since it is zero---if all quantum numbers are chosen to be put into a crossing state, then there are no $c_\alpha^\dagger(t)$ operators to commute with, so no $A(t)$ is generated.  We also dropped the $n=0$ part because $A_{\alpha_j}(t)|\Phi(t) \rangle = A_{\alpha_j}(t)|0 \rangle=0$ by assumption [Eq. \eqref{eq: A annihilates 0}].  In the second term, we dropped the $n=N-1$ and $n=N$ parts of the sum, since there must be at least two $c_\alpha^\dagger(t)$ operators in order to produce a $B(t)$ operator.  In the third term we dropped the $n=0$ part, since $\left( H - i \frac{d}{dt} \right) |\Phi(t) \rangle = \left( H - i \frac{d}{dt} \right)| 0 \rangle = 0$ [recall from Eq. \eqref{eq: H annihilates 0} that $H$ annihilates the empty state], and the $n=1$ part, since $|\Phi_{\alpha_j}(t)\rangle =0$.  [We could have dropped the $|\Phi_{\alpha_j}(t)\rangle$ contributions to the first two terms of \eqref{eq: Schrod proof three terms to cancel type B} but have left them in to simplify the notation.]

    We now relabel the summation variables in the first two terms of \eqref{eq: Schrod proof three terms to cancel type B} to find
    \bseq
    \begin{align}
        \text{first term of }\eqref{eq: Schrod proof three terms to cancel type B} &= \sum_{n=1}^{N-1} \sum_{\mathbf{m} \in \mathcal{I}_{n+1}(\mathbf{N} ) } \left( \sgnleft\mathbf{m} \right) \sum_{\bm{\ell} \in \mathcal{I}_1( \mathbf{m})  } \left( \sgnright\bm{\ell} \right)  c_{\alpha_{\mathbf{N} / \mathbf{m}}}^\dagger(t) A_{\alpha_{\ell_1} }(t) | \Phi_{\alpha_{\mathbf{m} / \bm{\ell}} }(t) \rangle\\
        &= \sum_{n=2}^N\sum_{\mathbf{m} \in \mathcal{I}_n(\mathbf{N} ) } \left( \sgnleft\mathbf{m} \right)   c_{\alpha_{\mathbf{N} / \mathbf{m}}}^\dagger(t) \sum_{\bm{\ell} \in \mathcal{I}_1( \mathbf{m})  } \left( \sgnright\bm{\ell} \right) A_{\alpha_{\ell_1} }(t) | \Phi_{\alpha_{\mathbf{m} / \bm{\ell}} }(t) \rangle
    \end{align}
    \eseq
    and
    \bseq
    \begin{align}
        \text{second term of } \eqref{eq: Schrod proof three terms to cancel type B} &= \sum_{n=0}^{N-2} \sum_{\mathbf{m} \in \mathcal{I}_{n+2}(\mathbf{N} ) } \left( \sgnleft\mathbf{m} \right) \sum_{\bm{\ell} \in \mathcal{I}_2( \mathbf{m})  } \left( \sgnright\bm{\ell} \right)  c_{\alpha_{\mathbf{N} / \mathbf{m}}}^\dagger(t) B_{\alpha_{\ell_1} \alpha_{\ell_2}}(t) | \Phi_{\alpha_{\mathbf{m} / \bm{\ell}} }(t) \rangle\\
        &= \sum_{n=2}^N\sum_{\mathbf{m} \in \mathcal{I}_n(\mathbf{N} ) } \left( \sgnleft\mathbf{m} \right)  c_{\alpha_{\mathbf{N} / \mathbf{m}}}^\dagger(t)  \sum_{\bm{\ell} \in \mathcal{I}_2( \mathbf{m})  } \left( \sgnright\bm{\ell} \right) B_{\alpha_{\ell_1} \alpha_{\ell_2} }(t) | \Phi_{\alpha_{\mathbf{m} / \bm{\ell}} }(t) \rangle. 
    \end{align}
    \eseq
    It is then clear from Eq. \eqref{eq: Phi condition type B} that the first two terms of \eqref{eq: Schrod proof three terms to cancel type B} exactly cancel the third.  This completes the proof.

    \section{Full calculation of nth crossing state}\label{sec: Full calculation of nth crossing state}
    
    We provide the detailed proof of our solution to the time-dependent Schrodinger equation of the one-lead IRL by verifying that the crossing states satisfy the appropriate inverse problems.  The crossing states for the multilead IRL and infinite-$U$ AIM that are stated in the main text can be verified similarly; see Ref. \cite{Culver_thesis} for details.
    
    We prove that the unsymmetrized crossing states of the one-lead IRL [defined by Eqs. \eqref{eq: n ansatz IRL}, \eqref{eq: single-celled F IRL}, and \eqref{eq: F product of cells IRL}] satisfy the appropriate family of inverse problems, namely [given $\partition{p}\in\mathbb{P}_q(\mathbf{n})$]
\bseq
\begin{align}
    \left(  H - i \frac{d}{d t} \right) |\chi_{  k_{\partition{p}}} (t) \rangle &= 
    \begin{cases}
         -B_{ k_{ n-1  } k_n }^{(\text{red})}(t)  |\chi_{ k_{\partition{p} / (n, n-1)} } (t) \rangle& q =2 \\
         -A_{k_n}(t) | \chi_{k_{\partition{p} / n} } (t) \rangle & 3 \le q \le n,
    \end{cases} \label{eq: aux state condition IRL}\\
    |\chi_{   k_{\partition{p}}}(t=0) \rangle &= 0,\label{eq: aux state initial condition IRL}\\
    |\chi_{k_{\partition{p}}}(t) \rangle &= |0 \rangle \qquad \text{when } \partition{p} \text{ is the empty list}.
\end{align}
\eseq
    Throughout, we reduce clutter by using the notation $\mathbf{n-1}=(1,\dots,n-1)$, $\mathbf{n-2}=(1,\dots,n-2)$, and $\mathbf{n-q}= (1,\dots, n-q)$ (note that the minus sign does \emph{not} mean removing an element from the list).
    
    In the compressed notation of Appendix \ref{sec: Notation for calculations}, the crossing state ansatz \eqref{eq: n ansatz IRL} reads
    \begin{multline}
        |\chi_{k_{\partition{p}}} (t)\rangle = \frac{1}{L^{n/2}}\int dx_\mathbf{n}\ F_{k_{\partition{p}}}(t,x_\mathbf{n}) \biggr[ \Theta(0<x_n<\dots < x_1< t) \psi^\dagger(x_n)\\
        + \frac{i}{v} \delta(x_n)\Theta(0<x_{n-1}<\dots < x_1<t) d^\dagger \biggr] \psi^\dagger(x_{\mathbf{n}/n})|0\rangle.
    \end{multline}
    This state vanishes at $t=0$ by construction [see the discussion below \eqref{eq: n=2 ansatz IRL}].  The main task is to confirm Eq. \eqref{eq: aux state condition IRL}.
    
    A straightforward calculation yields
    \begin{multline}
        \left( H - i \frac{d}{dt}\right) |\chi_{k_{\partition{p}}} (t)\rangle = \frac{1}{L^{n/2}}\Biggr\{  \int dx_\mathbf{n}\ \left[-i \left(\frac{\pd}{\pd t} + \sum_{j=1}^n \frac{\pd}{\pd x_j}  \right)  F_{k_{\partition{p}}}(t,x_\mathbf{n})  \right]
        \Theta(0<x_n<\dots<x_1 <t) \psi^\dagger(x_\mathbf{n})\\
        + \frac{i}{v}\int dx_{\mathbf{n-1}}\ \left[\left( -i \frac{\pd}{\pd t} -i \sum_{j=1}^{n-1} \frac{\pd}{\pd x_j}  + \pole \right)  F_{k_{\partition{p}}}(t,x_\mathbf{n-1},0)  \right]
        \Theta(0<x_{n-1}<\dots <x_1<t) d^\dagger \psi^\dagger(x_{\mathbf{n-1}})\\
        + \frac{i}{v} \left( -i + \frac{1}{2}U \right) \int dx_{\mathbf{n-2}}\ F_{k_{\partition{p}}}(t,x_\mathbf{n-2},0,0)
        \Theta(0<x_{n-2}<\dots <x_1 <t)d^\dagger \psi^\dagger(0) \psi^\dagger(x_{\mathbf{n-2}}) \Biggr\}|0\rangle. \label{eq: H-id/dt on chi IRL}
    \end{multline}
    To derive this, we have noted
    \beq
        \left(\frac{\pd}{\pd t} + \sum_{j=1}^n \frac{\pd}{\pd x_j}  \right)\Theta(0<x_n<\dots<x_1 <t) = \delta(x_n)\Theta(0<x_{n-1}<\dots<x_1 <t),
    \eeq
    which leads to a cancellation of unwanted impurity-electron terms [see Eq. \eqref{eq: convenient cancellation} for the $n=2$ case].  We have also used the averaging prescription to make the following replacement:
    \beq
        \delta(x_n) \Theta(0<x_n< \dots < x_1 <t) \to \frac{1}{2}\delta(x_n)\Theta(0<x_{n-1}< \dots < x_1 <t).
    \eeq
    The first in Eq. \eqref{eq: H-id/dt on chi IRL} term vanishes because $F_{k_{\partition{p}}}(t,x_\mathbf{n})$ is a function of coordinate differences only.  The second term vanishes because $F_{k_{\partition{p}}}(t,x_\mathbf{n-1},0)$ is $e^{-i \pole x_{n-q+1}}$ times a function of coordinate differences.  Thus, we are left with
\begin{multline}
    \left( H - i \frac{d}{dt}\right) |\chi_{k_{\partition{p}}} (t)\rangle = \frac{1}{L^{n/2}}\frac{i}{v} \left( -i + \frac{1}{2}U \right) \int dx_{\mathbf{n-2}}\ F_{k_{\partition{p} / \mathbf{p}(s) } }(t,x_\mathbf{n-q})\\
    \times F_{k_{\mathbf{p}(s)}}(t, x_{n-q+1},\dots, x_{n-2},0,0)
    \Theta(0<x_{n-2}<\dots <x_1 <t)d^\dagger \psi^\dagger(0) \psi^\dagger(x_{\mathbf{n-2}}) |0\rangle,
\end{multline}
where we used the product form of $F$ [Eq. \eqref{eq: F product of cells IRL}].  Let us compare this to the terms we are trying to cancel.  If $q=2$ [i.e., $\mathbf{p}(s) = (n-1,n)$], we have
\begin{multline}
    - B_{k_{n-1} k_n}^{(\text{red})}(t) |\chi_{ k_{\partition{p} / (n-1,n)} } (t) \rangle = \frac{1}{L^{n/2}}\frac{U}{v} \int dx_\mathbf{n-2}\ F_{ k_{\partition{p} / \mathbf{p}(s) } }(t, x_\mathbf{n-2})\\
    \mathcal{T}(k_{n-1}) \left( e^{-i k_{n-1} t} - e^{-i \pole t} \right)e^{-i k_n t } \Theta(0<x_{n-2}<\dots< x_1 <t)d^\dagger \psi^\dagger(0)\psi^\dagger(x_\mathbf{n-2})|0\rangle,
\end{multline}
and so the condition \eqref{eq: aux state condition IRL} holds if we have
\beq
    \left( 1 + i\frac{1}{2}U \right)F_{k_{n-1} k_n}(t,0,0) = U
    \mathcal{T}(k_{n-1}) \left( e^{-i k_{n-1} t} - e^{-i \pole t} \right)e^{-i k_n t },
\eeq
which has already been shown in the $n=2$ calculation [see Eq. \eqref{eq: third requirement n=2 IRL}].  If instead $q \ge3$, we have [again using Eq. \eqref{eq: F product of cells IRL}]
\begin{multline}
    - A_{k_n}(t) |\chi_{ k_{\partition{p} / n} }  (t) \rangle = \frac{1}{L^{n/2}}\frac{i}{v}U \int dx_\mathbf{n-1}\ F_{ k_{\partition{p} / n } }(t, x_\mathbf{n-2},0)\\
    \times 
    e^{-i k_n t} 
    \Theta(0<x_{n-2}<\dots< x_1 <t)d^\dagger \psi^\dagger(0)\psi^\dagger(x_\mathbf{n-2})|0\rangle = \\ \frac{1}{L^{n/2}}\frac{i}{v}U \int dx_\mathbf{n-1}\ F_{ k_{\partition{p} / \mathbf{p}(s) } }(t, x_\mathbf{n-q})F_{k_{\mathbf{p}(s) / n } }(t,x_{n-q+1},\dots, x_{n-2},0,0)\\
    \times 
    e^{-i k_n t} 
    \Theta(0<x_{n-2}<\dots< x_1 <t)d^\dagger \psi^\dagger(0)\psi^\dagger(x_\mathbf{n-2})|0\rangle,
\end{multline}
and so the condition \eqref{eq: aux state condition IRL} holds if we have
\beq
    \left( 1 + i\frac{1}{2}U \right)F_{k_{\mathbf{p}(s)}}(t,x_{n-q+1},\dots,x_{n-2},0,0 )= iU F_{k_{\mathbf{p}(s) / n } }(t,x_{n-q+1},\dots, x_{n-2},0,0) e^{-i k_n t}.
\eeq
This holds due to the definition \eqref{eq: single-celled F IRL} of the function $F$ for single-celled partitions.  We have thus verified Eq. \eqref{eq: aux state condition IRL}, completing the solution.

    \section{Further details in the evaluation of observables.}
    
    \subsection{Dot occupancy in the multilead IRL}\label{sec: Dot occupancy in the multilead IRL Appendix}
    
    We fill in the gap between Eq. \eqref{eq: nd^1 mcIRL} (the leading order correction to the occupancy at arbitrary time) and Eq. \eqref{eq: occupancy bare mcIRL} (the steady state limit).  To do this, we need to evaluate the long-time limit of $\text{Re}\ \Omega[t; \gamma_1' k_1', \gamma_2' k_2'; \gamma_1 k_1, \gamma_2 k_2]$, where $\Omega$ is given by Eq. \eqref{eq: Omega as integrals mcIRL}.
    
    Written in full, Eq. \eqref{eq: Omega as integrals mcIRL} from the main text reads
    \begin{multline}
    \Omega(t;\gamma_1' k_1',\gamma_2' k_2'; \gamma_1 k_1,\gamma_2 k_2) = 
    \frac{1}{2\gmax\Delta}
    \int_0^t dx_1\ \mathcal{T}_U \mathcal{T}(k_1)\left( e^{-i k_1 (t-x_1)} - e^{-i \pole(t-x_1)} \right)e^{-i k_2(t-x_1)} e^{-i \pole x_1 }\\
    \times 
    \biggr[\delta_{\gamma_1'}^{\gamma_2} e^{i k_1'(t-x_1)} 
    + \frac{1}{\gmax}i\mathcal{T}^*(k_1') \left( e^{i k_1'(t-x_1)} -e^{i \pole^*(t-x_1)} \right)   \biggr]i\mathcal{T}^*(k_2')\left( e^{i k_2' t} - e^{i \pole^* t} \right),
\end{multline}
where we can replace $\mathcal{T}_U\to U$ to get the first order expansion.  We can assume $k_1'+k_2' = k_1 +k_2$, since this is the only case we need for evaluating \eqref{eq: nd^1 mcIRL}.
    
Recalling that Im $\pole = - \Delta <0$, we see that there are several terms in the integrand that decay as $e^{-\Delta t}$ for large time; they can all be dropped in the limit.  The terms $e^{-i \pole(t-x_1)}$ and $e^{i \pole^* (t-x_1)}$, which each have absolute value $e^{- \Delta(t-x_1)}$, cannot immediately be neglected, since they are of order one when $x_1 \sim t$; however, the factor of $e^{-i \pole x_1}$ can combine with either one of these terms to yield the absolute value $e^{- \Delta t}$, which then can be neglected.  The remaining time-dependent terms are all phases and cancel by assumption ($e^{-i(k_1 + k_2 -k_1' -k_2')t} =1$), leaving
\bseq
\begin{align}
    \lim_{t\to\infty} \Omega(t;\gamma_1' k_1',\gamma_2' k_2'; \gamma_1 k_1,\gamma_2 k_2) &= \frac{1}{2\gmax\Delta}
    \int_0^\infty dx_1\ \mathcal{T}_U \mathcal{T}(k_1)e^{i (k_1+k_2- k_1' -\pole) x_1}  \biggr[\delta_{\gamma_1'}^{\gamma_2} 
    + \frac{1}{\gmax}i\mathcal{T}^*(k_1')   \biggr]i\mathcal{T}^*(k_2')\\
    &= -\frac{1}{4\gmax\Delta^2}
    \mathcal{T}_U \mathcal{T}(k_1)|\mathcal{T}(k_2')|^2   
    \biggr[\delta_{\gamma_1'}^{\gamma_2} 
    + \frac{1}{\gmax}i\mathcal{T}^*(k_1')   \biggr].
\end{align}
\eseq  
We thus obtain
\begin{multline}
    \lim_{t\to\infty} \sum_{\sigma \in \text{Sym}(2)} (\sgn \sigma) 2\text{Re}\left[ \Omega(t;\gamma_1 k_1,\gamma_2 k_2; \gamma_{\sigma_1} k_{\sigma_1},\gamma_{\sigma_2} k_{\sigma_2})  \right] =\\
    -\frac{|\mathcal{T}(k_2)|^2}{2\gmax\Delta^2} \text{Re} \left\{
    \mathcal{T}_U \left[  \left(1 + \frac{i}{\gmax}\mathcal{T}^*(k_1) \right) \left( \mathcal{T}(k_1) - \mathcal{T}(k_2\right) - \left( 1 -\delta_{\gamma_1}^{\gamma_2} \right)\mathcal{T}(k_1) \right]   \right\}.\label{eq: sum over Omega intermediate step mcIRL}
\end{multline}
Since the $\mathcal{S}$ matrix $\mathcal{S}(k) = 1- i\mathcal{T}(k)$ is a pure phase, $\mathcal{T}(k)$ satisfies a version of the optical theorem:
    \beq
        2 \text{Im}\left[ \mathcal{T}(k) \right] + |\mathcal{T}(k)|^2 =0.\label{eq: optical identity RLM}
    \eeq
    Repeated use of this identity simplifies \eqref{eq: sum over Omega intermediate step mcIRL} further; to leading order in $U$, we find
\beq
    \eqref{eq: sum over Omega intermediate step mcIRL} = \frac{U}{2\gmax \Delta^2} \Biggr\{  \left( 1 - \frac{1}{2\gmax} |\mathcal{T}(k_1)|^2\right) |\mathcal{T}(k_2)|^2 \text{Re}\left[\mathcal{T}(k_2) \right] 
    - \text{Re}\left[\mathcal{T}(k_1)\right] |\mathcal{T}(k_2)|^2\left(\delta_{\gamma_1}^{\gamma_2} -\frac{1}{2\gmax}|\mathcal{T}(k_2)|^2 \right) \Biggr\}.\label{eq: final integrand nd direct mcIRL}
\eeq
Note that we have a term independent of $k_1$; this leads to a linear divergence in bandwidth.  Equation \eqref{eq: occupancy bare mcIRL} in the main text follows from noting that $\int_{-D}^D \frac{dk_1}{2\pi}\ \fermifn(k_1) = (D+\mu)/(2\pi) + O\left(e^{-\frac{D-|\mu|}{T}} \right)$ [for a Fermi function $\fermifn(k)$ with temperature $T$ and chemical potential $\mu$] and recalling the expression \eqref{eq: occupancy mcRLM} for the noninteracting steady state occupancy $\langle n_d\rangle_{\text{s.s.}}^{(0)}$.

    \subsection{The current in the two-lead AIM}\label{sec: The current in the two-lead AIM}

    \paragraph{Noninteracting contribution to the current.} Here, we derive Equation \eqref{eq: current N electrons RLM} in the main text (the contribution to the current that does not involve any crossing states).  From Eq. \eqref{eq: Psi0Bar AIM}, we obtain
\bseq
\begin{align}
    \langle \widehat{I}_{\text{Sym}} \rangle^{(0,0)}  &= \frac{1}{[2\pi \delta(0)]^N} \lim_{\text{all } k_j'\to k_j} \left( E - E' \right) \frac{\pd}{\pd \overline{\phi}}\biggr \rvert_{\overline{\phi}=0} \sum_{\sigma\in\text{Sym}(N)}(\sgn \sigma) \prod_{m=1}^N \{ c_{\gamma_{\sigma_m} k_{\sigma_m}' a_{\sigma_m} ,\text{in}}, \overline{c}_{\gamma_m k_m a_m,\text{in} }^\dagger \} \\
    &= \frac{1}{2\pi \delta(0)} \sum_{m=1}^N \lim_{k_m' \to k_m} (k_m - k_m') \frac{\pd}{\pd \overline{\phi}}\biggr \rvert_{\overline{\phi}=0} \{ c_{\gamma_{m} k_{m}' a_m ,\text{in}}, \overline{c}_{\gamma_m k_m a_m,\text{in} }^\dagger \},\label{eq: Isteadystate with delta RLM}
\end{align}
\eseq
where we have noted that the ``in'' operators are Dirac normalized.  If any permutation other than the identity is chosen, then the result is zero in the limit of all $k_j'\to k_j$.

To evaluate \eqref{eq: Isteadystate with delta RLM}, we note that the even sector ``in'' operator is given by Eq. \eqref{eq: ckindagger RLM} (with barred fields and a spin index), and the odd sector ``in'' operator is a simple plane wave.  Thus
\beq
    \overline{c}_{o k a,\text{in}}^\dagger= \overline{c}_{ok a }^\dagger,\qquad  
    \overline{c}_{e k a,\text{in}}^\dagger =\overline{c}_{e k a}^\dagger + \int dx\ F_{k,\text{in}}(x) \left[ \Theta(0<x) \overline{\psi}_{ea}^\dagger(x) + \frac{i}{v} \delta(x) d_a^\dagger \right].\label{eq: barred in ops odd/even}
\eeq
We then obtain the ``in'' operators in the lead $1$/lead $2$ basis by rotation:
\beq
    \begin{pmatrix}
    \overline{c}_{1 k a ,\text{in}}\\
    \overline{c}_{2 k a, \text{in}}
    \end{pmatrix}
    = \mathcal{U}^\dagger
    \begin{pmatrix}
        \overline{c}_{o k a ,\text{in}}\\
        \overline{c}_{e k a, \text{in}}
    \end{pmatrix}.
\eeq
In particular, we have
\beq
    c_{\gamma k a , \text{in}} = \sum_{\dot{\gamma}=o,e} \mathcal{U}_{\gamma \dot{\gamma}}^\dagger c_{\dot{\gamma} k a },\qquad \overline{c}_{\gamma k a,\text{in}}^\dagger = \sum_{\dot{\gamma}=o,e} \overline{c}_{\dot{\gamma} k a,\text{in} }^\dagger \mathcal{U}_{\dot{\gamma} \gamma}  \qquad (\gamma=1,2),
\eeq
which yields
\bseq
\begin{align}
    &\{c_{\gamma k' a' ,\text{in} }, \overline{c}_{\gamma k a,\text{in} }^\dagger \} = \sum_{\dot{\gamma},\dot{\gamma}' = o,e} \mathcal{U}_{\dot{\gamma}' \gamma}^* \mathcal{U}_{\dot{\gamma} \gamma} \{c_{\dot{\gamma}' k'a' ,\text{in} }, \overline{c}_{\dot{\gamma} k a,\text{in} } \}  \qquad (\gamma=1,2)\\
    &= \frac{1}{2}\left[\sum_{\dot{\gamma} =o,e} \{c_{\dot{\gamma} k' a',\text{in} }, \overline{c}_{\dot{\gamma} k a,\text{in} } \} +(-1)^{\gamma -1}\left(  \{c_{o k' a',\text{in} }, \overline{c}_{e k a,\text{in} } \}  + \{c_{e k' a',\text{in} }, \overline{c}_{o k a,\text{in} } \} \right) \right].\label{eq: gammagamma inin anticomm}
\end{align}
\eseq
We proceed to evaluate the necessary anticommutators in the odd/even basis.  We note
\bseq
\begin{align}
    &\begin{pmatrix}
        \overline{c}_{oka}\\
        \overline{c}_{eka}
    \end{pmatrix}
    = \overline{\mathcal{U}}\mathcal{U}^\dagger
    \begin{pmatrix}
        c_{o k a}\\
        c_{e k a}
    \end{pmatrix}
    =
    \begin{pmatrix}
        \cos (\overline{\phi}/2) & - i \sin(\overline{\phi}/2)\\
        -i \sin(\overline{\phi} /2) & \cos(\overline{\phi}/2 )
    \end{pmatrix}
    \begin{pmatrix}
        c_{o k a}\\
        c_{e k a}
    \end{pmatrix},\\
    &\text{hence: } \frac{\pd}{\pd \overline{\phi}}\biggr \rvert_{\overline{\phi} = 0} 
    \begin{pmatrix}
        \overline{c}_{o k a}\\
        \overline{c}_{e k a}
    \end{pmatrix}
    =
    -\frac{i}{2}
    \begin{pmatrix}
        0 & 1 \\
        1 & 0
    \end{pmatrix}
    \begin{pmatrix}
        c_{o k a}\\
        c_{e k a}
    \end{pmatrix}
    = -\frac{i}{2}
    \begin{pmatrix}
        c_{e k a}\\
        c_{o k a}
    \end{pmatrix},\\
    &\text{and: } \frac{\pd}{\pd \overline{\phi}}\biggr \rvert_{\overline{\phi} = 0} \overline{\psi}_{ea}(x) = -\frac{i}{2} \psi_{oa}(x).
\end{align}
\eseq    
Applying these to Eq. \eqref{eq: barred in ops odd/even} yields
\bseq
\begin{align}
    \frac{\pd}{\pd \overline{\phi}}\biggr \rvert_{\overline{\phi} = 0} \overline{c}_{oka,\text{in}} &= -\frac{i}{2} c_{eka},\\
    \frac{\pd}{\pd \overline{\phi}}\biggr \rvert_{\overline{\phi} = 0} \overline{c}_{e k a,\text{in}}&= -\frac{i}{2} \left[ c_{oka} + \int dx\ F_{k,\text{in}}^*(x)\Theta(0<x) \psi_{oa}(x) \right].
\end{align}
\eseq
We have $\frac{\pd}{\pd \overline{\phi}} \rvert_{\overline{\phi} = 0} \{ c_{\dot{\gamma} k' a',\text{in}}, \overline{c}_{\dot{\gamma} k a,\text{in} }^\dagger \} =0 $ (for $\dot{\gamma}=o$ or $e$), since $\{ c_{o k' a'}, c_{e k a}^\dagger\} = \{ c_{e k' a'}, c_{o k a}^\dagger\} =0$.  Thus, the odd-odd and even-even contributions to Eq. \eqref{eq: gammagamma inin anticomm} vanish; the nonvanishing anticommutators under the $\overline{\phi}$ derivative are the odd-even and even-odd combinations.  For odd-even, we find
\bseq
\begin{align}
    \frac{\pd}{\pd \overline{\phi}}\biggr \rvert_{\overline{\phi} = 0} \{ c_{o k' a',\text{in}}, \overline{c}_{e k a,\text{in} }^\dagger \}&= \frac{i}{2} \left( \{ c_{o k' a'}, c_{o k a}^\dagger \} + \int dx\ F_{k,\text{in}}(x) \Theta(0<x) \{c_{ok'a'}, \psi_{o}^\dagger(x) \} \right)\\
    &= \frac{i}{2} \left[ 2\pi \delta(k-k') - i \mathcal{T}(k) \left( \pi\delta(k-k') + \text{P.V.}\frac{i}{k-k'} \right) \right]\delta_{a a'},
\end{align}
\eseq
where we have recalled that $F_{k,\text{in}}(x) = -i\mathcal{T}(k) e^{i k x}$; P.V. indicates the principal value.  Only the pole in $k-k'$ survives in the $k' \to k$ limit:
\beq
    \lim_{k' \to k} (k-k') \frac{\pd}{\pd \overline{\phi}}\biggr \rvert_{\overline{\phi} = 0} \{ c_{o k'a',\text{in}}, \overline{c}_{e ka,\text{in} }^\dagger \} = \frac{i}{2}\mathcal{T}(k) \delta_{aa'}.
\eeq
The even-odd contribution is found similarly:
\bseq
\begin{align}
    \lim_{k' \to k} (k-k') \frac{\pd}{\pd \overline{\phi}}\biggr \rvert_{\overline{\phi} = 0} \{ c_{e k' a',\text{in}}, \overline{c}_{o k a,\text{in} }^\dagger \} &= \lim_{k' \to k} (k-k') \frac{i}{2} \biggr( \{ c_{e k' a'}, c_{e k a}^\dagger \} \notag\\
    & \qquad \qquad  + \int dx\ F_{k',\text{in}}^*(x) \Theta(0<x)\{ \psi_{ea'}(x), c_{e k a}^\dagger \} \biggr)\\
    &= -\frac{i}{2}\mathcal{T}^*(k) \delta_{a a'}.
\end{align}
\eseq
From Eq. \eqref{eq: gammagamma inin anticomm}, we thus find (still for $\gamma=1,2$)
\bseq
\begin{align}
    \lim_{k' \to k} (k-k') \frac{\pd}{\pd \overline{\phi}} \biggr \rvert_{\overline{\phi}= 0} \{ c_{\gamma k' a', \text{in}}, \overline{c}_{\gamma k a, \text{in} }^\dagger \} &= \frac{1}{2}(-1)^{\gamma -1} \left( \frac{i}{2} \mathcal{T}(k) -\frac{i}{2} \mathcal{T}^*(k) \right)\delta_{aa'}\\
    &= \frac{1}{4} (-1)^{\gamma-1}|\mathcal{T}(k)|^2 \delta_{a a'},
\end{align}
\eseq
where we have used the optical identity \eqref{eq: optical identity RLM}.  Equation \eqref{eq: Isteadystate with delta RLM} then yields Eq. \eqref{eq: current N electrons RLM} in the main text.

    \paragraph{Contribution from the first crossing state.}
    We derive Eq. \eqref{eq: current up to first crossing AIM} in the main text.  First, we set up the calculation in the compact notation of Appendix \ref{sec: Notation for calculations}.  We write the Lippmann-Schwinger ``in'' state as
    \beq
        |\overline{\Psi}_{\text{in}}\rangle \equiv |\overline{\Psi}_{\gamma_\mathbf{N} k_\mathbf{N} a_\mathbf{N}, \text{in}} \rangle,
    \eeq
    where $\gamma_\mathbf{N} k_\mathbf{N} a_\mathbf{N}$ are the $N$ arbitrary quantum numbers characterizing the incoming plane waves.  In this notation, Eq. \eqref{eq: deriv formula second AIM} from the main text becomes
    \beq
        \langle \widehat{I}_{\text{Sym}}\rangle = 
        \mathcal{N}^{-1} \lim_{\text{all } k_j' \to k_j} \left(E - E'\right)    \frac{\pd }{\pd \overline{\phi} }\biggr\rvert_{\overline{\phi} = 0}\langle \Psi_{\gamma_\mathbf{N} k_\mathbf{N}' a_\mathbf{N},\text{in}} |\overline{\Psi}_{\gamma_\mathbf{N} k_\mathbf{N} a_\mathbf{N},\text{in} } \rangle, 
    \eeq
    and we wish to evaluate the right-hand side to the leading order in an expansion in crossings.  For the calculation, it is convenient to define the overlap of states with two electrons, arbitrary quantum numbers on both sides, and the zero crossing part subtracted off:
    \bseq
    \begin{align}
        \overline{\Omega}[\gamma_1' k_1' a_1',\gamma_2' k_2' a_2';\gamma_1 k_1 a_1,\gamma_2 k_2 a_2  ] &\equiv \langle \Psi_{\gamma_1' k_1' a_1' \gamma_2' k_2' a_2', \text{in}}|\overline{\Psi}_{\gamma_1 k_1 a_1 \gamma_2 k_2 a_2, \text{in}} \rangle\notag\\
        &\qquad - \langle 0| c_{\gamma_1' k_1' a_1',\text{in}} c_{\gamma_2' k_2' a_2',\text{in}}  \overline{c}_{\gamma_2 k_2 a_2,\text{in} }^\dagger\overline{c}_{\gamma_1 k_1 a_1,\text{in} }^\dagger |0\rangle \label{eq: OmegaBar def}\\
        &= \frac{1}{2} \overline{\Omega}_{(0,2)}[\gamma_1' k_1' a_1',\gamma_2' k_2' a_2';e k_1 a_1,e k_2 a_2  ] +\frac{1}{2} \overline{\Omega}_{(2,0)}[e k_1' a_1',e k_2' a_2';\gamma_1 k_1 a_1,\gamma_2 k_2 a_2  ] \notag\\
        &\qquad + \frac{1}{4} \overline{\Omega}_{(2,2)}[e k_1' a_1',e k_2' a_2';e k_1 a_1,e k_2 a_2  ],\label{eq: OmegaBar as sum of OmegaBars}
    \end{align}
    \eseq
    where
    \bseq
    \begin{align}
        &\overline{\Omega}_{(0,2)}[\gamma_1' k_1' a_1',\gamma_2' k_2' a_2';e k_1 a_1,e k_2 a_2  ] =
        \langle 0| c_{\gamma_1' k_1' a_1',\text{in}} c_{\gamma_2' k_2' a_2',\text{in}} | \overline{\Phi}_{e k_1 a_1 e k_2 a_2,\text{in}}\rangle,\label{eq: OmegeBar02 def}\\
        &\overline{\Omega}_{(2,0)}[e k_1' a_1',e k_2' a_2';\gamma_1 k_1 a_1,\gamma_2 k_2 a_2  ] = \langle \Phi_{e k_1' a_1' e k_2' a_2',\text{in}} | \overline{c}_{\gamma_2 k_2 a_2,\text{in} }^\dagger\overline{c}_{\gamma_1 k_1 a_1,\text{in} }^\dagger |0\rangle,\\
        &\overline{\Omega}_{(2,2)}[e k_1' a_1',e k_2' a_2';e k_1 a_1,e k_2 a_2  ] = \langle \Phi_{e k_1' a_1' e k_2' a_2',\text{in}} | \overline{\Phi}_{e k_1 a_1 e k_2 a_2,\text{in}}\rangle. 
    \end{align}
    \eseq
    The term $\overline{\Omega}_{(2,2)}$ contains a product of two crossing states (one from each wavefunction) and so will be dropped later, but it is convenient to keep it for the moment.  A short calculation shows the identity
    \beq
        \lim_{\substack{k_1' \to k_1 \\ k_2' \to k_2} } (k_1 + k_2 - k_1' -k_2') \Omega[\gamma_1' k_1' a_1',\gamma_2' k_2' a_2';\gamma_1 q_1 a_1,\gamma_2 q_2 a_2  ] = 0, \label{eq: identity for AIM prev}
    \eeq
    where the momenta $k_1,k_2,q_1$, and $q_2$ are arbitrary (and also the spins and lead indices).  Note that the bar has been removed \footnote{We expect a more general version of \eqref{eq: identity for AIM prev} to hold (with $n=2$ quantum numbers on each side replaced by general $n$), as it would then be a time-independent analog of a general identity we have shown in the time-dependent case in Ref. \cite{Culver_thesis}.}.
    
    Next, we apply Wick's theorem to the overlap of interest, with the wavefunction on each side truncated so that at most two quantum numbers can be in a crossing state.  We use the notation $\alpha_j \equiv \gamma_j k_j a_j$, $\alpha_j' \equiv \gamma_j k_j' a_j$ to reduce clutter, finding
    \begin{multline}
        \langle \Psi_{\alpha_\mathbf{N}',\text{in} } | \overline{\Psi}_{ \alpha_\mathbf{N},\text{in}} \rangle = \langle \Psi_{\alpha_\mathbf{N}',\text{in} }^0 | \overline{\Psi}_{ \alpha_\mathbf{N},\text{in}}^0 \rangle + \sum_{\mathbf{m},\mathbf{m}' \in \mathcal{I}_2(\mathbf{N})} (\sgnleft{\mathbf{m}})(\sgnleft{\mathbf{m}'}) \sum_{\sigma\in \text{Sym}(2) }(\sgn \sigma) \left( \prod_{j=1}^{N-2} \{ c_{\alpha_{(\mathbf{N} / \mathbf{m}')_{\sigma_j} }',\text{in} }, \overline{c}_{\alpha_{(\mathbf{N}/ \mathbf{m} )_j},\text{in} }^\dagger \} \right)\\
        \times \overline{\Omega}[\alpha_{\mathbf{m} '}' ; \alpha_{\mathbf{m} } ].
    \end{multline}
    The first term on the right-hand side, with zero crossings, has already been dealt with.  To get the current, we apply $\mathcal{N}^{-1}\lim_{\text{all } k_j' \to k_j} \left(E - E'\right)    \frac{\pd }{\pd \overline{\phi} }\rvert_{\overline{\phi} = 0}$ to both sides.  If the $\overline{\phi}$ derivative acts on one of the anticommutators, then we get zero due to the identity \eqref{eq: identity for AIM prev} (since $\overline{\Omega}$ gets replaced by $\Omega$).  Thus, we have to act the derivative on $\overline{\Omega}$; this removes the bar from all of the anticommutators, diagonalizing the sum ($\mathbf{m}' = \mathbf{m}$ and $\sigma = $identity) and yielding
    \begin{multline}
        \langle \widehat{I}_{\text{Sym}} \rangle = \langle \widehat{I}_{\text{Sym}} \rangle^{(0,0)} + \mathcal{N}^{-1} \sum_{\mathbf{m}\in \mathcal{I}_2(\mathbf{N})} [ 2\pi \delta(0)]^{N-2}   \lim_{\substack{ k_{m_1}'\to k_{m_1} \\ k_{m_2}'\to k_{m_2} } } (k_{m_1} + k_{m_2} - k_{m_1}' - k_{m_2}' )    \frac{\pd }{\pd \overline{\phi} }\biggr\rvert_{\overline{\phi} = 0} \overline{\Omega}[\alpha_\mathbf{m}'; \alpha_\mathbf{m} ]\\
        = \langle \widehat{I}_{\text{Sym}} \rangle^{(0,0)} + [2\pi\delta(0)]^{-2} \sum_{1\le m_1 < m_2 \le N}  \lim_{\substack{ k_{m_1}'\to k_{m_1} \\ k_{m_2}'\to k_{m_2} } } (k_{m_1} + k_{m_2} - k_{m_1}' - k_{m_2}' ) \\
        \times \frac{\pd }{\pd \overline{\phi} }\biggr\rvert_{\overline{\phi} = 0} \overline{\Omega}[\gamma_{m_1} k_{m_1}' a_{m_1},  \gamma_{m_2} k_{m_2}' a_{m_2};\gamma_{m_1} k_{m_1} a_{m_1}, \gamma_{m_2} k_{m_2} a_{m_2}],
    \end{multline}
    where we have written the compact notation in full and recalled that $\mathcal{N} = [2\pi \delta(0)]^N$.  From the definition, $\overline{\Omega}$ satisfies a symmetry property in its quantum numbers that allows us to replace the sum over $m_1<m_2$ by an unrestricted sum with an extra factor of $1/2$.  Comparing to Eqs. \eqref{eq: current as 00+02+20 AIM} and \eqref{eq: OmegaBar as sum of OmegaBars}, we can then read off
    \begin{multline}
        \langle \widehat{I}_{\text{Sym}} \rangle^{(\ell_1,\ell_2)} = [2\pi\delta(0)]^{-2} \frac{1}{4} \sum_{m_1,m_2=1}^N  \lim_{\substack{ k_{m_1}'\to k_{m_1} \\ k_{m_2}'\to k_{m_2} } } (k_{m_1} + k_{m_2} - k_{m_1}' - k_{m_2}' ) \\
        \times \frac{\pd }{\pd \overline{\phi} }\biggr\rvert_{\overline{\phi} = 0} \overline{\Omega}_{(\ell_1,\ell_2)}[\gamma_{m_1} k_{m_1}' a_{m_1},  \gamma_{m_2} k_{m_2}' a_{m_2};e k_{m_1} a_{m_1}, e k_{m_2} a_{m_2}],
    \end{multline}
    where $(\ell_1,\ell_2)$ is $(0,2)$ or $(2,0)$.  The $(0,2)$ case is Eq. \eqref{eq: I02 AIM intermediate step} from the main text.  It suffices to calculate the $(0,2)$ term, since the $(2,0)$ term is the complex conjugate.  We ignore the $(2,2)$ term, involving $\overline{\Omega}_{(2,2)}$, since it involves a product of two crossings.
    
    By antisymmetry, the overlap $\overline{\Omega}_{(0,2)}$ can be written as the antisymmetrization of some ``reduced'' overlap $\overline{\Omega}_{(0,2)}^{(\text{red})}$ as follows:
\begin{multline}
    \overline{\Omega}_{(0,2)}[\gamma_1' k_1' a_1',\gamma_2' k_2' a_2';e k_1 a_1,e k_2 a_2  ] = \sum_{\sigma,\sigma' \in \text{Sym}(2)} (\sgn \sigma)(\sgn \sigma')\\
    \times \overline{\Omega}_{(0,2)}^{(\text{red})} [\gamma_{\sigma_1'}' k_{\sigma_1'}' a_{\sigma_1'}',\gamma_{\sigma_2'}' k_{\sigma_2'}' a_{\sigma_2'}';e k_{\sigma_1} a_{\sigma_1},e k_{\sigma_2} a_{\sigma_2}  ],\label{eq: OmegaBar in terms of OmegaBarRed}
\end{multline}
where $\overline{\Omega}_{(0,2)}^{(\text{red})}$ is only defined modulo antisymmetrization.  To specialize the quantum numbers $\gamma_\mathbf{N} k_\mathbf{N} a_\mathbf{N}$ to the case of two filled Fermi seas, we replace the sums over quantum numbers according to Eq. \eqref{eq: sum replacement rule with spin}.  Relabelling summation indices, we then obtain
    
    \bseq
\begin{align}
    &\langle \widehat{I}_{\text{Sym}} \rangle^{(0,2)} = [2\pi \delta(0)]^{-2} \frac{1}{2} \sum_{\gamma_1,\gamma_2=1,2} \sum_{\substack{ k_1 \in \mathcal{K}_{\gamma_1}  \\ k_2 \in \mathcal{K}_{\gamma_2} }} \sum_{a_1, a_2} \sum_{\sigma \in \text{Sym}(2)}(\sgn \sigma) \lim_{\substack{ k_1'\to k_1 \\ k_2'\to k_2 } } (k_1 + k_2 - k_1'- k_2' )\notag\\
    &\qquad \qquad \qquad \qquad 
    \times \frac{\pd}{\pd \overline{\phi}}\biggr\rvert_{\overline{\phi} =0} \overline{\Omega}_{(0,2)}^{(\text{red})} [\gamma_1 k_1' a_1,\gamma_2 k_2' a_2;e k_{\sigma_1} a_{\sigma_1},e k_{\sigma_2} a_{\sigma_2}  ] \\
    &\overset{\text{therm. limit}}{\longrightarrow} \frac{1}{2}\sum_{\gamma_1,\gamma_2=1,2} \int_{-D}^D \frac{dk_1}{2\pi}\frac{dk_2}{2\pi}\ \fermifn_{\gamma_1}(k_1)\fermifn_{\gamma_2}(k_2) \sum_{a_1, a_2} \sum_{\sigma \in \text{Sym}(2)}(\sgn \sigma) \notag\\
    &\qquad \times \lim_{\substack{ k_1'\to k_1 \\ k_2'\to k_2 } } (k_1 + k_2 - k_1'- k_2' )
    \frac{\pd}{\pd \overline{\phi}}\biggr\rvert_{\overline{\phi} =0} \overline{\Omega}_{(0,2)}^{(\text{red})} [\gamma_1 k_1' a_1,\gamma_2 k_2' a_2;e k_{\sigma_1} a_{\sigma_1},e k_{\sigma_2} a_{\sigma_2}  ].  \label{eq: current02 therm limit in terms of Omegared AIM}
\end{align}
\eseq
We recall that the two electron crossing state is given by
\beq
    |\overline{\Phi}_{e k_1 a_1 ek_2 a_2,\text{in} }\rangle= |\overline{\chi}_{e k_1 a_1 ek_2 a_2,\text{in} }\rangle - |\overline{\chi}_{e k_2 a_2 ek_1 a_1,\text{in} }\rangle,
\eeq
where the unsymmetrized crossing state is given by Eq. \eqref{eq: ansatz n=2 AIM} with a bar over the electron fields:
\begin{multline}
    |\overline{\chi}_{e k_2 a_2 ek_1 a_1,\text{in} }\rangle = \int dx_1 dx_2\ F_{e k_1 a_1 e k_2 a_2}^{b_1 b_2}(x_1,x_2) \biggr[ \Theta(0< x_2 < x_1) \overline{\psi}_{e b_2}^\dagger(x_2)\overline{\psi}_{e b_1}^\dagger(x_1) \\
    + \frac{i}{v}\delta(x_2) \Theta(0< x_1) d_{b_2}^\dagger \overline{\psi}_{e b_1}^\dagger(x_1) 
    -\frac{1}{2 v^2} \delta(x_1)\delta(x_2) d_{b_2}^\dagger d_{b_1}^\dagger \biggr] |0\rangle,
\end{multline}
where the function $F$ is given in Eq. \eqref{eq: F n=2 AIM}.

We proceed to evaluate the reduced function $\overline{\Omega}_{(0,2)}^{(\text{red})}$.  Any terms that are finite in the limit of equal momenta ($k_j' \to k_j$) can be dropped; we are looking for a \emph{real} pole, such as $1/(k_1 +k_2 -k_1' -k_2')$ (as opposed to a pole off the real axis).  Such a pole is not present in the $\mathcal{T}$ matrix prefactors that appear explicitly in the ``in'' operators and crossing state (since Im $\pole = -\Delta \ne 0$), so it can only be produced by the position integral in the overlap itself (see below).  From Eqs. \eqref{eq: OmegaBar in terms of OmegaBarRed} and \eqref{eq: OmegeBar02 def} and the form of the crossing state, we can read off
\begin{multline}
    \overline{\Omega}_{(0,2)}^{(\text{red})} [\gamma_1' k_1' a_1',\gamma_2' k_2' a_2';e k_1 a_1,e k_2 a_2  ] = \int dx_1 dx_2\ F_{e k_1 a_1 e k_2 a_2}^{b_1 b_2}(x_1,x_2) \Theta(0<x_2<x_1)\\
    \times \{ c_{\gamma_1' k_1' a_1',\text{in}}, \overline{\Psi}_{eb_1}^\dagger(x_1)\} \{ c_{\gamma_2' k_2' a_2',\text{in}}, \overline{\Psi}_{eb_2}^\dagger(x_2)\} + (\text{regular}),
\end{multline}
where ``regular'' indicates omitted terms that are finite in the limit of equal momenta.  These are the terms involving the anticommutators $\{ c_{\gamma_1' k_1' a_1',\text{in}}, \overline{\Psi}_{eb_1}^\dagger(x_1)\} \{ c_{\gamma_2' k_2' a_2',\text{in}}, d_{b_2}^\dagger\}$ and $\{ c_{\gamma_1' k_1' a_1',\text{in}},d_{b_1}^\dagger\} \{ c_{\gamma_2' k_2' a_2',\text{in}}, d_{b_2}^\dagger\}$.  In the former case, we get a single integral over $x_1$ involving $F(x_1,0)$, which (due to the factor of $\pole$ in the exponent in $F$) produces a momentum denominator with a complex pole; in the latter case, we just get a constant again with complex poles.

To take the $\overline{\phi}$ derivative, we recall that $\frac{\pd}{\pd \overline{\phi}}  \rvert_{\overline{\phi}=0} \overline{\psi}_{e b}^\dagger(x) = \frac{i}{2}\psi_{o b}^\dagger(x)$.  This derivative can act on either one of the anticommutators, setting $\overline{\psi}^\dagger_{eb} =\psi_{eb}^\dagger$ in the other one.  In the calculation that follows, we do this derivative, then relabel variables in one term (using the fact that $\overline{\Omega}_{(0,2)}^{(\text{red})}$ is only defined up to antisymmetrization), then recall the anticommutators $\{ c_{\gamma' k' a',\text{in}}, \psi_{eb}^\dagger(x)\} = \frac{1}{\sqrt{2}}\mathcal{S}^*(k')\delta_{b a'}e^{-i k' x} $ [where $\mathcal{S}(k) = 1 -i \mathcal{T}(k)$ is the $\mathcal{S}$ matrix] and $\{ c_{\gamma' k' a',\text{in}}, \psi_{ob}^\dagger(x)\} = \frac{1}{\sqrt{2}}(-1)^{\gamma'-1}\delta_{ba'}e^{-i k' x}$, then put in the explicit form of $F$.  The ``regular'' part is dropped throughout.  Following these steps, we find
\bseq
\begin{align}
    &\frac{\pd}{\pd \overline{\phi}} \biggr \rvert_{\overline{\phi}=0} \overline{\Omega}_{(0,2)}^{(\text{red})} [\gamma_1' k_1' a_1',\gamma_2' k_2' a_2';e k_1 a_1,e k_2 a_2  ] = \frac{i}{2} \int dx_1 dx_2\ F_{ek_1 a_1 ek_2 a_2}^{b_1 b_2}(x_1,x_2)\notag\\
    &\qquad \times \Theta(0<x_2<x_1) \left[ \{ c_{\gamma_1' k_1' a_1',\text{in}}, \Psi_{eb_1}^\dagger(x_1)\} \{ c_{\gamma_2' k_2' a_2',\text{in}}, \Psi_{ob_2}^\dagger(x_2)\} + (o \leftrightarrow e) \right]\\
    &\overset{\text{up to antisymm.}}{=}\frac{i}{2} \int dx_1 dx_2\ F_{ek_1 a_1 ek_2 a_2}^{b_1 b_2}(x_1,x_2)
    \Theta(0<x_2<x_1)\notag\\
    &\qquad \times \left[ \{ c_{\gamma_1' k_1' a_1',\text{in}}, \Psi_{eb_1}^\dagger(x_1)\} \{ c_{\gamma_2' k_2' a_2',\text{in}}, \Psi_{ob_2}^\dagger(x_2)\} - (b_1\leftrightarrow b_2, x_1\leftrightarrow x_2) \right]\\
    &= \frac{i}{2} \int dx_1 dx_2\ F_{ek_1 a_1 ek_2 a_2}^{b_1 b_2}(x_1,x_2)
    \Theta(0<x_2<x_1)\notag\\
    &\qquad \times \frac{1}{\sqrt{2}}\mathcal{S}^*(k_1') \frac{1}{\sqrt{2}}(-1)^{\gamma_2'-1} \left[ \delta_{b_1 a_1'}\delta_{b_2 a_2'}e^{-i k_1' x_1}e^{-i k_2' x_2} - (b_1\leftrightarrow b_2, x_1\leftrightarrow x_2) \right]\\
    &= \frac{i}{2}\int dx_1 dx_2\ \left[ -  \mathcal{T}(k_1) \mathcal{T}(k_2) \frac{U \mathcal{T}\left[(k_1 + k_2 -U)/2  \right]}{4\Delta}\right] e^{i(k_1 +k_2)x_1}e^{-i \pole(x_1- x_2)}\notag\\
    &\qquad\times P_{- a_1 a_2}^{\ a_1' a_2'} \Theta(0<x_2<x_1) \frac{1}{2}\mathcal{S}^*(k_1') (-1)^{\gamma_2' -1}    \left( e^{-i k_1' x_1}   e^{-i k_2' x_2} + e^{-i k_2' x_1}   e^{-i k_1' x_2}\right).
\end{align}
\eseq
The integration over position yields a real pole:
\beq
    \int dx_1 dx_2\ e^{i (k_1 +k_2 -k_1' -\pole)x_1}e^{i(\pole-k_2')x_2} \Theta(0<x_2<x_1) = -\frac{\mathcal{T}(k_2')}{2\Delta} \text{P.V.} \frac{1}{k_1 + k_2 -k_1' -k_2'}.
\eeq
Thus, we obtain
\begin{multline}
    \lim_{\substack{ k_1'\to k_1 \\ k_2'\to k_2 } } (k_1 + k_2 - k_1'- k_2' )\frac{\pd}{\pd \overline{\phi}}\biggr\rvert_{\overline{\phi} =0} \overline{\Omega}_{(0,2)}^{(\text{red})} [\gamma_1' k_1' a_1',\gamma_2' k_2' a_2';e k_1 a_1,e k_2 a_2  ] = \\
    \frac{i}{32\Delta^2} (-1)^{\gamma_2' -1}  \mathcal{T}(k_1) \mathcal{T}(k_2) U \mathcal{T}\left[(k_1 + k_2 -U)/2  \right] \mathcal{S}^*(k_1')\left[\mathcal{T}(k_1') + \mathcal{T}(k_2')\right]P_{- a_1 a_2}^{\ a_1' a_2'},
\end{multline}
and then Eq. \eqref{eq: current02 therm limit in terms of Omegared AIM} yields [using $\mathcal{S}^*(k) = \mathcal{T}^*(k)/\mathcal{T}(k)$ and the spin sums $P_{- a_1 a_2}^{\ a_1 a_2}=1$ and $P_{- a_1 a_2}^{\ a_2 a_1}=-1$]
\begin{multline}
    \langle \widehat{I}_{\text{Sym}} \rangle^{(0,2)} \overset{\text{therm. limit}}{\longrightarrow} \frac{i}{32\Delta^2}\int_{-D}^D \frac{dk_1}{2\pi}\frac{dk_2}{2\pi}\  \left[ \fermifn_1(k_1) + \fermifn_2(k_1) \right]\left[\fermifn_1(k_2) - \fermifn_2(k_2) \right]\\
    \times \mathcal{T}^*(k_1) \mathcal{T}(k_2) U \mathcal{T}\left[(k_1 + k_2 -U)/2  \right] \left( \mathcal{T}(k_1) + \mathcal{T}(k_2) \right).
\end{multline}
By taking the adjoint and relabelling, we can read off that $\langle \widehat{I}_{\text{Sym}}\rangle^{(2,0)}= \langle \widehat{I}_{\text{Sym}}\rangle^{(0,2)*}$.  We thus obtain Eq. \eqref{eq: current up to first crossing AIM} in the main text.

\section{Equilibrium occupancy of the IRL from the literature}
\label{sec: Equilibrium occupancy of the IRL from the literature}

From Bethe ansatz, a series form is known for the occupancy $\langle n_d\rangle_{\text{equilibrium}}$ at zero temperature as a function of $\epsilon_d$ in the multilead IRL.  We show here that the leading order of this series agrees with our result for the leading order equilibrium occupancy reached as the long-time limit following a quench [Eq. \eqref{eq: ndeq univ mcIRL}]. 

It turns out that in the universal regime, the number of leads does not appear in the answer.  The calculation was first done by Ponomarenko \cite{Ponomarenko}.  In the one-lead case, Rylands and Andrei \cite{RylandsAndrei} calculated the occupancy including a Luttinger interaction, and Camacho \emph{et al}. \cite{CamachoSchmitteckertCarr} have verified that setting the Luttinger interaction to zero (which recovers the one-lead IRL) results in exact agreement between the answers of Ref. \cite{Ponomarenko} and Ref. \cite{RylandsAndrei}.  We transcribe the result from Ref. \cite{CamachoSchmitteckertCarr} with some minor changes in notation:
    \beq
        \langle n_d \rangle_{\text{equilibrium}} \equiv n_d(x_d) =
        \begin{cases}
            \frac{1}{2} - \sum_{n=0}^\infty h_n^<(\alpha) \left( \frac{x_d}{\pi h_{n=0}^<(\alpha)}\right)^{2n+1} & 0 \le x_d < 1\\
            \sum_{n=1}^\infty h_n^>(\alpha) \left( \frac{x_d}{\pi h_{n=0}^<(\alpha)}\right)^{-\frac{2n}{\alpha}} & x_d  \ge 1\\
            1- n_d(-x_d) & x_d <0,
        \end{cases}
    \eeq
    where $x_d = \epsilon_d/ T_K$ (note that $\epsilon_0$ in Ref. \cite{CamachoSchmitteckertCarr} is our $\epsilon_d$) and
    \bseq
        \begin{align}
            h_n^<(\alpha) &= \frac{1}{\sqrt{\pi}}\frac{(-1)^n}{n!} \frac{\Gamma(1+ \frac{\alpha}{2} (2n+1) }{\Gamma( 1+ \frac{\alpha-1}{2} (2n+1) ) }, \\
            h_n^>(\alpha) &= \frac{1}{2\sqrt{\pi}}\frac{(-1)^{n+1}}{n!}\frac{\Gamma(1/2  + n/\alpha)}{\Gamma(1- \frac{\alpha-1}{\alpha}n ) }.
        \end{align}
    \eseq
    The quantity $\alpha$ is an RG invariant, with $\alpha=2$ in the noninteracting case ($U=0$).  In principle the Bethe ansatz $\alpha$ could differ from our formula \eqref{eq: alpha mcIRL} for $\alpha$ in the main text, but we find that they are the same (at least to leading order in $U$).  To compare with our answer for the occupancy in the main text, we expand to linear order about $\alpha-2$.    For $|x_d| <1$, we obtain
    \beq
        n_d(x_d) = \frac{1}{2} - \sum_{n=0}^\infty \Biggr\{ \frac{(-1)^n}{\pi(1+ 2n)} + (\alpha-2) \frac{(-1)^n}{2\pi} \left[ 1 -2\ln 2 + \psi(2+2n) -\psi(3/2 +n) \right]
        + O\left((\alpha-2)^2\right)\Biggr\} x_d^{2n +1},
    \eeq
    where $\psi = \Gamma'$ is the digamma.  This sum yields \eqref{eq: ndeq univ mcIRL} in the main text once we identify $\alpha-2 = -\frac{2U}{\pi}$ at leading order [in agreement with Eq. \eqref{eq: alpha mcIRL}].  For $x_d>1$, we obtain
    \begin{multline}
        n_d(x_d)= \sum_{n=1,3,\dots} \Biggr\{ \frac{(-1)^{(n-1)/2}}{\pi n} 
        + (\alpha-2)\frac{(-1)^{(n-1)/2}}{2\pi}\Bigg[ 1 -\ln 2 - \ln x_d + \psi(1/2 + n/2)
        - \psi(1+ n/2) \Bigg] + O\left((\alpha-2)^2\right) \Biggr\}x_d^{-n}\\
        +\sum_{n=2,4,\dots} \left\{ (\alpha-2) \frac{(-1)^{n/2 - 1 }}{4}  + O\left((\alpha-2)^2\right) \right\}x_d^{-n}.
    \end{multline}
    The sum over the $\alpha=2$ part yields the standard noninteracting result $1/2 - (1/\pi)\arctan x_d$.  Numerical evaluation of the $\alpha-2$ correction term again agrees with Eq. \eqref{eq: ndeq univ mcIRL} in the main text.
    
    \section{Perturbative check: the current in the Anderson model}
    \label{sec: Perturbative check the current in the Anderson model}
    
    We calculate the steady state current in the Anderson model to leading order in $U$ using Keldysh perturbation theory, confirming the result \eqref{eq: current small U AIM} from the main text.  Rather than evaluate the current operator directly, we find it more convenient to use the Meir-Wingreen formula \cite{MeirWingreen}, which relates the steady state current to an impurity-impurity Green's function.
    
For this calculation, we allow a magnetic field on the dot; that is, the dot energy can be spin dependent:
\bseq
\begin{align}
    H^{(0)} &= -i \int_{-L/2}^{L/2} dx\ \sum_{\substack{\gamma=1,2\\a}}\psi_{\gamma a}^\dagger(x) \frac{d}{dx} \psi_{\gamma a}(x) + \sum_a \epsilon_a d_a^\dagger d_a+  \sum_{\substack{\gamma=1,2\\ a}}\left[ \frac{v}{\sqrt{2}}\psi_{\gamma a}^\dagger(0) d_a + \text{h.c.} \right],\\
    H^{(1)} &= U n_{\uparrow}n_{\downarrow},\\
    H &= H^{(0)} + H^{(1)}.
\end{align}
\eseq
Our conventions in this appendix depart in two ways from the rest of the paper.  First, all time-dependent operators are in either the Heisenberg picture (subscript $H$) or the interaction picture (no subscript), with the usual sign (i.e., $e^{i H t}$ or $e^{i H^{(0)} t}$ on the left, respectively).  Second, repeated spin indices are \emph{not} summed in the absence of a summation sign.
    
Let $\rho$ be the density matrix describing the two-leads each separately in thermal equilibrium, with no tunneling:
\beq
    \rho = \exp \left[ -\frac{1}{T_1}\sum_{\substack{|k|<D \\ a}}(k-\mu_1) c_{1ka}^\dagger c_{1ka} \right] \otimes \exp \left[ -\frac{1}{T_2} \sum_{\substack{|k|<D \\ a}}(k-\mu_2) c_{2ka}^\dagger c_{2ka} \right].
\eeq
The Fermi functions of the leads are $\fermifn_{\gamma}(k) = [e^{(k-\mu_\gamma)/T_\gamma} +1 ]^{-1}$ ($\gamma=1,2$).

The retarded Green's function with respect to the time-evolving density matrix $\rho(t) \equiv e^{-i H t }\rho e^{i H t}$ is given by
\beq
    \mathcal{G}_{a,a}^R(t; t_1,t_2) \equiv -i \Theta(t_1- t_2)\text{Tr} \left[\rho(t)   \{d_{a H}(t_1), d_{aH}^\dagger(t_2)    \} \right]/\text{Tr}\rho. \label{eq: ret fn wrt Psi(t)}
\eeq
This Green's function, and all others introduced below, is implicitly in the thermodynamic limit ($L\to\infty$ with fixed bandwidth $D$; note in particular that the system size goes to infinity before the evolution time $t$).  In the steady state, we get a function of the time difference only:\footnote{This is a nontrivial assumption---that a time-independent steady state \emph{is} reached.}
\beq
    \lim_{t\to\infty} \mathcal{G}_{a,a}^R(t; t_1 , t_2) \equiv \mathcal{G}_{a,a}^R(t_1 - t_2).
\eeq
The Meir-Wingreen formula, specialized to the Hamiltonian we consider here, is the following expression for the steady state current:
\beq
    \lim_{t\to\infty} I_{\text{Sym}}(t)=  - \Delta \sum_a \int_{-D}^D \frac{dw}{2\pi}\ \left[\fermifn_1(w) - \fermifn_2(w)\right] \text{Im}\left[  \widetilde{\mathcal{G}}_{a,a}^R(w) \right], \label{eq: MW formula}
\eeq
where $I_{\text{Sym}}(t) = \text{Tr} \left[ \rho(t) \widehat{I}_{\text{Sym}}\right]/\text{Tr}\rho$, $\widehat{I}_{\text{Sym}}= \frac{i}{2\sqrt{2}} v \left( \psi_{1 a}^\dagger(0) - \psi_{2 a}^\dagger(0) \right) d_a + \text{h.c.}$, and $\widetilde{\mathcal{G}}_{a,a}^R(w) \equiv \int dt'\ e^{i w t'}\mathcal{G}_{a,a}^R(t')$.  The derivation \cite{MeirWingreen} proceeds by applying Keldysh identities to Dyson equations for Green's functions \footnote{There are some slight differences between our setup and that of Ref. \cite{MeirWingreen}; we turn on the couplings at $t=0$ suddenly, while Ref. \cite{MeirWingreen} turns on the couplings at $t=-\infty$ gradually.  However, we have confirmed Eq. \eqref{eq: MW formula} by deriving it directly in our case using the method of Ref. \cite{MeirWingreen}.}. 
    
Equation \eqref{eq: ret fn wrt Psi(t)} is more conveniently written as
\beq
    \mathcal{G}_{a,a}^R(t;t_1,t_2) = G_{a,a}^R(t + t_1, t+ t_2),     
\eeq    
where $G_{a,a}^R(t_1,t_2) = -i \Theta(t_1- t_2)\text{Tr}\left[\rho\  \{d_{a H}(t_1), d_{aH}^\dagger(t_2)    \}  \right] / \text{Tr}\rho$ is the retarded Green's function defined relative to the initial density matrix $\rho$ (rather than the time-evolving density matrix).  For perturbative evaluation of $G_{a,a}^R(t_1,t_2)$ (with $0<t_2<t_1$ and with $t_1$ and $t_2$ later to be shifted by $t$), we introduce a Keldysh contour $C$ that runs from $0$ to $t_1$ (the $+$ branch) and back (the $-$ branch).  The Keldysh Green's functions with respect to $\rho$ are defined by
\beq
    G_{a,a}^{\alpha_1 \alpha_2}(t_1, t_2) = -i \text{Tr}\left[ \rho\  T_C\ d_{a H}(t_1^{(\alpha_1)}) d_{aH}^\dagger(t_2^{(\alpha_2)}) \right] / \text{Tr}\rho, 
\eeq
where $\alpha_1$ and $\alpha_2$ are branch indices ($\pm)$ and $T_C$ is the path-ordering symbol.  The retarded Green's function is given by $G_{a,a}^R = G_{a,a}^{T} - G_{a,a}^{<}$, where $G_{a,a}^T = G_{a,a}^{+ +}$ and $G_{a,a}^< = G_{a,a}^{+-}$.   By some standard manipulations, we obtain another form more suitable for perturbation theory:
\beq
    G_{a,a}^{\alpha_1 \alpha_2}(t_1, t_2) = -i \text{Tr}\left[ \rho\  T_C\ U_C d_{a}(t_1^{(\alpha_1)}) d_a^\dagger(t_2^{(\alpha_2)}) \right] / \text{Tr}\rho,\label{eq: Keldysh Green's function} 
\eeq
where the impurity operators evolve in the interaction picture and $U_C$ is the interaction picture propagator:
\beq
    U_C = T_C \exp\left[ -i \int_C dt'\ H^{(1)}(t')  \right].
\eeq
Our first task is to expand $G_{a,a}$ to first order in $U$ in terms of the Keldysh Green's functions of the two-lead RLM, which are defined as follows:
\beq
    G_{a,a}^{(0)\alpha_1 \alpha_2}(t_1, t_2) = -i \text{Tr}\left[ \rho\  T_C\ d_{a}(t_1^{(\alpha_1)}) d_a^\dagger(t_2^{(\alpha_2)}) \right] / \text{Tr}\rho.
\eeq
We write the interaction term in the Hamiltonian as $H^{(1)} = U d_a^\dagger d_a d_{\overline{a}}^\dagger d_{\overline{a}}$, where $\overline{a}$ is the opposite spin to $a$ ($\overline{a}=\downarrow$ if $a=\uparrow$ and vice-versa).  In the first order correction to the Green's function, the $\overline{a}$ impurity operators contract with each other, yielding a factor of $-i G_{\overline{a}, \overline{a}}^{(0)<}(t',t')$.  We find
\beq
    G_{a,a}(t_1,t_2) = G_{a,a}^{(0)}(t_1,t_2) - i U \int_C dt'\ G_{\overline{a},\overline{a}}^{(0)<}(t',t') G_{a,a}^{(0)}(t_1,t') G_{a,a}^{(0)}(t',t_2) + O(U^2),\label{eq: Keldysh Green's function leading order}
\eeq
where Keldysh branch indices have been suppressed.  We specialize the left-hand side to the retarded Green's function, then use a Langreth rule to obtain
\beq
    G_{a,a}^R (t_1,t_2) = G_{a,a}^{{(0)} R} (t_1,t_2)- i U \int dt'\ G_{\bar{a}, \bar{a}}^{{(0)} <}(t',t')G_{a, a}^{{(0)} R}(t_1,t')G_{a, a}^{{(0)} R}(t', t_2).
\eeq
Note that we have replaced $\int_0^{t_1}dt' \to \int_{-\infty}^{\infty} dt'\equiv \int dt'$, since the retarded Green's functions restrict $t'$ to the interval $0< t_2< t'< t_1$.  The retarded Green's function of the RLM depends only on the difference of times (see below), so we find
\bseq
\begin{align}
    &\mathcal{G}_{a,a}^R(t_1 - t_2) = \notag\\
    & \lim_{t\to \infty} \left[ G_{a,a}^{{(0)} R} (t+ t_1, t+ t_2)- i U \int dt'\ G_{\bar{a}, \bar{a}}^{{(0)} <}(t',t')G_{a, a}^{{(0)} R}(t+ t_1,t')G_{a, a}^{{(0)} R}(t', t+ t_2)  \right]\\
    &= G_{a,a}^{(0)R}(t_1 - t_2) - i U \left[ \lim_{t\to\infty} G_{\bar{a}, \bar{a}}^{{(0)} <}(t, t) \right] \int dt'\ G_{a, a}^{{(0)} R}(t_1- t')G_{a, a}^{{(0)} R}(t'- t_2).
\end{align}
\eseq
The time integral is a convolution, so the Fourier transform yields
\beq
    \widetilde{\mathcal{G}}_{a,a}^R(w) = \widetilde{G}_{a,a}^{(0)R}(w) - i U \left[ \lim_{t\to\infty} G_{\bar{a}, \bar{a}}^{{(0)} <}(t, t) \right] \left[ \widetilde{G}_{a, a}^{{(0)} R}(w)\right]^2.
\eeq
Thus, the only Green's functions that we need from the two-lead RLM are the following:
\bseq
\begin{align}
    G_{a,a}^{(0)R}(t_1,t_2) &= -i \Theta(t_1-t_2) e^{-i \pole_a (t_1-t_2)} \equiv G_{a,a}^{(0)R}(t_1- t_2),\\
    \widetilde{G}_{a,a}^{(0)R}(w) &= \frac{1}{w- z_a} = \frac{\mathcal{T}_a(w)}{2\Delta},\\
    - i \left[ \lim_{t\to\infty} G_{\bar{a}, \bar{a}}^{{(0)} <}(t, t) \right] &= \lim_{t\to\infty}  \text{Tr}\left[\rho\ e^{i H^{(0)} t} d_{\bar{a}}^\dagger d_{\bar{a}} e^{-i H^{(0)}t} \right] /\text{Tr}\rho\\
    &=  \frac{1}{2}\int_{-D}^{D} \frac{dk}{2\pi}\left[\fermifn_1(k) + \fermifn_2(k)\right] \frac{|\mathcal{T}_{\bar{a}}(k)|^2}{2\Delta},\label{eq: Glesser limit}
\end{align}
\eseq
where $z_a = \epsilon_a - i \Delta$ and $\mathcal{T}_a(k) = 2\Delta/(k- z_a)$.  These Green's functions can be found by conventional means or by using the time-dependent operators of Sec. \ref{sec: Noninteracting case: The resonant level model}.  Equation \eqref{eq: Glesser limit} is the two-lead case of the more general Eq. \eqref{eq: occupancy mcRLM} from the main text.
    
Using the ``optical'' identity \eqref{eq: optical identity RLM}, we then obtain
\beq
    \text{Im}\left[ \widetilde{\mathcal{G}}_{a,a}^R(w) \right] = - \frac{|\mathcal{T}_a(w)|^2}{4 \Delta} -\frac{U}{16\Delta^3} \int_{-D}^D \frac{dk}{2\pi}\ \left[\fermifn_1(k) - \fermifn_2(k) \right] |\mathcal{T}_{\bar{a}}(k)|^2 |\mathcal{T}_a(w)|^2 \text{Re}\left[ \mathcal{T}_a(w)\right].
\eeq
Then from \eqref{eq: MW formula}, we find the steady state current to first order in $U$:
\begin{multline}
    \lim_{t\to\infty}I_{\text{Sym}}(t) = \int_{-D}^D \frac{dw}{2\pi}\ \left[\fermifn_1(w) - \fermifn_2(w) \right]\sum_a \frac{1}{4}|\mathcal{T}_a(w)|^2 \\
    + \frac{U}{16 \Delta^2} \int_{-D}^D \frac{dk}{2\pi}\frac{dw}{2\pi}\ \left[\fermifn_1(k) + \fermifn_2(k) \right]\left[\fermifn_1(w) - \fermifn_2(w) \right]  \sum_a |\mathcal{T}_{\bar{a}}(k)|^2 |\mathcal{T}_a(w)|^2 \text{Re}\left[\mathcal{T}_a(w) \right]. 
\end{multline}
This agrees with Eq. \eqref{eq: current small U AIM} from the main text once we take the dot energy to be spin independent [$\epsilon_a = \epsilon$, hence $\mathcal{T}_a(k) = \mathcal{T}(k)$].  As stated in the main text, we can also obtain the answer with spin dependence using the wavefunction method.

\end{widetext}

\bibliography{references}    

\end{document}